\documentclass[aps, twocolumn, reprint, superscriptaddress, longbibliography]{revtex4-1}
\usepackage[utf8]{inputenc}
\usepackage{amsmath}
\usepackage{amssymb}
\usepackage{hyperref}
\usepackage{graphicx}
\usepackage{epstopdf}
\usepackage{dcolumn}
\usepackage{booktabs}
\usepackage{bm}
\usepackage{xspace}
\usepackage{verbatim}
\usepackage[table]{xcolor}

\usepackage{tabularx}
\usepackage{enumitem}
\usepackage{braket,dsfont,soul}
\usepackage{array}
\usepackage{multirow}
\setlist{nolistsep}

\definecolor{mygray}{gray}{0.75}
\definecolor{lightBlue}{HTML}{e5f7ff}

\usepackage{color}

\newcommand{\mt}[1]{\textcolor{black}{#1}}
\newcommand{\vm}[1]{\textcolor{black}{#1}}
\newcommand{\RY}[1]{\textcolor{black}{#1}}

\DeclareMathOperator{\Tr}{Tr}
\hypersetup{colorlinks=true,linkcolor=blue,citecolor=blue, filecolor=blue,urlcolor=blue,breaklinks=true}

\AtBeginDocument{%
    \newwrite\bibnotes
    \def\bibnotesext{Notes.bib}
    \immediate\openout\bibnotes=\jobname\bibnotesext
    \immediate\write\bibnotes{@CONTROL{REVTEX41Control}}
    \immediate\write\bibnotes{@CONTROL{%
    apsrev41Control,author="08",editor="1",pages="1",title="0",year="1"}}
     \if@filesw
     \immediate\write\@auxout{\string\citation{apsrev41Control}}%
    \fi
}%

\begin{document}

\title{Review: Quantum Metrology and Sensing with Many-Body Systems}

\author{Victor Montenegro}
\affiliation{Institute of Fundamental and Frontier Sciences, University of Electronic Science and Technology of China, Chengdu 611731, China}
\affiliation{Key Laboratory of Quantum Physics and Photonic Quantum Information, Ministry of Education, University of Electronic Science and Technology of China, Chengdu 611731, China}
\affiliation{College of Computing and Mathematical Sciences, Department of Applied Mathematics and Sciences, Khalifa University, 127788 Abu Dhabi, United Arab Emirates}

\author{Chiranjib Mukhopadhyay}
\affiliation{Institute of Fundamental and Frontier Sciences, University of Electronic Science and Technology of China, Chengdu 611731, China}
\affiliation{Key Laboratory of Quantum Physics and Photonic Quantum Information, Ministry of Education, University of Electronic Science and Technology of China, Chengdu 611731, China}

\author{Rozhin Yousefjani}
\affiliation{Institute of Fundamental and Frontier Sciences, University of Electronic Science and Technology of China, Chengdu 611731, China}
\affiliation{Key Laboratory of Quantum Physics and Photonic Quantum Information, Ministry of Education, University of Electronic Science and Technology of China, Chengdu 611731, China}
\affiliation{Qatar Center for Quantum Computing, College of Science and Engineering, Hamad Bin Khalifa University, Doha, Qatar}

\author{Saubhik Sarkar}
\affiliation{Institute of Fundamental and Frontier Sciences, University of Electronic Science and Technology of China, Chengdu 611731, China}
\affiliation{Key Laboratory of Quantum Physics and Photonic Quantum Information, Ministry of Education, University of Electronic Science and Technology of China, Chengdu 611731, China}

\author{Utkarsh Mishra}
\affiliation{Department Of Physics and Astrophysics, University of Delhi, Delhi 110007, India}

\author{Matteo G. A. Paris}
\affiliation{Universita di Milano, I-20133 Milano, Italy}

\author{Abolfazl Bayat}
\email{abolfazl.bayat@uestc.edu.cn}
\affiliation{Institute of Fundamental and Frontier Sciences, University of Electronic Science and Technology of China, Chengdu 611731, China}
\affiliation{Key Laboratory of Quantum Physics and Photonic Quantum Information, Ministry of Education, University of Electronic Science and Technology of China, Chengdu 611731, China}
\affiliation{Shimmer Center, Tianfu Jiangxi Laboratory, Chengdu 641419, China}

\begin{abstract}
Quantum systems, fabricated across various spatial scales from nano to micrometers, are very delicate and naturally sensitive to the variations of their environment. These features make them excellent candidates for serving as sensors with wide range of applications. 
\vm{Indeed, the exceptional precision of quantum sensors arises from their compact size and inherent sensitivity, enabling measurements with unprecedented accuracy within highly localized regions. A key advantage of quantum sensors lies in their resource efficiency, as their achievable precision can scale super-linearly with respect to resources, such as system size, in contrast to the linear scaling characteristic of classical sensors. This phenomenon, commonly referred to as quantum-enhanced sensitivity, fundamentally depends on exploiting uniquely quantum mechanical features, including superposition, entanglement, and squeezing.}
Originally, quantum sensing was formulated for particles prepared in a special form of entangled states. {\color{black} Yet, certain realization of these probes may be susceptible to decoherence and interaction between particles may also be detrimental to their performance.} An alternative framework for quantum sensing has been developed through exploiting quantum many-body systems, where the interaction between particles plays a crucial role. In this review, we investigate different aspects of the latter approach for quantum metrology and sensing. Many-body probes have been used for sensing purposes in both equilibrium and non-equilibrium scenarios. 
{\color{black} Quantum criticality, as a well-studied subject in many-body physics, has been identified as a resource for achieving quantum-enhanced sensitivity in both of these scenarios.} In equilibrium, various types of criticalities, such as first order, second order, topological, and localization phase transitions have been exploited for sensing purposes.
In non-equilibrium scenarios, quantum-enhanced sensitivity has been discovered for Floquet, dissipative, and time crystal phase transitions.  
While each type of these criticalities, either in equilibrium or non-equilibrium scenarios, has its own characteristics, the presence of one feature is crucial for achieving quantum-enhanced sensitivity and that is energy/quasi-energy gap closing. In non-equilibrium quantum sensing, time becomes another parameter which can affect the sensitivity of the probe. Typically, the sensitivity enhances as the probe evolves in time. 
In this review, we provide an overview on recent progresses on different aspects of quantum metrology and sensing with many-body systems. 
\end{abstract}

\maketitle

\tableofcontents

\section{Introduction}

The rapidly emerging quantum technologies are expected to revolutionize our lives in the coming decades~\cite{dowling2003quantum}. In a broad sense, quantum technologies are developed in three directions, namely  quantum simulation and computation~\cite{lloyd1996universal,georgescu2014quantum}, quantum communication~\cite{zeilinger1999experiment,gisin2007quantum}, and quantum sensing~\cite{degen2017quantum}. Thanks to advancements in fabricating ultra-precise quantum devices, the progress in all these three aspects have  significantly been accelerated. Quantum sensing~\cite{degen2017quantum} is one of the pillars of quantum technologies which allow for developing a new generation of sensors for detecting gravitational, magnetic and electric fields as well as estimating physical quantities with unprecedented precision, well beyond the capability of conventional classical sensors. The application prospect of quantum sensors is very wide and includes mining~\cite{crawford2021quantum}, environmental monitoring ~\cite{lanzagorta2015quantum}, gravimetry~\cite{qvarfort2018gravimetry, qvarfort2021optimal, armata2017quantum, montenegro2025heisenberg, szigeti2020highprecision, kritsotakis2018optimal, gietka2019supersolid}, biological imaging~\cite{wu2016diamond,taylor2016quantum,aslam2023quantum}, space exploration~\cite{kaltenbaek2021quantum}, radar and lidar technology~\cite{maccone2020quantum,karsa2024quantum,Casariego2023propagating, reichert2024heisenberg}, searching for fundamental particles~\cite{brady2022entangled, brady2023entanglement,shi2023ultimate,backes2021aquantum,kang2024nearquantum,wang2022limitsonaxions,Jiang2021search} --- the list goes on. There are several reasons that make quantum sensors superior over their classical counterparts. {\color{black} First, the even smaller sizes of quantum sensors, which can be fabricated at the atomic scales, allows for measuring environmental parameters within an extremely localized spatial extensions, such as biological cells~\cite{Schirhagl2014nitrogenvacancy,zhang2021towards}.} Second, the natural delicacy of quantum superposition allows quantum sensors to detect very weak signals~\cite{hong2024femtotesla,Jiang2021searchfor}. {\color{black} Third, quantum features, such as superposition, entanglement and squeezing, can be harnessed by quantum probes to achieve quantum-enhanced precision outperforming their classical counterparts using the same amount of resources~\cite{giovannetti2004quantum,Giovannetti2011advances,Giovannetti2006quantummetrology}. Fourth, quantum sensors have been developed on a wide variety of physical platforms, showing their flexibility in fabrication, control and application. Such platforms includes ultra-cold atoms~\cite{kasevich1992measurement, peters1999measurement, bongs2019taking, el2020aedge, stray2022quantum, peters2001high, fixler2007atom, tino2021testing, bloom2014optical, hinkley2013atomic, takamoto2005optical}, ion traps~\cite{leibfried2004toward, maiwald2009stylus, biercuk2010ultrasensitive, sawyer2012spectroscopy, brownnutt2015ion, baumgart2016ultrasensitive, baumgart2016ultrasensitive, huntemann2016single}, atomic vapors~\cite{budker2007optical, kominis2003subfemtotesla, dang2010ultrahigh, balabas2010polarized, shah2007subpicotesla, fernholz2008spin, wasilewski2010quantum, xia2006magnetoencephalography}, nuclear magnetic resonance systems~\cite{kitching2011atomic, fang2012advances,Jiang2021search, wang2022limitsonaxions, wu2022enhanced, jiang2024long}, solid state defects in diamond~\cite{taylor2008high, clevenson2015broadband, jensen2014cavity, le2012efficient, chernobrod2005spin, balasubramanian2008nanoscale, dolde2011electric, zhao2012sensing, barson2017nanomechanical, wu2021nanodiamond, vetter2022zero, zhou2014quantum, xie2020dissipative, schirhagl2014nitrogen,patel2020subnanotesla}, superconducting circuits~\cite{bal2012ultrasensitive, danilin2018quantum, wang2019heisenberg, yu2025experimental}, photonic setups~\cite{holland1993interferometric, mitchell2004super, pezze2007phase, higgins2007entanglement, nagata2007beating, ono2013entanglement, Xiao2024Non}, and optomechanical devices~\cite{hu2024picotesla, gavartin2012ahybrid, liu2022roomtemperature, westerveld2021sensitive, sansa2020optomechanical,  li2018characterization, fardianmelamed2025infrared,  fogliano2021ultrasensitive, pikovski2012probing, chowdhury2023membrane, gosling2024sensing, forstner2012cavity, mccormick2019quantum, aspelmeyer2014cavity, moser2013ultrasensitive, guzman2014high, krause2012high, chaste2012nanomechanical, forstner2014ultrasensitive, degen2009nanoscale, rugar2004single, li2021cavity, liu2021progress, xia2023entanglement}, the list goes on.
}

In order to quantify the precision of a sensor for estimating an unknown parameter $\theta$, one needs to specify a proper figure of merit. In case of maximal-likelihood estimation theory, the imprecision of the estimation can be quantified by the variance of an \emph{unbiased} max-likelihood estimator. This variance is bounded through Cram\'{e}r-Rao inequality~\cite{nla.cat-vn81100, Rao1992} by $1/MI(\theta)$, where $M$ is the number of samples and $I(\theta)$ is a quantity called Fisher information, which will be defined and discussed later. Dependence on $M$ is a direct consequence of central limit theorem, therefore the most important quantity which bounds the precision of a sensor is the Fisher information. In fact, in estimation theory, Fisher information is the conventional figure of merit for evaluating the performance of a sensor. Every sensing procedure exploits some resources such as time, particle number, system size, etc. The resource efficiency of a given sensor is determined by the scaling of Fisher information with respect to those resources. In other words, such scaling shows how the precision is enhanced by using more resources. In the absence of quantum features, the Fisher information at best scales linearly with the resource, which is known as standard quantum limit (or classical shot-noise limit). However, exploiting quantum features {\color{black} such as quantum coherence,  entanglement, and squeezing may result in super-linear scaling of the Fisher information with respect to their resource.} This is known as quantum-enhanced sensitivity and is one of the key properties that makes quantum sensors superior over their classical counterparts. 

\vm{Seminal ideas in quantum metrology can be traced back to the 1980s, when debates arose over whether quantum-mechanical radiation-pressure fluctuations disturb position measurements in an interferometry setup~\cite{caves19809quantum}, as well as quantum mechanical noise in linear amplifiers~\cite{caves1982quantum} and interferometry~\cite{caves1981quantum}. For further insights, see the theoretical review by Caves et al.~\cite{caves1980onthemeasurement}, along with an experimental progress review by Bocko et al.~\cite{bocko1996onthemeasurement}. Moreover, in 1994, Samuel L. Braunstein and Carlton M. Caves~\cite{braunstein1994statistical} formulated a more general uncertainty principle based on a Riemannian metric on the space of quantum-mechanical density operators. Later, quantum-enhanced sensitivity was proposed in an interferometric setup by V. Giovannetti, S. Lloyd, and L. Maccone~\cite{giovannetti2004quantum}.} In this paper, the authors show that a special form of entangled states, such as Greenberger–Horne–Zeilinger (GHZ) states, can enhance the precision of detecting a phase shift  quadratically with respect to the probe size. Optical setups have since been used to experimentally confirm this effect quite extensively~\citep{mitchell2004super, pezze2007phase, ono2013entanglement}. Even for matter-based platforms, the effect was tested experimentally in ion-trap systems~\cite{leibfried2004toward} and then extended to various physical platforms like superconducting qubits~\cite{wang2019heisenberg} and nitrogen vacancy centers~\cite{bonato2016optimized}. 
{\color{black} The interferometry-based quantum sensing has several advantages, including: (i) the quantum-enhanced precision can be achieved for all range of parameters; and (ii) the measurement which results in quantum-enhanced precision is fixed and does not depend on the unknown parameter. Nonetheless, this approach also has inherent limitations, as it relies on GHZ-type entangled states,} which are difficult to create and, in general, might be susceptible to decoherence and particle loss. In addition, in this strategy a very special form of unitary operation, namely phase shift, is assumed for encoding the information in the state of the probe. In fact, it has been shown that any disturbance to the considered unitary operation diminishes the precision~\cite{de2013quantum}. All these challenges may prevent the original interferometric sensing procedure to scale up for large system sizes or restrict it to special forms of interactions.  

\begin{figure}
    \centering
    \includegraphics[width=\linewidth]{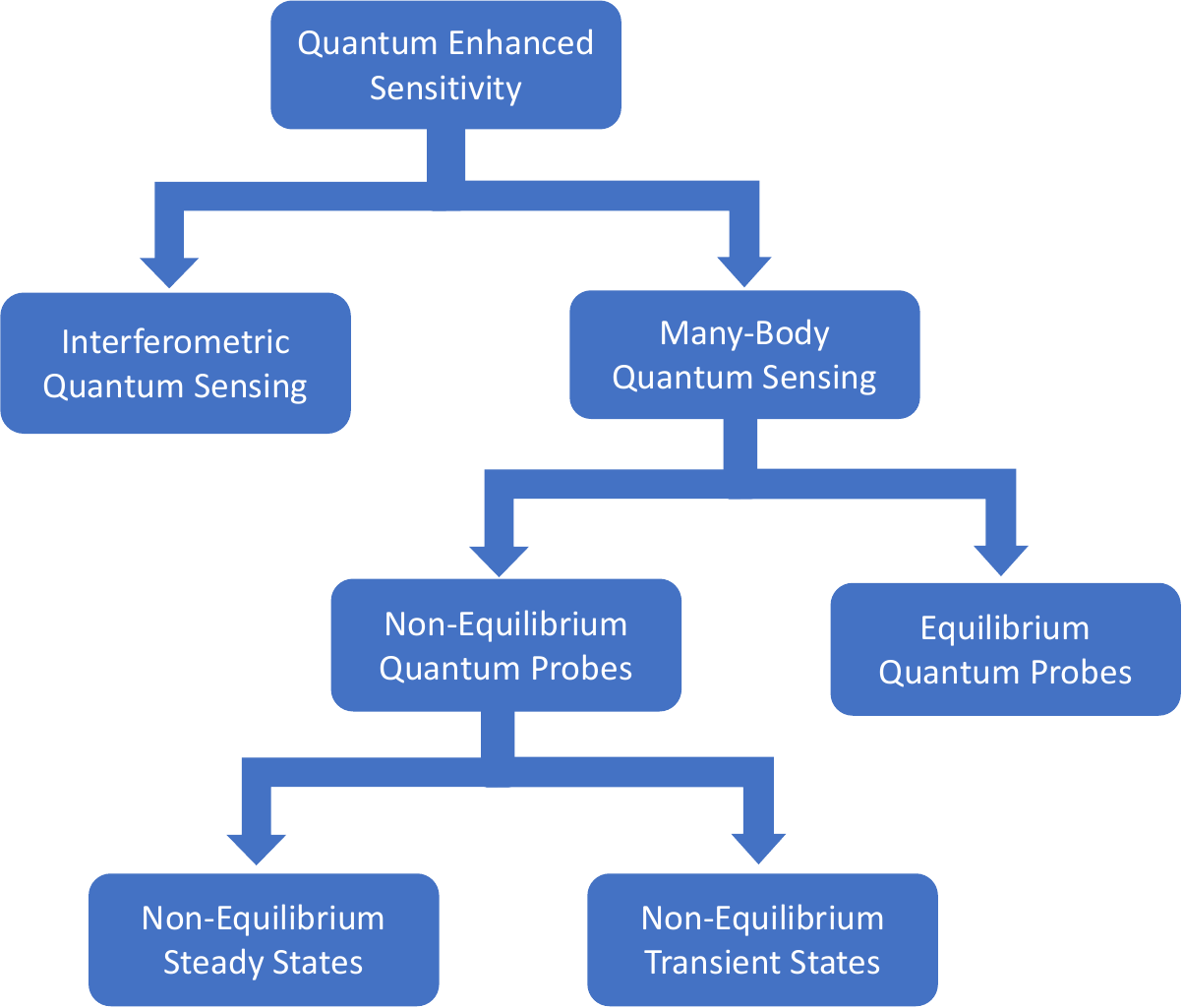}
    \caption{\textbf{Schematic for achieving quantum-enhanced sensing.} The main advantage of quantum sensors over their classical counterparts is manifested through quantum-enhanced sensitivity. Various strategies have been discovered to achieve such enhancement. Originally, interferometric quantum sensors  were proposed and realized for measuring a phase shift with quantum-enhanced precision. Later, the capacity of quantum many-body systems for achieving quantum-enhanced sensitivity was identified which is the subject of this review. Many-body systems have been used for quantum sensing in both equilibrium (e.g.~criticality in the ground state) and non-equilibrium modes. In the non-equilibrium case, we can further divide the sensors into non-equilibrium steady states and transient dynamics.}
    \label{fig:schematic_Quant_Enhanced}
\end{figure}

An alternative method to interferometric quantum sensing has emerged in many-body systems. There are several distinct features between the two methods for sensing external parameters: (i) while the interferometric sensing requires  entangled GHZ-type state preparation, in many-body sensors entanglement is either inherently present in the spectrum of the system or generated freely during the dynamics; (ii) while interaction between particles {\color{black} is usually destructive in interferometric sensing~\cite{de2013quantum}, it plays a crucial role in many-body sensors;} (iii) unlike GHZ-type states, many-body systems are more resistive against decoherence and particle loss; and (iv) in contrast to the interferometric sensors, the optimal measurement may depend on the unknown parameter which is supposed to be measured. While features (i)-(iii) show the benefits of many-body sensors over their interferometric counterparts, the feature (iv) shows that taking advantage of many-body systems for sensing is not free of challenge. 
These \textit{pros} and \textit{cons} motivate further investigations of many-body systems for developing a new generation of sensors\vm{. See, for instance, Ref.~\cite{baak2022classical} for an analytically solvable many-body problem in Hamiltonian parameter estimation, both for classical and quantum metrology}. In general, there are two different approaches to exploit many-body systems for metrology purposes with quantum-enhanced sensitivity. The first approach, which typically requires equilibrium states, such as the ground state, exploits quantum criticality for achieving quantum-enhanced sensitivity. In the second approach, however, non-equilibrium  dynamics of a many-body system is exploited to accumulate information about the parameters of interest. Further divisions can be considered for non-equilibrium probes as they might be used at non-equilibrium steady states or transient dynamics. All these methods for achieving quantum-enhanced sensitivity are summarized in Fig.~\ref{fig:schematic_Quant_Enhanced}. 
\RY{The advantages and limitations of each of these methods are comprehensively summarized in Table~\ref{table:final-summary}.
Additionally, the frequently used notations in this review article are summarized in Table~\ref{tab:notation} for convenience.}

Quantum criticality has already been identified as a resource for sensing in both equilibrium and non-equilibrium many-body systems. Originally, quantum phase transition in the transverse Ising model has been explored for metrology purposes~\cite{zanardi2006ground}. It has been then extended to various types of criticalities including, second-order~\cite{zanardi2006ground,zanardi2007mixed,gu2008fidelity,zanardi2008quantum,invernizzi2008optimal,gu2010fidelity,gammelmark2011phase,skotiniotis2015quantum,rams2018limits,wei2019fidelity,chu2021dynamic,liu2021experimental,montenegro2021global,mirkhalaf2021criticality,di2021critical,mihailescu2024multiparameter},  first-order~\cite{raghunandan2018high, mirkhalaf2018supersensitive, yang2019engineering, sarkar2024exponentially}, topological~\cite{budich2020non,sarkar2022free,koch2022quantum,yu2022experimental,Sarkar2024Critical}, 
and Stark phase transitions~\cite{he2023stark,yousefjani2023longrange,yousefjani2025nonlinearity,sarkar2024noisy}. 
Apart from these distinct types of criticalities at equilibrium, non-equilibrium quantum phase transitions can also demonstrate quantum-enhanced sensitivity. This includes dissipative~\cite{fernandez2017quantum,baumann2010dicke,baden2014realization,klinder2015dynamical,rodriguez2017probing,fitzpatrick2017observation,fink2017observation,ilias2022criticality,Ilias2023Criticality,Alipour2014Quantum}, 
Floquet~\cite{mishra2021driving,mishra2022integrable} and  time crystal~\cite{montenegro2023quantum,Cabot2024Continuous,lyu2020eternal,Iemini2023,yousefjani2024discrete} phase transitions.
Each type of quantum criticality comes with its own features and characteristics. For instance, second-order quantum phase transitions take place in the ground state of a many-body system and are accompanied by spontaneous symmetry breaking and long-range correlations which are described by Landau-Ginzburg formalism~\cite{sachdev1999quantum}. On the other hand, topological phase transitions neither show symmetry breaking nor follow the Landau-Ginzburg theorem~\cite{xu2012unconventional} and Stark localization takes place across the entire spectrum and not just the ground state~\cite{van2019bloch,schulz2019stark}. While different types of quantum criticalities have their own characteristics, one feature is common among all of them for showing quantum-enhanced sensitivity which is energy/quasi-energy \emph{gap closing}. This is an interesting observation which guides us in searching for finding new potential quantum sensors with the capability of achieving quantum-enhanced precision. Interestingly, the realization of criticality-based quantum sensors are becoming viable in various physical platforms, including nuclear magnetic resonance systems~\cite{liu2021experimental},Rydberg atoms~\cite{ding2022enhanced} and superconducting devices~\cite{yu2025experimental}.

\begin{table*}[t]
\begin{tabular}{|cll|}

\hline
\multicolumn{1}{|c|}{\cellcolor[HTML]{CBCEFB}{\color[HTML]{000000} \textbf{Sensing Strategy}}} 
& \multicolumn{1}{c|}{\cellcolor[HTML]{C3EBC3}\textbf{Advantages}}   
& \multicolumn{1}{c|}{\cellcolor[HTML]{FFCCC9}\textbf{Limitations}} 
\\ \hline

\multicolumn{3}{|c|}{\cellcolor[HTML]{C0C0C0} \textbf{Interferometric Quantum Sensors}}    
\\ \hline

\multicolumn{1}{|c|}{\textbf{\begin{tabular}[c]{@{}c@{}}
External parameter is \\ induced by phase shift \\ $\hat{U}(\theta)=e^{i\theta \hat{G}}$
\end{tabular}}} 

& \multicolumn{1}{l|}{\begin{tabular}[c]{@{}l@{}}
1. Heisenberg scaling possible
\\ 2. Universal optimal measurement 
\end{tabular}} 

& \begin{tabular}[c]{@{}l@{}}
1. Difficult preparation 
\\ 2. Potentially sensitive to decoherence \\
\noindent and particle loss
\\ 3. Works only for phase shift operations
\end{tabular}
\\ \hline

\multicolumn{3}{|c|}{\cellcolor[HTML]{C0C0C0}\textbf{Many-Body Quantum Sensors at Equilibrium}}   
\\ \hline

\multicolumn{1}{|c|}{\textbf{\begin{tabular}[c]{@{}c@{}}
Ground or Thermal \\ State based for Hamiltonians \\ of the form $\hat{H}{=}\hat{H}_0{+} \hat{H}_1(\theta)$ 
\end{tabular}}}  
  
& \multicolumn{1}{l|}{\begin{tabular}[c]{@{}l@{}}
1. Quantum-enhanced sensitivity achievable\\ by gap-closing phase transitions 
\\2. Quantum-enhanced sensitivity possible
\\ 3. Robust against imperfections
\end{tabular}} 

& \begin{tabular}[c]{@{}l@{}}
1. Fine tuning is needed (local sensing)
\\ 2. Optimal measurement parameter-value \\ dependent
\\ 3. Preparation might be resource consuming
\end{tabular}    
\\ \hline

\multicolumn{3}{|c|}{\cellcolor[HTML]{C0C0C0}\textbf{Non-Equilibrium Many-Body Quantum Sensors}} 
\\ \hline

\multicolumn{1}{|c|}{\textbf{\begin{tabular}[c]{@{}c@{}}
Non-Equilibrium Steady \\ State based
\end{tabular}}}    

& \multicolumn{1}{l|}{\begin{tabular}[c]{@{}l@{}}
1. Initialization does not matter
\\ 2. Quantum-enhanced sensitivity possible
\\ 3. Works even with partial accessibility  
\\ 4. Dissipation can also be harnessed
\end{tabular}} 
    
& \begin{tabular} [c]{@{}l@{}}
1. Fine tuning is needed (local sensing)
\\ 2. Optimal measurement parameter-value
\\ dependent
\\ 3. Reaching steady state is time consuming
\end{tabular}    
\\ \hline

\multicolumn{1}{|c|}{\textbf{\begin{tabular}[c]{@{}c@{}}
Transient State based
\end{tabular}}}  
  
& \multicolumn{1}{l|}{\begin{tabular}[c]{@{}l@{}}
1. Easy initialization \\ 2. Quantum-enhanced sensitivity possible\\
3. Time control for better precision  \\
4. May operate across the entire phase \\
(no fine tuning)\\
5. Sequential measurements are adaptable
\end{tabular}}  
   
& \begin{tabular}[c]{@{}l@{}}
1. Decoherence is a limiting factor
\\ 2. Optimal measurement parameter-value\\ dependent
\end{tabular}    
\\ \hline

\end{tabular}
\caption{\textbf{Quantum-enhanced sensing strategies}. A comparison between different means of quantum sensing is provided in this Table. This review is mainly focused on quantum many-body probes at equilibrium and non-equilibrium configurations.  }
\label{table:final-summary}
\end{table*}


\begin{table*}[t]   
   \small
   \centering
   \begin{tabular}{l|l}
   \toprule\toprule
   \textbf{Notation} & \textbf{Meaning}  \\ 
   \hline
   \midrule
   $I(\theta)$ & Classical Fisher information (or just Fisher information) with respect to unknown parameter $\theta$.\\
   $I_Q(\theta)$ & Quantum Fisher information with respect to unknown parameter $\theta$.\\
   $\tilde{\Theta}$ & Statistical estimator corresponding to unknown parameter $\theta$. \\
   $\partial_\theta$ & Partial derivative with respect to unknown parameter $\theta$. \\
   $F(\rho_1,\rho_2)$ & Quantum fidelity between two density matrices $\rho_1$ and $\rho_2$, given by $F(\rho_1,\rho_2)= \sqrt{\text{Tr}\sqrt{\sqrt{\rho_1}\rho_2 \sqrt{\rho_1}}}$.\\
   $\mathrm{Var}[\cdot]$ & Variance of a random variable. \\
   $\hat{O}$ & Any quantum mechanical operator.\\
   $\boldsymbol{\theta}$ & Vector of multiple parameters.\\
   $N_{\rm ex}$ & Number of particles (or excitations) in a many-body system.\\
   $L$ and $N$ & Depending on the context, both have been used for the size of a many-body system. \\
   $\hat{\sigma}_{\cdot}$ & $2\times2$ Pauli matrices. \\
   $n_\mathrm{seq}$ & Number of consecutive steps in sequential measurement sensing protocols. \\
   $T$ & Temperature.\\
   $B$ and $h$ & Magnetic fields. \\
   \hline
   \bottomrule
   \end{tabular}
   \caption{\vm{Notation used throughout the review, unless specified otherwise.}}
   \label{tab:notation}
\end{table*}

Quantum criticality is in general a property of equilibrium states such as the Gibbs thermal state (which becomes the ground state at zero temperature limit)~\cite{sachdev1999quantum,vojta2003quantum}. Preparing such equilibrium states can be challenging in practice, demanding ultra-low temperatures or extremely long preparation times. 
{\color{black} One can in fact analyze the ratio of the Fisher information and the time cost to discern the practical quantum advantage. 
The time-dependent bound on the Fisher information~\cite{boixo2007generalized} provides valuable insights on the prediction of such advantage.
As will be discussed later within a broader resource analysis segment, this leads to outcomes specific to the critical system.}
In addition, criticality-enhanced quantum precision is only achievable around the phase transition point. This requires fine tuning and thus is mostly beneficial in the context of local sensing, where significant prior information about the parameter of interest is available. 
Beyond criticalities, the sensing capability of many-body systems have also been investigated in their strongly correlated 
phases, even in the presence of thermal fluctuations~\cite{mihailescu2023thermometry,mihailescu2024quantum,mihailescu2024multiparameter}.
Moreover, as an alternative approach, one can also use non-equilibrium dynamics of many-body systems for sensing unknown parameters. These setups are more experimental friendly as their initial state can be a simple product state and entanglement is generated naturally during the dynamics. Non-equilibrium dynamics in many-body systems can be induced through different methods, including quantum quench~\cite{mitra2018quantum,daley2012measuring,calabrese2006time,calabrese2011quantum,shi2024universal, gietka2022understanding}, measurement quench~\cite{pouyandeh2014measurement,bayat2018measurement,skinner2019measurement,li2018quantum}, external driving~\cite{chu2004beyond,gritsev2017integrable,oka2019floquet} and adiabatic evolution~\cite{mihailescu2024quantum}. Apart from the probe size, in non-equilibrium quantum sensors, time is also a key resource which plays an important role and has to be included in the definition of standard quantum limit and quantum-enhanced sensitivity. In general, Fisher information scales as  $I\sim t^\alpha L^\beta $, where $t$ is time, $L$ is the system size and $\alpha$ and $\beta$ are two exponents~\cite{ilias2022criticality}. In non-equilibrium sensors, the standard quantum limit is defined as $\alpha{=}\beta{=}1$. Any situation with $\alpha>1$ or $\beta>1$ is called quantum-enhanced sensitivity with the special case of Heisenberg limit, defined as $\alpha{=}\beta{=}2$. In certain systems in which the unitary dynamics preserves the number of excitations, the Fisher information may scale with the number of excitations too. In fact, in Ref.~\cite{manshouri2024quantum} it was shown that in excitation-preserving dynamics Fisher information scales as $I\sim t^\alpha L^\beta N_{\rm ex}^\gamma$, where  $N_{\rm ex}$ is the number of excitations and $\gamma$ is its corresponding exponent. This is an interesting observation which shows how the sensitivity increases from single excitation subspace, i.e.~$N_{\rm ex}{=}1$, to half filling, i.e.~$N_{\rm ex}{=}L/2$. In addition, Ref.~\cite{gietka2023squeezing} proposes using spin-orbit coupled quantum gases which they claim can be utilized to overcome the Heisenberg limit, and Ref.~\cite{gietka2022understanding} demonstrates that Fisher information can exhibit exponential scaling as a physical effect by quenching dynamics far beyond the critical point. Unlike criticality-based quantum sensors, the criteria for achieving quantum-enhanced precision is not yet well-characterized in non-equilibrium quantum many-body probes.  For instance, we know that localized phases are not good for sensing~\cite{he2023stark,yousefjani2023longrange,yousefjani2025nonlinearity,sahoo2024localization} but what features in extended phases are crucial for quantum-enhanced sensitivity are not yet identified. 

In practice, no quantum device is perfectly isolated from its environment and thus decoherence is an inevitable part of any quantum protocol. Consequently, any quantum dynamics is actually an open system evolution. Different formalisms have been developed for addressing open system dynamics, among them are: (i) master equation~\cite{Carmichael1993An}; and (ii) non-Hermitian Hamiltonians~\cite{Rotter2009A}. In the former, the evolution of the system and its environment is described by a unitary operation. By tracing out the environment, the evolution of the system, typically within Born-Markov approximations, is thus characterized by the Lindblad (or Gorini-Kossakowski-Sudarshan-Lindblad) master equation~\cite{breuer2002theory}. In this formalism a quantum system initially prepared in a pure state may indeed evolve to a mixed state due to the action of jump operators --- a signature of information leakage (irreversible process) to the environment. In non-Hermitian formalism, the environment is monitored for tracking a trajectory in which no jump operator action takes place~\cite{Plenio1998The, Daley2014Quantum}. As a result, an initially pure quantum state remains pure throughout the dynamics, despite being evolved effectively as an open quantum system. Other mechanisms that address or profit from dissipation include decoherence-free subspaces~\cite{lidar1998decoherence, PhysRevLett.79.1953},  by coupling the quantum resources to a common environment that can be measured at least in part~\cite{braun2011heisenberg} and coherent averaging~\citep{fraisse2015coherent} more generally leading to quantum-enhanced measurements without entanglement~\citep{braun2018quantum}, dynamical decoupling~\cite{viola1999dynamical, khodjasteh2005fault}, and quantum reservoir engineering~\cite{PhysRevLett.77.4728}. The latter technique involves adding an engineered Lindbladian term to the open system dynamics. This approach can be designed to steer the system into a desired pure steady-state, effectively \textit{cooling} the system to that state~\cite{myatt2000decoherence, pielawa2007generation, pielawa2010engineering, groszkowski2022reservoir}. Both master equation and non-Hermitian formalisms have been investigated for quantum sensing. Indeed, a general joint (system plus environment) quantum Fisher information upper bound has been formulated~\cite{escher2011general}. Hence, one would expect that such open quantum system dynamics may also achieve quantum-enhanced sensitivity provided that the system goes through a phase transition, which is again accompanied by an eigenvalue gap closing. In the master equation formalism, the phase transition is determined for the steady state of the system at the point that the eigenvalues of the Liouvillian operator shows gap closing~\cite{minganti2018spectral}. An example of such systems with quantum-enhanced sensitivity is boundary time crystals~\cite{montenegro2023quantum}. In the non-Hermitian systems, quantum-enhanced sensitivity has been shown to emerge at the exceptional points~\cite{wiersig2020review}, where two or more eigenvalues and corresponding eigenstates coalesce, as well as in the cases of different types of gap closing in the complex energy plane~\cite{Sarkar2024Critical}.  

In this review, we explore the capacity of many-body systems for serving as quantum sensors in both equilibrium and non-equilibrium regimes. Our paper is complementary to a few previous review papers on different aspects of quantum sensing. In Ref.~\cite{paris2009quantum}, the theoretical aspects of quantum estimation theory are comprehensively discussed.  On the other hand, Ref.~\cite{degen2017quantum} mostly focuses on experimental implementation of quantum sensors. More recent reviews have been dedicated for exploring Fisher information and its properties~\cite{meyer2021fisher,liu2020quantum} and entanglement-enhanced metrology~\cite{huang2024entanglement}. In addition, recently an essay has also been dedicated to atomic, molecular, and optical platforms for realizing quantum sensors~\cite{ye2024essay}. While quantum many-body sensors have received significant attention in recent years, these aforementioned recent progresses have not been covered in the previous review articles. This makes it timely to provide a review on the progress towards quantum many-body sensors.

\section{Elements of Parameter Estimation}\label{sec:elements}

Precise measurements of relevant quantities are essential in science.
In physics, high-precision measurements lead us to test the theory of special relativity~\cite{Einstein1923-EINTPO, einstein-german, Michelson333, PhysRev.59.223, Hafele-1, Hafele-2}, scrutinize the theory of general relativity via gravitational wave detection~\cite{Einstein1923-EINTPO, PhysRevLett.116.061102}, and validate the Standard Model of physics by confirming the existence of the elusive Higgs particle~\cite{PhysRevLett.13.321, PhysRevLett.13.508, PhysRevLett.13.585, TheCMSCollaboration2022}. Thus, pursuing novel schemes and strategies for achieving high-precision measurements are highly desirable for advancing science. However, measuring a relevant quantity is only sometimes a straightforward procedure. 
Furthermore, quantities such as coupling strengths, temperature, and in general, those lacking corresponding quantum observables, such as entanglement~\cite{genoni2008optimal,breda2010experimental} or purity~\cite{brunelli2012estimation}, can only be determined indirectly. Therefore, one requires to infer a relevant quantity of the system by indirect means. The art of inferring such a relevant quantity by means of measuring another is known as parameter estimation~\cite{sidhu2020geometric, d2000parameter, paris2009quantum}\vm{---see recent studies on the connection between quantum speed limits and quantum parameter estimation \cite{maleki2023speed, herb2024quantum, delcampo2013quantum}. For a comprehensive review of this relationship, see Ref.~\cite{deffner2017quantum}.}

\begin{figure}[b]
    \centering
    \includegraphics[width=\linewidth]{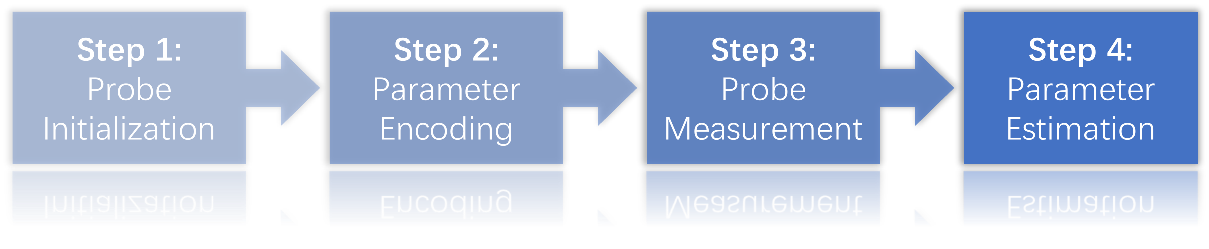}
    \caption{\textbf{Sketch of parameter estimation steps:} Standard scheme for estimating an unknown parameter $\theta$. An initial quantum state $\rho$ dynamically encodes the parameter $\theta$ to be estimated as $\rho(\theta)$. To extract information about $\theta$, a set of measurements via positive operator-valued measure (POVM) $\{\hat{\Pi}_x\}$, are performed on the probe with random outcomes $x$. The resulting measurement outcomes are fed into a classical estimator $\tilde{\Theta}$ to infer the unknown parameter $\theta$.}
    \label{fig_parameter_estimation_steps}
\end{figure}

The central aim of parameter estimation theory is to infer the unknown parameters of interest as precisely as possible. Any parameter estimation scheme is composed of four basic steps, see Fig.~\ref{fig_parameter_estimation_steps}:
\begin{enumerate}
\item Choosing an adequate probe,
\item Encoding the unknown parameters dynamically into the probe,
\item Extracting information about the parameters of interest by performing measurements onto the probe,
\item Inferring the true (unknown) parameters via a classical estimator.
\end{enumerate}

All the above steps need optimization to successfully estimate the parameters. It is important to note that the actual values of the parameters are always unknown by definition, and one must infer them indirectly. One common methodology to quantify the uncertainty in estimating a single unknown parameter $\theta$ is accessing the system of interest through measurements with observable $\hat{O}$, which leads to estimating the error of $\theta$ using the error propagation formula~\cite{sidhu2020geometric}:
\begin{equation}
\delta \theta = \frac{\Delta \hat{O}}{\left|\partial_\theta\langle \hat{O}\rangle\right|}, \hspace{1cm} \frac{\partial}{\partial \theta} := \partial_\theta \label{eq:error-propagation}
\end{equation}
where $\Delta \hat{O}$ and $\langle \hat{O} \rangle$ are the standard deviation and the expectation value of the outcomes associated with $\hat{O}$, respectively. The denominator in Eq.~\eqref{eq:error-propagation} indicates that in order to decrease the error in the estimation of $\theta$ (i.e., $\delta \theta \ll 1$), one needs to increase the denominator, i.e., $\left|\partial_\theta\langle \hat{O}\rangle\right| \gg 1$. To achieve this, it is necessary to find a set of outcomes with an expectation value $\langle \hat{O}\rangle$ that varies significantly with respect to $\theta$. Conversely, a set of outcomes that varies weakly with $\theta$ (i.e., flat slope $|\partial_\theta\langle \hat{O}\rangle| \ll 1$) will result in a substantial error (i.e., $\delta \theta \gg 1$). While Eq.~\eqref{eq:error-propagation} offers a heuristic explanation of a necessary ingredient for estimating a parameter, a more rigorous and operational framework must be provided to determine a formal figure of merit for quantifying estimation performance.
In fact,  Eq.~\eqref{eq:error-propagation} does not take into account the entire probability distribution of the outcomes and cannot provide the framework to determine a proper figure of merit for quantifying the performance of an estimation strategy.

\subsection{Global and local estimation theory}
A suitable strategy to estimate a parameter usually depends on the amount of {\em a priori} information one has on the parameter itself. If no {\em a priori} information is available, the estimation strategy, i.e.~the choice of the measure to be performed on the probe and the procedure employed to process data, should be {\em optimal in average} i.e.~should be equally good for any value of the parameter. 
The set of techniques and methods used to choose and optimize the figure of merit in this case is known as {\em \vm{global estimation} theory}~\cite{helstrom1976quantum,holevo2011probabilistic}\vm{, see Ref.~\cite{mukhopadhyay2025current} for a recent review on global estimation theory.} Once a \vm{global estimation} has been performed an experimenter is left with some {\em a priori } information to be improved, i.e.~one should look for figures of merits that select the best estimation strategy for a specific value or range of values of a parameter. This optimization problem is addressed by {\em local estimation theory}~\cite{Casella98,vantrees2004detection,helstrom1976quantum} which allows one to achieve the ultimate bound on precision.  The quantum version, namely the local quantum estimation theory~\cite{braunstein1994statistical,paris2009quantum} is the fundamental tool of quantum probing, sensing and metrology and is the main subject of the present review, with the exception of Section \ref{sec:global-sensing}, where various aspects of \vm{global estimation} theory are discussed.

\subsection{Single-parameter classical estimation bound}

Cram\'{e}r and Rao~\cite{nla.cat-vn81100, Rao1992} first introduced such operational and formal framework for quantifying the performance in estimating a parameter, with the variance with respect to such parameter as the natural figure of merit. For the single-parameter case, where $\theta$ is the only unknown parameter to be determined, one gets the Cram\'{e}r-Rao inequality (for \textit{unbiased} estimators) as:
\begin{equation}
\text{Var}[\tilde{\Theta}] \geq \frac{1}{M I(\theta)},\label{eq:cramer-rao-bound-single}
\end{equation}
where $M$ is the number of measurements (trials), $\tilde{\Theta}:=\tilde{\Theta}(x)$ is known as the \textit{estimator} of $\theta$, with $x$ being a set of measurement outcomes (discrete or continuous). Note that the estimator $\tilde{\Theta}$ is a function that only depends on the measurement outcomes $x$, and it maps the set of measurement outcomes to a parameter space with an estimated value of $\theta$. The expression $I(\theta)$ is known as Fisher information (FI), denoted as:
\begin{equation}
I(\theta) = \int \frac{1}{p(x|\theta)} (\partial_\theta p(x|\theta))^2 dx,\label{eq:cfi}
\end{equation}
for a continuous set of measurement outcomes, or $I(\theta) = \sum_i p(x_i|\theta)^{-1} (\partial_\theta p(x_i|\theta))^2$ for the discrete case. Here $p(x|\theta)$ is the conditional probability of obtaining outcome $x$ given $\theta$. Equation \eqref{eq:cramer-rao-bound-single} sets the precision one can achieve with a particular measurement for a single trial. The equality is reached by using an optimal estimator. 

The Fisher information $I(\theta)$ quantifies the sensitivity of the conditional probabilities with respect to the parameter to be estimated. To enhance sensing precision, it is necessary to increase $I(\theta)$ by identifying conditional probability distributions that exhibit more significant changes as the unknown parameter varies. This results in $\partial_\theta p(x|\theta) \gg 1$, leading to $I(\theta)\gg 1$. The question of how to identify such advantageous scenarios is addressed in the following section.

\subsection{Optimizing the classical bound: the QFI}

The Fisher information defined in Eq.~\eqref{eq:cfi} is known as the classical Fisher information (CFI) and is unique for a given measurement basis~\cite{Morozova1991, meyer2021fisher}. Probability distributions that significantly vary with respect to the unknown parameter are preferred, as they yield larger values of CFI and, consequently, better parameter sensitivity. Thus, finding such probability distributions is of paramount importance. Classically, searching for them is an arduous task, as optimization needs to be performed over all possible measurement bases. 
Nonetheless, this optimization task can be achieved for quantum probes, leading to a tighter bound on the variance of the estimator~\cite{HELSTROM1967101, paris2009quantum, 1055103, 1055173, braunstein1994statistical}. 
Consider a quantum probe $\rho$ encoding the unknown parameter $\theta$ as $\rho(\theta)$ and a set of general quantum measurements $\{\hat{\Pi}_x\}$. $\hat{\Pi}_x$ is a positive operator valued measure (POVM) with outcome $x$, such that $\int_x\hat{\Pi}_xdx=\mathbb{I}$. \vm{Note that we have not specified a particular encoding for the parameter. In fact, throughout this review, while we will encounter several cases where the encoding is governed by a unitary transformation, any encoding that ensures $\rho(\theta)$ remains a valid quantum state is permissible. This includes scenarios where the unknown parameter is not a Hamiltonian parameter, as investigated in thermometry~\cite{mehboudi2019thermometry, mehboudi2022fundamental, yang2024sequential, montenegro2020mechanical, zhang2022nonmarkovian, ullah2023lowtemperature, sone2018quantifying}, or more generally, when quantum probes are in thermal equilibrium~\cite{abiuso2025fundamental}. It also extends to cases such as the estimation of the spectral density of a quantum reservoir~\cite{wei2021nonmarkovian} or situations involving unknown parameters in non-unitary dynamics~\cite{yang2023extractable, wei2021threshold}.}

The conditional probability of obtaining the outcome $x$ for a given $\theta$ obeys the Born rule:
\begin{equation}
p(x|\theta) = \text{Tr}[\hat{\Pi}_x \rho(\theta)].
\end{equation}
Assuming regularity conditions, and since the trace and the derivative are linear operators, the derivative of the probability distribution $\partial_\theta p(x|\theta)$ translates into taking the derivative of the quantum state $\partial_\theta \rho(\theta)$, specifically $\partial_\theta p(x|\theta) = \text{Tr}[\hat{\Pi}_x \partial_\theta \rho(\theta)]$. Here, the derivative of the quantum state satisfies the Lyapunov equation given by:
\begin{equation}
\partial_\theta \rho(\theta) = \frac{1}{2}\{\hat{L}(\theta), \rho(\theta)\},\label{eq:derivative-rho}
\end{equation}
with $\{*, \triangle\}=*\triangle + \triangle *$ being the anticommutator between operators $*$ and $\triangle$, and $\hat{L}(\theta)$ is a self-adjoint operator called \textit{Symmetric Logarithmic Derivative} (SLD). Following Ref.~\cite{paris2009quantum}, the CFI $I(\theta)$ can be upper bounded as
\begin{equation}
I(\theta) \leq I_Q(\theta) := \text{Tr}[\rho(\theta) \hat{L}(\theta)^2].\label{eq:quantum-fisher-information}
\end{equation}
$I_Q(\theta)$ denotes the quantum Fisher information (QFI), leading to the quantum Cram\'{e}r-Rao bound~\cite{paris2009quantum, 1055103, 1055173, braunstein1994statistical, sidhu2020geometric, albarelli2020perspective, liu2019quantum}:
\begin{equation} \label{eq:cramer-rao-classical-quantum}
\text{Var}[\tilde{\Theta}] \geq \frac{1}{I(\theta)} \geq \frac{1}{I_Q(\theta)}.
\end{equation}

Note that the derivative of the quantum state satisfies other expressions, e.g., the \textit{Right Logarithmic Derivative} (RLD)~\cite{albarelli2020perspective}. This gives rise to different quantum lower bounds---
see Ref.\cite{meyer2021fisher} for a thorough discussion on the uniqueness of CFI and QFI, and Ref.\cite{albarelli2020perspective} for an overview of different achievable bounds.

The QFI of Eq.~\eqref{eq:quantum-fisher-information} accounts for the ultimate precision limit for the estimation of $\theta$. Note that the QFI, as the maximization of the CFI over all possible measurements, depends on the quantum statistical model $\rho(\theta)$ and the SLD $\hat{L}(\theta)$ only, and it does not depend explicitly on a particular measurement basis. The optimal POVM can be composed with the set of projectors over the eigenstates of $\hat{L}(\theta)$ with an associated optimal quantum estimator\cite{paris2009quantum}.

To saturate the quantum Cram\'{e}r-Rao bound in Eq.~\eqref{eq:cramer-rao-bound-single}, one needs both the optimal POVM (as determined by the support of $\hat{L}(\theta)$) and an optimal classical estimator for post-processing data, such as the maximum-likelihood estimator in the limit of large data sets~\cite{Olivares-2009, pezze2007phase, Hradil-1996, LeCam-1986, nla.cat-vn81100, braunstein1994statistical, HELSTROM1967101} (refer to Refs.~\cite{rubio2019quantum, rubio2020bayesian} for Bayesian analysis in the scenario of limited data). The optimal POVM ensures that the CFI equals the QFI, i.e., $I(\theta) = I_Q(\theta)$. \vm{Thus, in practical scenarios, any observable measurement used to extract information about an unknown parameter must be compared against the fundamental sensing benchmark set by the QFI (For example, see Refs.~\cite{troiani2018universal, montenegro2025heisenberg} for a theoretical demonstration of QFI saturation via magnetization measurements and near-saturation through homodyne techniques~\cite{montenegro2020mechanical}, as well as Ref.~\cite{yu2022quantum} for experimental work achieving near QFI saturation in the phase estimation of a solid-state spin system). Crucially, the optimal POVM always exists and plays a key role as it guarantees that the quantum Cram\'{e}r-Rao for single-parameter estimation can always be attained, provided that both the estimator and the measurement basis are are chosen optimally.}

The SLD operator $\hat{L}(\theta)$ is an essential ingredient in the quantum parameter estimation framework, which can readily be obtained from the solution of the Lyapunov equation as~\cite{paris2009quantum}:
\begin{equation}
\hat{L}(\theta) = 2 \int_0^\infty e^{-t \rho(\theta)} \partial_\theta \rho(\theta) e^{-t \rho(\theta)} dt,\label{eq:L-Lyapunov}
\end{equation}
being basis independent, and
\begin{equation}
\hat{L}(\theta) = 2 \sum_{n,m} \frac{\langle \phi_m |\partial_\theta \rho(\theta)| \phi_n \rangle}{\epsilon_n + \epsilon_m} |\phi_m \rangle \langle \phi_n |\label{eq:L-eigendecomposition}
\end{equation}
which is obtained from Eq.~\eqref{eq:L-Lyapunov} using the spectral decomposition of $\rho(\theta)$, that is:
\begin{equation}
\rho(\theta) = \sum_i \epsilon_i |\phi_i\rangle\langle\phi_i|,
\end{equation}
where $\epsilon_i$ and $|\phi_i\rangle$ are the $i$-th eigenvalue and eigenvector of $\rho(\theta)$, respectively. From Eq.~\eqref{eq:L-eigendecomposition}, it is straightforward to obtain the QFI in this basis as follows:
\begin{equation}
I_Q(\theta) = 2 \sum_{n,m} \frac{|\langle \phi_m |\partial_\theta \rho(\theta)| \phi_n \rangle|^2}{\epsilon_n + \epsilon_m}, \hspace{0.5cm} \epsilon_n + \epsilon_m \neq 0.\label{eq:qfi-eigenvectors}
\end{equation}

It has been shown that it is also possible to separate the classical from the quantum contributions of the QFI by observing the dependence of the eigenvalues and eigenvectors with respect to the parameter $\theta$. This leads to the following formula for the QFI~\cite{paris2009quantum}:

\mt{\begin{equation}
I_Q(\theta)=\bar{I}(\theta) + 2\sum_{n\neq m} \epsilon_{nm} |\langle \phi_m | \partial_\theta \phi_n \rangle|^2,\label{eq:qfi-explicit}
\end{equation}
where $\bar{I}(\theta)=\sum_i (\partial_\theta \epsilon_i)^2/\epsilon_i$ is the CFI of the distribution of the eigenvalues of $\rho(\theta)$, and $\epsilon_{nm}$ may be written as}~\cite{paris2009quantum}
\begin{equation}
\epsilon_{nm} = \frac{(\epsilon_n - \epsilon_m)^2}{\epsilon_n + \epsilon_m} + \text{any antisymmetric term}.
\end{equation}
The second term of Eq.~\eqref{eq:qfi-explicit} shows the detailed contribution arising from the dependence of the eigenstates of the quantum probe with respect to the parameter $\theta$.

The QFI of Eqs.~\eqref{eq:quantum-fisher-information}-\eqref{eq:qfi-eigenvectors} applies to any quantum state. However, for pure quantum states $\rho(\theta)=|\psi(\theta)\rangle\langle\psi(\theta)|$ such that $\text{Tr}[\rho(\theta)^2]=1$, SLD $\hat{L}_{\theta} = 2 \partial_\theta \rho(\theta)$  and the QFI adopts a simpler form given by~\cite{paris2009quantum, sidhu2020geometric, liu2019quantum}:
\begin{equation}
I_Q(\theta)=4 ( \langle \partial_\theta \psi | \partial_\theta \psi \rangle - \langle \partial_\theta \psi | \psi \rangle \langle \psi | \partial_\theta \psi \rangle ).\label{eq:qfi-pure-states}
\end{equation}

Particular instances of relevant quantum states, for example, two-qubit states in Bloch representation, a single-qubit dephasing along the $z$-axis in a magnetic field along the $z$-axis, and X-shape two-qubit states, can be found in~\cite{liu2019quantum}.

\mt{Other formulas to evaluate the QFI for specific quantum statistical model have been found, which are useful in several situations of interest, e.g.~unitary families~\cite{liu2015quantum}, Bloch-sphere representation~\cite{zhong2013fisher,chapeau2015optimized}, qubit X-states~\cite{maroufi2021analytical}, and Gaussian states~\cite{PhysRevA.88.040102,monras2013phase,PhysRevA.89.032128,sparaciari2015bounds,PhysRevA.93.052330,sparaciari2016gaussian,bakmou2020multiparameter}.}

\subsection{Multi-parameter estimation}\label{sec:elements-multi}

Consider the situation where $d$ unknown parameters $\pmb{\theta}=(\theta_1, \theta_2,\ldots,\theta_d)$ need to be estimated. In this multi-parameter case, the Cram\'{e}r-Rao inequality generalizes to (for the sake of simplicity $M = 1$)~\cite{sidhu2020geometric, paris2009quantum, liu2019quantum}
\begin{equation}
\text{Cov}[\tilde{\pmb{\Theta}}] \geq I(\pmb{\theta})^{-1},\label{eq:classical-cramer-rao-multi}
\end{equation}
where $\text{Cov}[\tilde{\pmb{\Theta}}]$ is the covariance matrix with elements
\begin{equation}
[\text{Cov}[\tilde{\pmb{\Theta}}]]_{ij} = \langle \tilde{\Theta}_i \tilde{\Theta}_j \rangle - \langle \tilde{\Theta}_i \rangle \langle \tilde{\Theta}_j\rangle,
\end{equation}
and $I(\pmb{\theta})$ is the CFI matrix with elements~\cite{liu2019quantum}
\begin{equation}
[I(\pmb{\theta})]_{ij} = \int p(x|\pmb{\theta}) \left( \partial_{\theta_i} \text{ln}[ p(x|\pmb{\theta}) ]  \right)\left( \partial_{\theta_j} \text{ln}[ p(x|\pmb{\theta}) ]  \right) dx,
\end{equation}
where the diagonal elements fulfill $\text{Var}[\tilde{\Theta}_i] \geq [I(\pmb{\theta})^{-1}]_{ii}$.

The matrix inequality in Eq.~\eqref{eq:classical-cramer-rao-multi} means that $(\text{Cov}[\tilde{\pmb{\Theta}}] - I(\pmb{\theta})^{-1})$ is a positive semidefinite matrix. While the above is the general case, dealing with this matrix form is challenging.
Hence, it is convenient to transform the matrix inequality of Eq.~\eqref{eq:classical-cramer-rao-multi} into a scalar bound by introducing a positive and real weight matrix $\mathcal{W}$ such that~\cite{albarelli2020perspective}:
\begin{equation}\label{eq:Cramer_Rao_Multi_weight}
\text{Tr}[\mathcal{W} \text{Cov}[\tilde{\pmb{\Theta}}]]\geq\text{Tr}[\mathcal{W} I(\pmb{\theta})^{-1}].
\end{equation}
Similarly to the single-parameter scenario, it is possible to obtain a tighter bound of Eq.~\eqref{eq:classical-cramer-rao-multi} leading to the multi-parameter quantum Cram\'{e}r-Rao bound~\cite{paris2009quantum, liu2019quantum}:
\begin{equation}
\text{Cov}[\tilde{\pmb{\Theta}}] \geq I(\pmb{\theta})^{-1} \geq I_Q(\pmb{\theta})^{-1},\label{eq:quantum-cramer-rao-multi}
\end{equation}
where the matrix elements of QFI matrix $I_Q(\pmb{\theta})$ are given by~\cite{paris2009quantum, liu2019quantum}
\begin{equation}
[I_Q(\pmb{\theta})]_{ij} = \frac{1}{2}\text{Tr}[\rho(\pmb{\theta}) \{\hat{L}_i(\pmb{\theta}),\hat{L}_j(\pmb{\theta})\} ].\label{eq:matrix-elements-qfi-matrix}
\end{equation}
As before, one can convert the matrix inequality in Eq.~\eqref{eq:quantum-cramer-rao-multi} into a scalar bound using $\mathcal{W}$ as follows:
\begin{equation}
\text{Tr}[\mathcal{W} \text{Cov}[\tilde{\pmb{\Theta}}]]\geq\text{Tr}[\mathcal{W} I(\pmb{\theta})^{-1}] \geq \text{Tr}[\mathcal{W} I_Q(\pmb{\theta})^{-1}].
\end{equation}
As a particular case, one can consider $\mathcal{W}$ to be the identity. This choice prioritizes the precision of $\pmb{\theta}$ as the sum of the variances of the unknown parameters
\begin{equation}
\sum_{i=1}^d \text{Var}[\tilde{\Theta}_i] \geq \sum_{i=1}^d [I(\pmb{\theta})^{-1}]_{ii}:=\text{Tr}[I(\pmb{\theta})^{-1}]\geq \text{Tr}[ I_Q(\pmb{\theta})^{-1}].
\end{equation}

While the above tighter estimation bound includes the SLDs $\hat{L}_i(\pmb{\theta})$ and $\hat{L}_j(\pmb{\theta})$, other bounds can also be achieved by defining the \textit{Right Logarithmic Derivative}~\cite{1055103} and the Holevo approach~\cite{Holevo:1414149, Holevo1977CommutationSO}. However, the precision bound using the SLDs already contains the same information found in both the Holevo and the RLDs bounds~\cite{albarelli2020perspective}. For this reason, it is common to use Eq.~\eqref{eq:quantum-cramer-rao-multi} with their corresponding matrix elements defined in Eq.~\eqref{eq:matrix-elements-qfi-matrix} as the ultimate precision limit for the multi-parameter estimation case.

One of the main differences between the single- and multi-parameter scenarios is that, while the CFI can be achieved in both scenarios, in principle, the QFI is attainable only in the single-parameter case and not for the jointly multi-parameter estimation~\cite{albarelli2020perspective}.
This can be understood because, generally, there is no single symmetric logarithmic derivative $\hat{L}_i(\pmb{\theta})$ that defines an optimal measurement basis suitable for all the unknown parameters. To quantify this notion of incompatibility between parameters~\cite{ragy2016compatibility}---those that cannot be estimated optimally with a single set of measurements---the incompatibility matrix $\pmb{D}$ has been introduced~\cite{albarelli2020perspective, carollo2019quantumness}, also known as mean Uhlmann curvature~\cite{carollo2019quantumness}, with elements:
\begin{equation}
D_{ij}=-\frac{i}{2}\text{Tr}[\rho(\pmb{\theta}) [\hat{L}_i(\pmb{\theta}),\hat{L}_j(\pmb{\theta})]],
\end{equation}
here $[\circ,\bullet]{=}{\circ\bullet} {-} {\bullet\circ}$.
The case $D_{ij}=0$ for all $i,j$ is known as the compatibility condition~\cite{ragy2016compatibility}. \textcolor{black}{In a recent work~\cite{yousefjani2025nonlinearity}, it was shown that when the above compatibility condition is met, it leads to quantum-enhanced precision in the simultaneous sensing of multiple parameters.} \vm{Moreover, quantum critical sensing can be applied to multi-parameter estimation~\cite{mihailescu2024multiparameter, yang2024quantum, yan2024quantum}. A key advantage of using quantum many-body probes undergoing quantum phase transitions is that they have been argued to help mitigate the fundamental incompatibility that arises when estimating multiple parameters simultaneously~\cite{difresco2022multiparameter}.} Lastly, a genuine quantum incompatibility measure with a geometric character has also been proposed~\cite{belliardo2021incompatibility}, along with a comprehensive study of incompatibility under general $p$-local measurements~\cite{chen2021hierarchical}.

The expressions given by Eqs.~\eqref{eq:quantum-fisher-information}-\eqref{eq:qfi-eigenvectors} to obtain QFI and SLD for single parameter estimation, and generalizations thereof to the multiparameter case all involve diagonalizing the quantum state $\rho$ explicitly to find out its spectral decomposition. For quantum many-body density matrices whose dimensions increase exponentially with system-size, this generally presents a significant computational challenge. To overcome this, Ref.~\citep{vsafranek2018simple} noted that the Lyapunov equations can be expressed as a set of linear equations in the vectorization picture~\citep{laub2005matrix,simoncini2016computational}, and thereby provided the following formula for the QFI matrix
\begin{eqnarray}
    \nonumber [I_Q(\pmb{\theta})]_{ij} &=& 2 \text{vec}[\partial_{\theta_i} \rho(\pmb{\theta})]^{\dagger} \Big( \overline{\rho(\pmb{\theta})} \otimes I \\
    &+&I \otimes \rho(\pmb{\theta})\Big)^{-1}\text{vec}[\partial_{\theta_j}\rho(\pmb{\theta})],\label{qfi_vec_expression}
\end{eqnarray}
\noindent where $\text{vec}[.]$ denotes the vectorized form of a matrix and $\overline{\rho(\pmb{\theta})}$ denotes complex conjugation of $\rho(\pmb{\theta})$. The corresponding SLDs are given as 
\begin{equation}
    \text{vec}[\hat{L}_i(\pmb{\theta})] = 2  \left( \overline{\rho(\pmb{\theta})} \otimes \hat{I} + \hat{I} \otimes\rho(\pmb{\theta})\right)^{-1}\text{vec}[\partial_{\theta_i}\rho(\pmb{\theta})]
\end{equation}

\noindent In case the quantum state $\rho$ is not full-rank, the inverse is to be replaced by the Moore-Penrose pseudoinverse. Thus, the problem of fully diagonalizing a large matrix is replaced by matrix inversion. In optical systems, states are often expressed in the coherent state basis, which is a non-orthogonal basis. Accordingly, an extension of Eq.~\eqref{qfi_vec_expression} for general non-orthogonal bases was achieved in Ref.~\citep{fiderer2021general}. 

\subsection{Geometry of parameter estimation}

Close links between Fisher information quantities and statistical distances have been found~\cite{PhysRevD.23.357, braunstein1994statistical}. In terms of the CFI, which quantifies the sensitivity of the probability distributions $p(x|\theta)$ concerning the unknown parameter $\theta$, the notion of distinguishability between different probability distributions among $M$ trials arises naturally (see Bengtsson and $\dot{Z}$yczkowski~\cite{bengtsson2017geometry}, and Kok and Lovett~\cite{kok2010introduction}). It turns out that distinguishing between two probabilities $p(x|\theta_1)$ and $p(x|\theta_2)$, taking $\text{d}s$ ($\text{d}s$ being the infinitesimal distance on the probability simplex space) along a line element $\text{d}\theta$, results in~\cite{bengtsson2017geometry, kok2010introduction}
\begin{equation}
\left(\frac{\text{d}s}{\text{d}\theta}\right)^2 = \sum_{\mu} \frac{1}{p^\mu}\frac{(\text{d}p^\mu)^2}{\text{d}\theta^2} = \sum_{\mu} \frac{1}{p^\mu}\left(\frac{\partial p^\mu}{\partial\theta}\right)^2 := I(\theta),\label{eq:simplex-2}
\end{equation}
which is the CFI of Eq.~\eqref{eq:cfi}. Further analysis leads to:
\begin{equation}
(\delta \theta)^2 \geq \frac{1}{MI(\theta)},
\end{equation}
where $\delta \theta$ is the segment of the path in the probability simplex---rather than the variance of $\tilde{\theta}$. Note that the above has been derived solely from a distance measure (a metric) by distinguishing probability distributions on the simplex space. It has also been shown that for distinguishing two probability distributions, one can use the relative entropy (not a metric) for this purpose, and consequently, a close connection between the CFI matrix and the relative entropy emerges~\cite{sidhu2020geometric}.

In quantum mechanics, on the other hand, the notion of distinguishing between normalized vectors in a complex Hilbert space can be addressed by the fidelity $F(\psi_1,\psi_2)=|\langle \psi_1|\psi_2\rangle|$ (not a metric). Indeed, this defines the Wootters distance~\cite{PhysRevD.23.357, braunstein1994statistical, Hilgevoord1991}
\begin{equation}
\text{d}s^2_\text{W}=(\arccos[F(\psi(\theta),\psi(\theta + \delta \theta))])^2,
\end{equation}
with the Fubini-Study metric given by~\cite{PhysRevA.94.043839}
\begin{equation}
h_{\text{FS}}=\frac{\langle \partial_\theta \psi|\partial_\theta \psi\rangle}{\langle \psi | \psi \rangle} - \frac{\langle \partial_\theta \psi| \psi\rangle \langle \psi| \partial_\theta \psi\rangle }{\langle \psi | \psi \rangle^2},
\end{equation}
where $\psi:=\psi(\theta)$. By setting $\langle \psi|\psi \rangle = 1$, it corresponds to the QFI for pure states (see Eq.~\eqref{eq:qfi-pure-states}), that is
\begin{equation}
I_Q(\theta) = 4 h_{\text{FS}}.
\end{equation}

Extending the Fubini-Study metric to density matrices, one finds that the Bures metric is proportional to the QFI matrix~\cite{PhysRevA.94.043839} and is given by
\begin{equation} \label{eq:QFI_Fidelity}
I_Q(\theta)=4 h_\text{Bures}=8 \lim_{\delta\theta\rightarrow 0} \frac{1 - F(\rho(\theta),\rho(\theta+\delta\theta))}{(\delta\theta)^2},
\end{equation}
where $\delta\theta$ is an infinitesimal increment of $\theta$ and the fidelity is the square root of the Uhlmann fidelity~\cite{UHLMANN1976273}
\begin{equation}
F(\rho_1,\rho_2)=\text{Tr}[\sqrt{ \sqrt{\rho_1}\rho_2\sqrt{\rho_1}}].
\end{equation}
For pure states $\rho_1=|\psi_1\rangle \langle \psi_1|$ and $\rho_2=|\psi_2 \rangle \langle \psi_2|$, the Uhlmann fidelity reduces to $F(\rho_1,\rho_2)=|\langle \psi_1|\psi_2\rangle|$.

\subsection{Alternatives to QFI for metrology}

Gauging the precision of parameter estimation through the quantum Cram\'{e}r-Rao bound is undoubtedly the most popular approach in the literature, thanks to its geometrical properties and its utility as a signature of multipartite entanglement. However, the Cram\'{e}r-Rao bound suffers from two practical difficulties. 
Firstly, although this bound is asymptotically tight for single parameter estimation, it may perform quite poorly in the non-asymptotic regime and especially if the likelihood function is highly non-Gaussian. This is a known issue even in classical parameter estimation theory, and several alternate bounds have been proposed in statistical literature~\citep{van2007bayesian}. Perhaps the most well-known is the Ziv-Zakai bound~\citep{ziv1969some}, which is obtained by turning the continuous parameter estimation problem to a hypothesis testing problem for each coarse-grained interval in the domain of the parameter. By noting that hypothesis-testing at each interval amounts to a quantum state discrimination problem, Tsang in Ref.~\citep{tsang2012ziv} constructed a quantum version of the Ziv-Zakai bound. Although such bounds are not generally tight, they can be shown to perform significantly better for finite set of observations in highly non-Gaussian environments~\citep{chang2022evaluating}. Extensions of the quantum Ziv-Zakai bound to multiparameter estimation~\citep{berry2015quantum} and noisy environments~\citep{chang2022evaluating} have also been investigated. Lu and Tsang also considered the quantum generalization of the classical Weiss-Weinstein bounds~\citep{lu2016quantum}. In yet another direction, Liu and Yuan~\citep{liu2016valid} obtained a tighter version of the Bayesian quantum Cram\'{e}r-Rao bound valid for both biased and unbiased estimators. For a function $f(\theta)$ of the quantum parameter $\theta$ which is estimated via a (generally) biased estimator of bias $b(\theta)$ with a prior probability distribution $p(\theta)$ on a quantum state $\rho (\theta)$, Liu and Yuan's bound reads 

\begin{equation}
    \text{Var}[f(\theta)] \geq \int p(\theta) \left[ \frac{\left[ f'(\theta) + b'(\theta) \right]^2}{I_Q(\theta)} + b^{2} (\theta) \right] d\theta,
\end{equation}
\noindent where $I_Q(\theta)$ is the QFI for $\rho(\theta)$ with respect to the parameter $\theta$. One can note that this bound reduces to the biased version of the Bayesian Quantum Cram\'{e}r-Rao bound for a uniform prior.
Secondly, and of significant relevance to many-body sensing, the QFI is not fully expressible in terms of experimentally easily accessible one and two-point correlation functions\citep{gabbrielli2018multipartite}. This was addressed recently in Ref.~\citep{lepori2021improving}, where an alternative functional dependent only on one and two-point correlators was proposed. More economic sensing with fewer copies was studied in Ref.~\cite{zhang2024inferring}. 

\section{Interferometric Quantum Sensing}

\vm{V. Giovannetti, S. Lloyd, and L. Maccone~\cite{giovannetti2004quantum} addressed the problem of estimating the unknown phase $\theta$ imprinted in a quantum state through a unitary operation of the form $\hat{U}(\theta)=e^{i\hat{H}\theta}$, where $\hat{H}$ is a Hermitian operator with real eigenvalues $E_k$ and their corresponding eigenvectors $|E_k\rangle$, such that $\hat{H}|E_k\rangle=E_k|E_k\rangle$.} This implies that the eigenvalues of the unitary operator $\hat{U}(\theta)$ are given by $u_k=e^{iE_k\theta}$ while the eigenvectors do not depend on $\theta$ and remain the same as $\hat{H}$, namely $\hat{U}(\theta)|E_k\rangle=u_k|E_k\rangle$. 
The unknown parameter $\theta$ can be encoded in a quantum state through 
$|\Psi(\theta)\rangle=\hat{U}(\theta)|\Psi_0\rangle$ for a given initial state $|\Psi_0\rangle$. In this case, one can easily show that the QFI of $|\Psi(\theta)\rangle$ is directly proportional to the variance of $\hat{H}$
\begin{equation} \label{eq:QFI_Variance}
 I_Q(\theta)=4\left(\langle \Psi(\theta)| \hat{H}^2 | \Psi(\theta)\rangle-  \langle \Psi(\theta)| \hat{H} | \Psi(\theta)\rangle^2 \right).
\end{equation}
One can maximize the above QFI with respect to the initial state. In fact, the maximum of $I_Q$ is achieved for the initial state $|\Psi_0\rangle=\left(|E_{\rm min}\rangle+|E_{\rm max}\rangle\right)/\sqrt{2}$,  where $|E_{\rm min}\rangle$ ($|E_{\rm max}\rangle$) is the eigenvector of $\hat{H}$ with the corresponding smallest (largest) eigenvalue $E_{\rm min}$ ($E_{\rm max}$). Thus, the corresponding QFI becomes
\begin{equation}\label{eq:QFI_GHZ_Eigvalue}
    I_Q=(E_{\rm max}{-}E_{\rm min})^2.
\end{equation}
To analyse the scaling of $I_Q$, in Eq.~(\ref{eq:QFI_GHZ_Eigvalue}), with respect to the system size one has to consider the details of the interaction in the Hamiltonian $\hat{H}$. For a given interaction of $\hat{H}=\sum_{\{j_1,\cdots,j_k\}} \hat{h}_{j_1\cdots j_k}^{(k)}$ where $\hat{h}_{j_1\cdots j_k}^{(k)}$ represents a $k$-body interaction between particles at sites $j_1$ to $j_k$, the QFI, at best, scales as $I_Q\sim N^{2k}$~\cite{boixo2007generalized}, see e.g. Ref.~\cite{luis2016nonlinear} and \vm{Ref.~\cite{beau2017nonlinear} for its generalization to many-body open systems.} 
{\color{black} In this review, we consider this quadratic scaling as Heisenberg limit, although, in the literature, there is not much agreement on the definition of the Heisenberg limit. For instance,  sometimes  $I_Q \sim N^{2k}$,  for $k$-body interacting  $\hat{H}$, may be defined as Heisenberg limit. Indeed, the most fundamental limit for interferometry-based sensing is given in Eqs.~(\ref{eq:QFI_Variance}) and (\ref{eq:QFI_GHZ_Eigvalue}).}  

{\color{black} Interferometry-based quantum sensing, presented above, has clear advantages. First, the quantum-enhanced sensitivity, i.e. super-linear scaling of the QFI with respect to the system size, is achievable for all values of the unknown parameter $\theta$. Second, the measurement basis which achieves this quantum-enhanced precision is independent of the unknown parameter $\theta$ which implies that no prior information about the unknown parameter is needed to achieve quantum-enhanced precision. These two features clearly make interferometry-based quantum sensors very useful. Although, this approach has its own limitations too.  The first limitation comes from the fact that} the quantum-enhanced sensitivity with interferometric approach relies on creating GHZ-type entangled states or equivalently N00N~\cite{mitchell2004super} or squeezed states~\cite{ma2011quantum,maccone2020squeezing}. Preparation of these special forms of entangled states is not an easy task and is resource consuming. In fact, delocalization of quantum correlation is found to be essential for fast generation of metrologically useful entangled states~\cite{chu2024quantum}. In Ref.~\cite{chu2023strong}, a strong metrological limit for the preparation time of such large-scale entanglement has been established through studying the Lieb-Robinson light cone. {\color{black} The second limitation of interferometry-based sensing is that it is only applicable to the situations where the unknown parameter is encoded through a phase shit operator, namely $\hat{U}(\theta)$. This obviously does not represent the most general encoding method and thus not every quantum probe lies in this category. The third limitation is the spacial form of the initial state, namely the GHZ-type state  $|\Psi_0\rangle=\left(|E_{\rm min}\rangle+|E_{\rm max}\rangle\right)/\sqrt{2}$ which are also known as Schr\"odinger cat states. These states are hard to implement in practice and, in general, prone to decoherence which  transforms the superposition into a classical mixture with no quantum advantage for sensing. In fact, robust and scalable realization of  GHZ states is an active area of research with several recent developments~\cite{pogorelov2021compact, zhuang2021many, moses2023a, bluvstein2024logical, bao2024creating, cao2024multi}. }
The interferometric  quantum sensing has been implemented experimentally in various platforms~\cite{mitchell2004super,pezze2007phase,ono2013entanglement,spagnolo2012phase, Okamoto-2012, Okamoto-2017}.

\section{Equilibrium Quantum Sensing: Second Order Quantum Phase Transition}\label{sec:2nd-order-pt}

It is interesting to consider a more  general unitary operation  $\hat{U}^{\prime}(\theta)=e^{i(\theta \hat{H}+\hat{H}^\prime})$ for encoding the parameter $\theta$, as $|\Psi(\theta)\rangle=\hat{U}^{\prime}(\theta)|\Psi_0\rangle$. For such an unitary operation, the sensing performance is again quantified by QFI of the parameter $\theta$ denoted as $I^{\prime}_Q(\theta)$. Note that as far as $[\hat{H},\hat{H}^{\prime}]\neq 0$, both the eigenvectors and eigenvalues of $\hat{U}^{\prime}(\theta)$ depend on $\theta$.  In Ref.~\cite{de2013quantum} it has been shown that for all choices of initials states $|\Psi_0\rangle$ and perturbing interactions $\hat{H}^{\prime}$ we always have $I_Q(\theta)\ge I^{\prime}_Q(\theta)$. This means that in the interferometric phase sensing scenario, the presence of perturbing interaction $\hat{H}^{\prime}$ is always destructive. For the case of sensing an external field, $\hat{H}^{\prime}$ can represent interaction among the particles of the probe. In such cases, the interaction between the constituents of the probe can reduce the performance of the system~\cite{boixo2007generalized,de2013quantum,skotiniotis2015quantum,pang2014quantum}.

An alternative method to interferometric sensing is to harness the interaction between the particles in strongly correlated many-body systems. The interplay between different parts of the Hamiltonian can induce phase transition in the system, also known as quantum criticality. Many-body systems exhibiting quantum criticality have  been identified as resources for quantum-enhanced sensitivity. Recently, C.~Hotter et al.~proposed a unified protocol that combines critical and conventional approaches to metrology~\cite{hotter2024combining}.  A more fundamental limit to many-body quantum metrology was recently obtained in Ref.~\cite{abiuso2025fundamental}. In this section, we review various forms of quantum criticality for  sensing and show that a common feature among all these systems is the emergence of a sort of ``gap-closing" in their energy spectrum. This is an important observation as it clearly shows that any phase transition which is accompanied by gap-closing is a potential resource for quantum sensing. 

As we have discussed in the previous section, despite the theoretical appeal, interferometry-based quantum sensing may face practical limitations. An alternative approach for achieving quantum-enhanced sensitivity is to exploit quantum criticality. The most conventional manifestation of quantum criticality is the second-order phase transition which describes a large set of phenomena.
Second-order quantum phase transition occurs at zero temperature, where quantum fluctuations lead to spontaneous symmetry-breaking at the critical point. Interestingly, this opens up the possibility of quantum-enhanced sensing near criticality~\citep{zanardi2006ground}. Consider a many-body probe which is described by the Hamiltonian 
\begin{equation}\label{eq:general_2nd_order}
\hat{H} =  \theta \hat{H}_1  + \hat{H}_2,
\end{equation}
\noindent with $\hat{H}_1$ representing the action of the target parameter $\theta$ on the probe, and $\hat{H}_2$ representing the internal physics of the probe itself. The interplay between $\hat{H}_1$ and $\hat{H}_2$ often leads to rich phase diagrams with the strength of the parameter $\theta$ determining the relevant phase. When unhindered by thermal fluctuations, a second order quantum phase transition may occur at a specific value of $\theta=\theta_c$, where the ground state of the system undergoes a global transformation. There are several common features in second-order quantum phase transitions: (i) spontaneous symmetry breaking;  (ii) the emergence of a scale invariance behavior for all observables in the system; (iii) the appearance of a diverging length scale $\xi$ which behaves as $\xi{\sim} |\theta-\theta_c|^{-\nu}$, where $\nu$ is a critical exponent; and (iv) an anti-crossing between the ground and the first excited state whose energy gap vanishes  at the transition point as the system sizes reaches thermodynamic limit. 
As we will discuss, a significant enhancement in sensitivity is often associated with the critical points in the phase diagram.

\begin{figure}[t]
    \centering
    \includegraphics[width=0.99\linewidth]{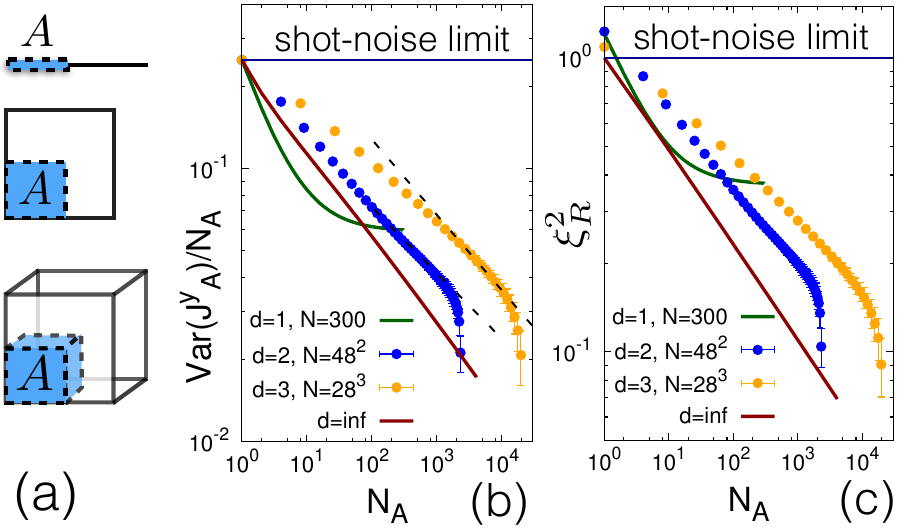} 
    \caption{\textbf{Sub-shot-noise scaling with critical quantum transverse field Ising chain probes for various dimensions}. In (b) and (c) $x$-axis is $N_A$, the number of spins inside any subsystem $A$ as depicted in (a); $y$-axis is (b) variance of the estimator of $y$-component of collective spin operator $\hat{J}^y= \sum_{i}\hat{S}^y_i$ normalized by $N_A$, and (c) collective spin squeezing parameter $\xi_R^2$. Figure taken from Ref.~\cite{frerot2018quantum}.}
\label{Fig_frerot_qcm}
\end{figure}

At this point, a distinction should be made between two different approaches to many-body critical quantum sensing.  In the first approach, the ground state of a many-body system, potentially at criticality,  is used as the \emph{input} state for an interferometry-based quantum sensing. The critical ground state is a delicate superposition of several states which can reveal significant variance with respect to certain observables and thus maximizes the QFI in an interferometry-based sensing, as given in Eq.(\ref{eq:QFI_Variance}). For instance, in Ref.~\citep{ma2011quantum}
many-body spin squeezed states are used as inputs to interferometric phase estimation protocols. Atom-interferometry based quantum metrology has grown substantially in recent years, and their detailed description lies outside the scope of the present review. We direct the interested reader to Ref.~\citep{pezze2018quantum} for a comprehensive review. However, even for interferometric phase sensors with equilibrium probe states of genuinely many-body quantum systems, Frerot and Roscilde~\citep{frerot2018quantum}  showed the possibility of quantum criticality enhanced sensing. 
They considered the problem of estimating a phase angle $\theta$ for the collective spin operator $\hat{\boldsymbol{J}} = \sum_{i} \hat{\boldsymbol{S}}_i$, with $\hat{\boldsymbol{S}}_i = (\hat{S}^x_i, \hat{S}^y_i, \hat{S}^z_i)$ being the vectorized spin operator at site $i$, such that the unitary $\hat{U} = \exp (i\theta \hat{J}^{x,y,z})$ represents the action of the interferometer along the relevant axis. By considering the ground state of a transverse field Ising chain  as the input of an interferometer, they showed that at the critical point, the scaling of precision of estimation of $\theta$ for a $N$-atom transverse field Ising chain in $d$-dimensions is given by 
\begin{equation}
    {(\delta \theta)}^2 \geq N^{-1-\frac{\eta-1}{d_m}},
\end{equation}
\noindent where $\eta$ is the scaling exponent of the long-range correlation function of the order parameter, as defined by $\langle \hat{\sigma}_i^x \hat{\sigma}_{i+r}^x\rangle \propto r^{2-d-\eta}$, at critical point for a $d$-dimensional model, and $d_m = \min (d,d_c)$, with upper critical dimension $d_c = 3$ for the transverse field Ising chain. The results, plotted in Fig.~\ref{Fig_frerot_qcm}, attest that sub-shot-noise scaling is indeed achievable with such probes. Similar results have also been found in Ref.~\cite{li2024quantum} in which interferometric quantum-enhanced sensitivity, using the critical ground state of a many-body system, can be exploited for detecting quantum phase transitions. 

The second approach, which is the main topic of this section, is that of Hamiltonian parameter estimation, where the many-body ground state is directly measured to gain access to the physical parameter in question. In the thermodynamic limit, where the system size $N{\rightarrow}\infty$, the QFI diverges~\cite{venuti2007quantum,schwandt2009quantum,albuquerque2010quantum,gritsev2009universal,
gu2008fidelity,greschner2013fidelity}
\begin{equation}\label{eq:QFI_diverging}
I_{Q}(\theta) \sim |\theta-\theta_c|^{-\alpha},
\end{equation}
where $\alpha{>}0$ is the corresponding critical exponent.
In finite-size systems, at the transition point $\theta{=}\theta_c$ the QFI shows algebraic dependence on the system size 
\begin{equation}\label{eq:QFI_finit_size}
I_{Q}(\theta{=}\theta_c) \sim N^{\beta}.
\end{equation} 
The two expected asymptotic behaviors in Eqs.~(\ref{eq:QFI_diverging}) and (\ref{eq:QFI_finit_size}) can be merged in a single ansatz  
\begin{equation}\label{eq:QFI_N_ansatz}
I_{Q}(\theta) \sim \dfrac{1}{N^{-\beta} + A|\theta-\theta_c|^{\alpha}},
\end{equation}
with $A$ being a constant. Note that in the thermodynamic limit (i.e.~$N{\rightarrow}\infty$) one retrieves Eq.~(\ref{eq:QFI_diverging}) and for finite system sizes at $\theta{=}\theta_c$ one recovers Eq.~(\ref{eq:QFI_finit_size}). 
The emergence of scale invariance behavior at the second-order quantum phase transition implies that the QFI follows a conventional finite size scaling ansatz~\cite{pathria2016statistical}  
\begin{equation}\label{eq:Finit_size_scaling_ansatz}
I_{Q}(\theta){=}N^{\alpha/\nu}g\left(N^{1/\nu}(\theta-\theta_c)\right),
\end{equation}
where $g(\cdot)$ is an arbitrary function {\color{black} which depends on the model and the quantity of interest}. This ansatz is routinely  used for determining critical exponents $\alpha$ and $\nu$ using finite size systems. Plotting the rescaled QFI, namely $N^{-\alpha/\nu}I_Q(\theta)$, versus $N^{1/\nu}(\theta-\theta_c)$ collapses all the curves for different system sizes for the right choice of critical parameters, namely $(\theta_c,\alpha,\nu)$.
Since the two ansatzes in Eq.~(\ref{eq:QFI_N_ansatz}) and Eq.~(\ref{eq:Finit_size_scaling_ansatz}) describe the same quantity, they are not independent. In fact, by factorizing $N^{\beta}$ from Eq.~(\ref{eq:QFI_N_ansatz}), one can show that the identicality of both ansatzes imposes the following identity between the critical exponents 
\begin{equation}\label{eq:exponents}
\beta=\frac{\alpha}{\nu}.
\end{equation}
This shows that the three critical exponents are not independent from each other and knowing two of them is enough for determining the other.
While the identity in Eq.~(\ref{eq:exponents}) is valid for one-dimensional systems, its generalization to $d$ dimensional probes is obtained as $\beta=\frac{\alpha}{d\nu}$~\cite{schwandt2009quantum,albuquerque2010quantum,rams2018limits}. 
This is a remarkable discovery which shows that unlike the interferometry-based sensing scenarios, e.g.~using GHZ states, the precision of criticality-based quantum sensing is not bounded by the Heisenberg scaling. In fact, quantum-enhanced sensitivity with $\beta>2$ can be achieved if $\alpha>2d\nu$. It is worth mentioning that for all realization of the second-order phase transition in the ground state, one can show that $\alpha{=}2$~\cite{schwandt2009quantum, albuquerque2010quantum}. However, some non-ground state continuous phase transitions behave like the second-order phase transition with $\alpha$ different from 2~\cite{he2023stark}. Therefore, for keeping generality of the paper, we keep $\alpha$ as a general parameter in our notations.

{\color{black} It is worth emphasizing that in criticality-based quantum sensing, quantum-enhanced precision can only be harnessed in the notion of local sensing, where significant prior information about the unknown parameter is available. Two primary reasons for this are: (i) as discussed above, quantum-enhanced precision can only be achived when the probe is operating near its criticality, namely $\theta{\sim}\theta_c$; and (ii) the optimal measurement basis, in general, depends on the unknown parameter $\theta$ and thus prior information is required to perform a measurement which is nearly optimal. In other words, when the parameter $\theta$ is roughly known and the purpose of sensing is to estimate its value with more precision, one can tune the probe to operate near its criticality and then perform a measurement which is optimal based on the prior value of the parameter. In such scenario, one can obtain the exact value of the parameter $\theta$ with achieving quantum-enhanced precision.   }  

\subsection{Free fermionic many-body probes }

The above discussion on second-order quantum phase transitions is very general. It will be more insightful to go through some examples and explicitly determine the scaling of QFI, namely $I_Q\sim N^\beta$, at the vicinity of the critical point. Free fermionic systems are an important class of many-body systems in condensed matter physics. In particular, they are analytically solvable which makes them an excellent toy model for understanding several complex phenomena, such as criticality. In addition, many spin chain systems can also be mapped to free fermion systems through Jordan-Wigner transformation and thus become solvable~\cite{sachdev1999quantum}. A general free fermion model with  $N$ particles can be described by Hamiltonian $\hat{H}(\theta)$ which in the Nambu formalism notation is written as~\citep{mbeng2020quantum} 
\begin{equation}
    \hat{H}(\theta) = \hat{\Phi}^{\dagger} \hat{H}^{\prime}(\theta) \hat{\Phi},
\end{equation}
where $\hat{\Phi}^{\dagger} = \left( \hat{c}_1^{\dagger}  \hat{c}_2^{\dagger} \dots \hat{c}_N^{\dagger} \hat{c}_1 \hat{c}_2 \dots \hat{c}_N\right)$  is a $2N\times 1$ vector with $\hat{c}_j(\hat{c}_j^{\dagger})$ being bare-particle annihilation (creation) operators and $\hat{H}^{\prime}$ being the $2N\times 2N$ Hermitian matrix containing information about $\theta$. The collection of normalized eigenvectors $\lbrace |E_k\rangle \rbrace$ of $\hat{H}^{\prime}$ are expressible in terms of $N\times N$-matrices $U$ and $V$ as 
\begin{equation}
    \lbrace |E_k\rangle \rbrace = \begin{pmatrix}
        U & \bar{V} \\
        V & \bar{U}
    \end{pmatrix},
\end{equation}
\noindent with $\bar{\circ}$ denoting complex conjugation and each column representing one eigenvector. The QFI for the ground state with respect to parameter $\theta$ can then be expressed as~\citep{mukhopadhyay2023modular} 
\begin{eqnarray}
I_{Q} & = \text{Tr}\left[\left(U^{\dagger}\frac{\partial U}{\partial \theta}\right)^2 - \frac{\partial^{2} U}{\partial \theta^{2}} \right] + \text{Tr}\left[\left(V^{\dagger}\frac{\partial V}{\partial \theta}\right)^2 - \frac{\partial^{2} V}{\partial \theta^{2}} \right] \nonumber \\
& + 2 \text{Tr}\left[U^{\dagger}\frac{\partial U}{\partial \theta} V^{\dagger} \frac{\partial V}{\partial \theta}\right]
\end{eqnarray}

\noindent This closed form of QFI can be used to study the scaling of QFI in free fermion systems. Another useful perspective for calculation of QFI for systems governed by generic quadratic fermionic Hamiltonians was provided by Carollo~\emph{et al}~\citep{carollo2018symmetric}, where the SLD of fermionic Gaussian states associated with such quadratic Hamiltonians was considered. In this case, instead of the Dirac-representation of fermions $\lbrace \hat{c}_{i}, \hat{c}_{j}^{\dagger}\rbrace=\delta_{ij}, \lbrace \hat{c}_{i}, \hat{c}_{j}\rbrace=0$, it is more convenient to work with the Hermitian Majorana representation given by fermion operators $\omega$ such that $\hat{\omega}_{2j-1} = \hat{c}_{j} + \hat{c}_{j}^{\dagger}$, and $\hat{\omega}_{2j} = i (\hat{c}_{j} - \hat{c}_{j}^{\dagger})$. Fermionic Gaussian states with $n$-fermionic modes and parametrized by the sensing parameter $\theta$, can then be defined as states of the form $\rho(\theta) = \exp(-\frac{i}{4} \hat{\boldsymbol{\omega}}^{T} \Omega(\theta) \hat{\boldsymbol{\omega}})/\Tr\left[\exp(-\frac{i}{4} \hat{\boldsymbol{\omega}}^{T} \Omega(\theta) \hat{\boldsymbol{\omega}})\right]$, where $\Omega(\theta)$ is a $2n\times 2n$ real anti-symmetric matrix and $\hat{\boldsymbol{\omega}} = (\hat{\omega}_1, \hat{\omega}_2,...,\hat{\omega}_{2n})$ are the $2n$ Majorana fermions. Owing to the above Gaussian structure, such states can be completely described in terms of the second moment, i.e., the two-point correlation matrix $\Gamma_{jk}(\theta) = \frac{1}{2}\Tr[ \rho(\theta) (\hat{\omega}_j \hat{\omega}_k {-}\hat{\omega}_k \hat{\omega}_j)] {=}\tanh{(i\Omega_{jk}(\theta)/2)}$.  Following the established mathematical approach for photonic Gaussian states~\citep{serafini2017quantum}, Carollo~\emph{et al} used the ansatz that the SLD in this case is also given by the quadratic form 
\begin{equation}
    \hat{L}(\theta) = \frac{1}{2} \hat{\boldsymbol{\omega}}^{T} K(\theta) \hat{\boldsymbol{\omega}} + \xi^{T}(\theta) \hat{\boldsymbol{\omega}} + \eta(\theta)
\end{equation}
with $\xi = 0$, and $\eta = \frac{1}{2}\text{Tr}\left[K(\theta)\Gamma(\theta)\right]$. The $2n\times 2n$ matrix $K$ was ultimately shown to have been given in terms of the correlation matrix $\Gamma(\theta)$ and its eigenvalues $\lbrace\gamma_k(\theta)\rbrace$ with corresponding eigenvectors $|k(\theta)\rangle$ as 
\begin{equation}
    \langle j(\theta)|K(\theta)|k(\theta)\rangle = \frac{{\left(\frac{\partial \Gamma}{\partial \theta}\right)}_{jk}}{\gamma_j(\theta) \gamma_k(\theta) - 1}.
\end{equation}
Recalling that QFI is just the expectation value of the square of the SLD operator and armed with the ease of calculating such moments for Gaussian systems, it is easy to calculate the QFI from this expression. Although strictly speaking this form of QFI technically only holds for full-rank states, one can nonetheless add an arbitrarily small white noise to the system to regularize the QFI. 

\subsection{Bosonic many-body probes}

Apart from Fermionic systems, there are also Bosonic many-body quantum systems, e.g., interacting Bosonic atoms in a lattice, which are canonically described by the Bose-Hubbard Hamiltonian as  

\begin{equation}
    H = -J \sum_{i} \left(\hat{b}_{i}^{\dagger} \hat{b}_{i+1} + \hat{b}_{i+1}^{\dagger} \hat{b}_{i}\right) -\mu \sum_{i} \hat{n}_{i} + \frac{U}{2} \sum_{i} \hat{n}_{i} \left(\hat{n}_i -1\right)
\end{equation}
\noindent where $\hat{b}_{i} (\hat{b}_{i}^{\dagger})$-s are the \emph{Bosonic} annihilation (creation) operators, $\hat{n}_{i}{=}\hat{b}_{i}^{\dagger} \hat{b}_{i}$ is the number operator at site $i$, and $J, U, \mu$ are the hopping, onsite interaction, and chemical potential paremeters, respectively. The Bose-Hubbard model can be implemented by loading cold bosonic atoms in Harmonic oscillator traps~\citep{jaksch1998cold,jaksch2005cold,lewenstein2007ultracold}. At zero temperature, the phase diagram consists of a Mott insulating phase ($J \ll U$) and a Superfluid phase ($J \gg U$). The quantum phase transition between these phases from the Mott insulator phase to the Superfluid phase involves the spontaneous breaking of the $U(1)$ symmetry of the Bose-Hubbard Hamiltonian. Quantum sensing in these atomic systems combines the theoretical formalism of photon-interfermometry based metrology with reasonable ease of producing quantum states and robustness against particle loss.  Considering a single-mode description of the resulting interacting BEC, where atom-atom interactions are described by a non-linear correction to the Harmonic trap potential, Gross \emph{et al} obtained the first experimental demonstration of quantum-enhanced metrology in cold Bosonic atoms~\cite{gross2010nonlinear}. Dunningham \emph{et al} proposed the use of number-correlated squeezed BEC states~\citep{dunningham2002interferometry}, which are more robust to particle loss than $N00N$ states. Another scheme by Dunningham and Burnett suggested that one can recreate the action of a (multiport) beam-splitter in an interferometer by simply tweaking the potential barrier between sites, thus enabling quantum-enhanced sensitivity~\citep{dunningham2004sub,cooper2009scheme}. Heisenberg-limited quantum analogs of inertial guidance systems, also known as \emph{atomic gyroscopes}, have been proposed in optical ring lattices~\citep{cooper2010entanglement}.  A multiconfigurational Hartree type self-consistent theoretical method for calculation of  quantum metrological quantities for interacting trapped Bosonic systems has also been proposed~\cite{baak2024self} . Precision estimation of chemical potential has also been studied in Bosonic quantum gases~\cite{marzolino2013precision}.  While strictly not many-body systems, criticalities associated with superradiant phase transitions engineered in optical cavities with Rabi or Jaynes-Cummings type interactions have also received renewed attention for the possibility of quantum-enhanced sensing ~\cite{garbe2020critical}. Moreover, effect of engineering anisotropies between coupling strengths of rotating wave and counter-rotating atom-cavity interactions of such models have also been studied in detail~\cite{wang2016the,Zhu2023criticality}.  
 
\subsection{Spin chain probes with criticality-enhanced sensitivity}

Spin chains are among the key models in many-body physics. They can be exploited as quantum probes in both static and dynamic scenarios. In this section, using explicit examples of critical spin chains, we study criticality in these models as a resource for quantum sensing. The most conventional type of criticality in spin chains is the second-order phase transition which has been described above in a general way. Some of these systems indeed support a second-order quantum phase transition, however this is not the only manifestation of criticality in spin systems.   Consider a simple one-dimensional quantum Ising model in a transverse magnetic field $h$ with the following Hamiltonian
\begin{equation}
\hat{H} =  h \underbrace{\sum_{i=1}^{N} \hat{\sigma}^{z}_{i}}_{\hat{H}_1}+J \underbrace{\sum_{i=1}^{N-1}  \hat{\sigma}^{x}_{i} \hat{\sigma}^{x}_{i+1}}_{\hat{H}_2}.
\label{eq:Ising_Transverse_Ham}
\end{equation}
By factoring $J$ and defining $\theta{=}h/J$, the Hamiltonian takes the general form of Eq.~(\ref{eq:general_2nd_order}). 
One can analytically solve the transverse Ising model Eq.~(\ref{eq:Ising_Transverse_Ham}) through Jordan-Wigner transformation~\citep{pfeuty1970one} which connects spin operators to fermionic ones as $\hat{c}_i=\prod_{j=1}^{i-1}(-\hat{\sigma}^z_i)\hat{\sigma}^{-}_i$ (and equivalently $\hat{c}_i^\dagger=\prod_{j=1}^{i-1}(-\hat{\sigma}^z_i)\hat{\sigma}^{+}_i$), where $\hat{\sigma}^{\pm}{=}(\hat{\sigma}^x\pm i\hat{\sigma}^y)/2$. The corresponding QFI for the dimensionless parameter $\theta$ has been analytically calculated for the ground state~\citep{zanardi2008quantum,invernizzi2008optimal} and has been shown to scale quadratically with system size, namely $I_Q\sim N^2$ near the critical point $h/J \rightarrow \pm 1$. 
Recently, a modular approach effectively augmenting these systems with periodic coupling defects~\citep{mukhopadhyay2023modular} has been proposed, which exploits the fact that multiple criticalities can be created in such a construction, all of which inherit the critical scaling behavior of the original chain. Alternatively, even for uniform chains, extending the ensemble measurement setup to real-time feedback control, i.e., adaptive strategies, allows one to overcome shot-noise scaling across the phase diagram ~\cite{salvia2023critical}. 

Other quantum spin models have also been extensively studied in the literature for determination of quantum-enhanced scaling. In addition to their relevance for quantum sensing, such QFI-based analysis often reveals the nature of multipartite entanglement in many-body quantum systems. These include the 1D quantum XY models~\citep{liu2013quantum,skotiniotis2015quantum,ye2016scaling} with Dzyaloshinsky-Moriya interactions~\citep{ozaydin2015quantum, ye2018quantum,ozaydin2020parameter}, XXZ models~\citep{liu2015quantum} which show Kosterlitz-Thouless phase transition~\citep{zheng2015probing}, Heisenberg-type rotationally symmetric models~\citep{ye2020quantum,lambert2019estimates}, and even applied to the quantum Hall probes~\cite{giraud2024quantum}. See Ref.~\citep{gu2010fidelity} for more references and earlier reviews. In addition to spinless fermion models, metrological advantage was also investigated for short-range spinful Fermi gases and QFI was shown to be an order parameter in such systems~\citep{lucchesi2019many}. Higher dimensional spin chains have also been shown to improve magnetometry with arbitrary spin-number transverse Ising models~\cite{singh2024dimensional}. Effect of lattice imperfections and disorder was considered in Ref.~\citep{wei2019fidelity}, whereby QFI analysis reveals that the two paradigmatic quantum disordered 1D lattices, namely Anderson ~\citep{anderson1958absence} and Aubry-Andre models~\citep{aubry1980analyticity} belong to different universality classes. Experimental demonstration of quantum-enhanced critical sensing has also been recently performed~\citep{liu2021experimental} with a small NMR system. 
Extensions to multiparameter estimation by calculating the QFI matrix have also been considered~\citep{bakmou2019quantum,di2022multiparameter} along with experimental demonstration for NMR systems~\citep{jiang2021multiparameter}. 
Departure from $T=0$, i.e., ground state assumption has been considered in several directions. One direction is to consider metrology in finite-temperatures. Similar to other markers of quantumness, presence of a quantum critical region fanning out from the critical point at $T=0$ has been detected with QFI~\citep{gabbrielli2018multipartite}.  Another popular direction of departure, namely considering driven quantum many-body sensors instead of probes at equilibrium has been discussed in a later section of this review.

\subsection{Quantum many-body probes with long-range interactions}

Since local interactions are useful in quantum-enhanced sensing, it is natural to ask whether criticality in models with longer-range interactions are also beneficial for metrology tasks~\cite{lakkaraju2023better}. Unfortunately, exact results are rare for such systems and approximate methods, such as Density Matrix Renormalization Group (DMRG), are often unsatisfactory for generic Hamiltonians with long-range interactions because of an area law violation. 
However, QFI has indeed been studied in many systems showing interesting physics. Perhaps the most studied model is the Lipshin-Meshkov-Glick (LMG) Hamiltonian for N sites in 1D, which generalizes the anisotropic XY Hamiltonian by dropping the nearest-neighbour assumption. 
The Hamiltonian can be written as
\begin{equation}
    \hat{H}_{P} = \frac{J}{N} \sum_{i,j,i < j} \left( \hat{\sigma}_{i}^{x}\hat{\sigma}_{j}^{x} + \gamma \hat{\sigma}_{i}^{y}\hat{\sigma}_{j}^{y} \right) + h \sum_{i} \hat{\sigma}_{i}^{z},
\end{equation}
\noindent where $\gamma$ is the anisotropy parameter, $J$ is exchange coupling,  and $h$ is the magnetic field. This model has been solved through Bethe ansatz~\citep{pan1999analytical,morita2006exact}, or alternatively by a Holstein-Primakoff transformation~\citep{dusuel2004finite,dusuel2005continuous}. Considering $J$ as the unit of interaction, this model has three estimable parameters, the magnetic field strength $h$, anisotropy $\gamma$, and the temperature $T$ when system is in thermal equilibrium. At zero temperature, this system shows quantum criticality at $h{=}h_c{=}J$. 
In Ref.~\citep{kwok2008quantum}, it has been shown that the QFI with respect to $h$ diverges at criticality as  $I_Q\sim (h-h_c)^{-2}$ (namely $\alpha{=}2$) in the symmetric phase ($h/J \geq 1 $) and $I_Q\sim (h_c-h)^{-1/2}$ (namely $\alpha{=}1/2$) in the broken phase ($0 \leq h/J \leq 1$). Remarkably, these exponents for QFI do not depend on the anisotropy $\gamma$. While at the critical point $h_c$, one obtains $I_Q(h_c)\sim N^{4/3}$, indicating quantum-enhanced sensitivity, two-site reduced density matrices results in $I_Q(h_c)\sim N^{2/3}$~\citep{ma2008reduced}. In~\citep{salvatori2014quantum}, the authors considered the estimation of $\gamma$ and $T$. 
In the thermodynamic limit near criticality, the leading contribution to QFI for both scales as $I_Q(h_c) \sim T^2$ at both phases, thus making this system attractive for relatively high-temperature thermometry. At this point, one should note that this also reflects a long-standing challenge for low-temperature thermometry, i.e., when $T\rightarrow0$, the QFI tends to vanish and the estimation error diverges. Other approaches like engineering non-Markovian interactions bosonic probes have been proposed to overcome this \cite{zhang2022non} divergence, although we are not aware of any such attempt in this many-body spin-chain context. Moreover, QFI for estimating anisotropy $\gamma$ in the thermodynamic limit shows a divergence at $\gamma = 1$ and scales differently in the ordered phase ($h > 1$) compared to the symmetry-broken phase ($0 \leq h < 1$) following the formula 

\begin{equation}
    I_{Q}(\gamma) \sim \begin{cases}
    \frac{9}{4(h-1)^2} + \frac{25 \beta^2}{12} + O(h) & h > 1 \\
    \frac{9}{4(\gamma -1)^2} - \frac{25 \beta^2 (h-1)}{6 (\gamma -1)} + O(h) & 0 \leq h < 1.
    \end{cases}\label{lmg_wrt_gamma}
\end{equation} 

\noindent Garbe \emph{et al}~\citep{garbe2022critical} have recently considered the influence of various driving protocols on QFI-based sensing in LMG and other fully connected systems.
In LMG and other fully connected models, all interactions are equally strong,  in contrast to models where all interactions except the nearest or next nearest are ignored. Thus, one may wonder what happens for models where long-range interactions are present, but the interaction strength decays following a power law~\citep{fernandez2018heisenberg}. Such models become extremely relevant for ion-trap based platforms, where the decay exponent can be fixed by tuning the lasers~\cite{monroe2021programmable}. In Ref.~\citep{yang2022super}, the authors considered a long-range Kitaev chain, which is an extension of the $N$-site tight-binding model, and whose Hamiltonian is given by 

\begin{align}
\hat{H} & =-\frac{J}{2}\sum_{j=1}^{N}(\hat{c}_{j}^{\dagger}\hat{c}_{j+1}+\hat{c}_{j+1}^{\dagger}\hat{c}_{j})-\mu\sum_{j=1}^{N}(\hat{c}_{j}^{\dagger}\hat{c}_{j}-\frac{1}{2})\nonumber \\
 & +\frac{\Delta}{2}\sum_{j=1}^{N-1}\sum_{l=1}^{N-j}\kappa_{\eta,\,l}(\hat{c}_{j}\hat{c}_{j+l}+\hat{c}_{j+l}^{\dagger}\hat{c}_{j}^{\dagger}),\label{eq:LRK}
\end{align}
\noindent where $\kappa_{\eta,l}$ is the coefficient for interaction between two sites $l$ distance apart, $J$ is the nearest neighbour coupling and $\mu$ is the chemical potential. The power-law decay is a specific case when $\kappa_{\eta,\,l} \propto 1/l^{\eta}$. The exponent $\eta$ controls the range of interaction. For example, $\eta{=}0$ represents a fully connected graph in which all pairs interact equally while $\eta\rightarrow \infty$ represents nearest neighbor interaction. The following expression for QFI for estimating the parameter $\Delta$ was obtained in~\citep{yang2022super}

\begin{equation}
I_{Q}(\Delta)\sim\begin{cases}
N^{2}(\ln N)^{2(1-\eta)} & \eta\in[0,\,1)\\
N^{2}(\ln\ln N)^{2} & \eta=1\\
N^{2} & \eta>1
\end{cases}.\label{eq:scaling0-double-int}
\end{equation}

\noindent Clearly, QFI shows quantum-enhanced sensitivity if the long-distance couplings decay slower than Coulombic, i.e.~$\eta\leq 1$. 

\section{Equilibrium Quantum Sensing: Localization Transition}

So far, second-order quantum phase transition in the ground state of many-body systems, manifested in different setups, has been investigated as a resource of enhanced precision sensing. In contrast, localization transitions, exemplified through Anderson and many-body localizations, affect the entire spectrum of the system and thus their impact is more drastic. Originally disorder was used for suppressing the effect of particle tunneling and thus inducing localization in which particles' wave functions extend only locally across a few sites. Localization takes place in both non-interacting (Anderson localization) and interacting (many-body localization) systems. The many-body localization transition is energy dependent such that each eigenstate localizes at a different value of disorder~\cite{luitz2015many,guo2021observation,yousefjani2023mobility}. Localization transition can also be induced by Floquet dynamics~\cite{yousefjani2023floquet} pseudo-random potentials~\cite{zhang2018universal,vznidarivc2018interaction} or Stark fields~\cite{schulz2019stark,morong2021observation}. In this section, we investigate the application of both single- and multi-particle systems under the effect of such localization effects for quantum sensing. 

\subsection{Pseudo-random localization transition}
A paradigmatic model which exhibits localization-delocalization transitions has been studied for quantum sensing purpose~\cite{sahoo2024localization}. It  is   represented  by  one-dimensional fermionic lattice in presence of a quasi-periodically modulated onsite potential, i.e., the potential varies from site-to-site. The Hamiltonian of the system is written as
\begin{eqnarray}\label{eq:Ham}
 \hat{H} &=& - J \sum_{i}(\hat{c}^{\dagger}_{i}\hat{c}_{i+1}+h.c)+V \sum_{i} \cos(2\pi i \omega) \hat{c}^{\dagger}_{i}\hat{c}_{i} 
 \cr 
 &+&U \sum_i \hat{n}_i \hat{n}_{i+1},
\end{eqnarray}
where  $J$ is tunneling, $\omega$ is an irrational number, $U$ is interaction strength, and $V$ is the unknown parameter which we wish to estimate. In addition, the operator  $\hat{c}^{\dagger}_{i} (\hat{c}_i)$ is fermionic creation (annihilation) operator at site $i$ and $\hat{n}_i=\hat{c}^{\dagger}_{i}\hat{c}_{i}$ accounts for particle number. The non-interacting case, i.e.~$U=0$, is known as  the  Aubry-Andr\'e-Harper (AAH)  model~\cite{PhysRev.76.1269}. 
In the AAH model with a single particle, there is an energy independent localization transition at a finite modulation strength, $V_c=2$. For a given $V$, all the states are either localized (for $V > V_c$) or extended (for $V < V_c$)~\cite{PhysRevB.14.2239}.
For finite size systems, proper scaling emerges at the transition for system sizes $F_n$ with either odd or even sequences from the Fibonacci series and for $\omega$ to be approximated by $\omega_n{=}F_n/F_{n+1}$. Here $F_n$ and $F_{n+1}$ are two consecutive Fibonacci numbers with the property $\omega{=}\lim_{n\to \infty}(F_n/F_{n+1}) \to (\sqrt{5}-1)/2$, which is the so-called golden ratio.

In addition to the single-particle case, one may consider the half-filled case in the presence of interaction $U$. The localization transition persists in the ground state in the presence of interaction~\cite{PhysRevLett.115.180401,mastropietro2017localization}. Moreover, further theoretical works have argued in favour of many-body localization (MBL)~\cite{PhysRevLett.119.075702,10.21468/SciPostPhys.1.1.010,PhysRevB.96.104205}, albeit with different universality properties than the uncorrelated disordered systems. 

\begin{figure}
    \includegraphics[width=\linewidth]{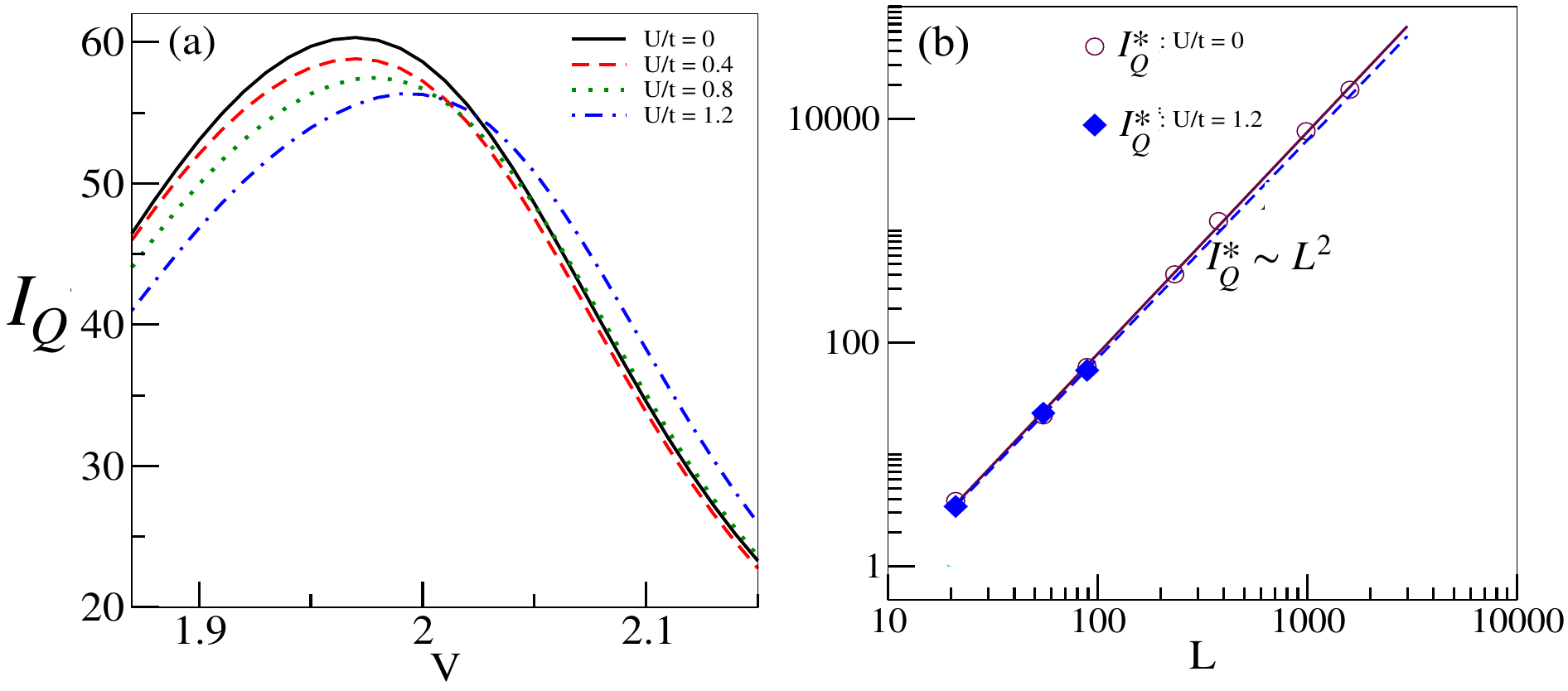}
    \caption{\textbf{Quasi-random disordered probe}. (a) QFI as a function of $V$ for different interaction stregths $U$ (see Eq.~\ref{eq:Ham}) for particular case at half-filling ($L=89$ and $n_f=45$). The plot shows $I_Q$ as a function of $V$ for varied interaction strengths (up to the interaction strength comparable to the kinetic energy). (b) The maximum QFI, $I_Q^{*}$,  as a function of $L$ for $U{=}0$. The number of fermions $n_f$ for $L$ = $21, 55, 89, 233, 377, 987$, and $1597$ are $11, 28, 45, 116, 189, 494,$ and $798$, respectively. The $I_Q^{*}$ nearly saturates the Heisenberg limit, i.e., $I_Q^{*}{\sim}L^2$ for both, the bare system and a system with moderate interaction ($U{=}1.2)$. Figure is adopted from~\cite{sahoo2024localization}.}
    \label{fig:fig1}
\end{figure}

Interestingly, the AAH model can be of potential use for quantum sensing of the strength of the quasi-modulated potential, $V$~\cite{sahoo2024localization}.
Figure \ref{fig:fig1} presents $I_Q$ as a function of $V$ for different interaction strength $U$ and fixed $L$. Two observations are noted: First, the peak tends to shift towards a higher $V^*$ ($V^*$ is the value of $V$ where $I_Q$ is maximum in finite system of size $L$) with increasing $U$. This is expected as many-body localization transition is supposed to occur at $V_c > 2$ in the presence of interaction; and second, the value of $I_Q$ tends to slightly decrease at $V^*$ with increasing $U$. The scaling of the peak of the QFI which occurs at $V=V^*$, namely $I_Q^*=I_Q(V^*)$ is presented in Fig.~2(b) which reveals scaling of the form $I_Q^*(U{=}0){=} L^{1.98(2)}$ in the non-interacting limit. Despite the lack of enough data points, it is evident that the effects of interaction on the scaling exponent remain pretty small in the range of weak to intermediate range.

\subsection{Stark localization}\label{Section:Stark_Probe}

Stark localization caused by an applied gradient field across a lattice has been exploited for ultra-precise sensing~\cite{he2023stark,yousefjani2023longrange,yousefjani2025nonlinearity,sahoo2024enhanced}.
The extended-localized transition in the limit of large one-dimensional systems takes place at an infinitesimal gradient field. 
This allows for sensing weak fields with great precision.
This section is devoted to Stark probes and their capability to serve as a weak-gradient field sensor.
\\

\begin{figure}[t]
\includegraphics[width=\linewidth]{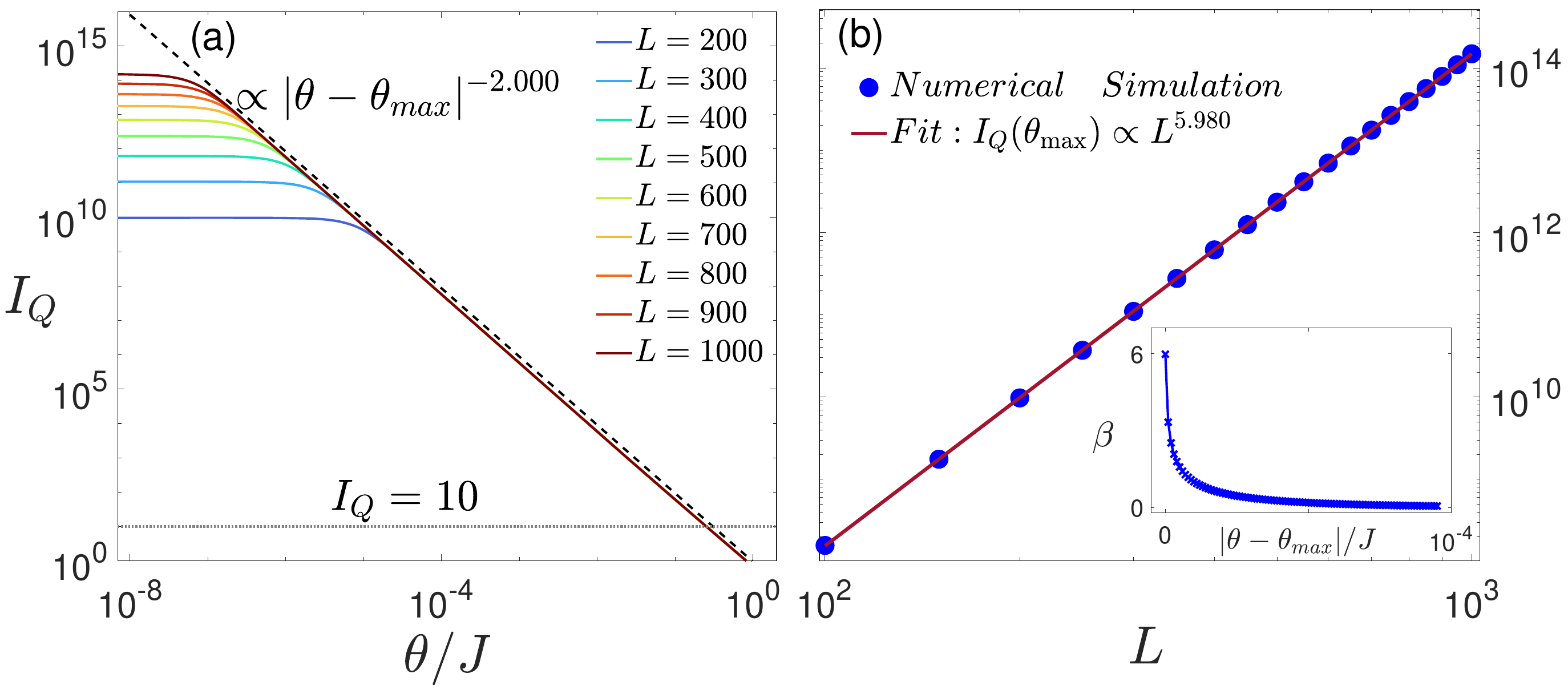}
\caption{\textbf{Single-particle Stark probe;} (a) QFI as a function of the Stark filed $\theta$ for single-particle Stark probe, prepared in the ground state of Eq.~(\ref{Eq:SingleParticle_Stark_Hamiltonian}) with various sizes. The dashed line is the best fitting function to describe the algebraic behavior of the QFI in the localized regime when $L{\rightarrow}\infty$.
(b) the maximum of the QFI at $\theta_{\max}$ versus probe size $L$. The numerical results (markers) are well-fitted by function $I_{Q}{\propto}L^{\beta}$  (line) with $\beta{=}5.98$. The inset shows the scaling behavior far from the criticality. Figure is adapted from~\cite{he2023stark}.}\label{fig1:Stark_probe}
\end{figure}

For a single particle that tunnels to its neighbors in a one-dimensional lattice affected by a linear gradient field with strength $h$, the Hamiltonian is given by
\begin{equation}\label{Eq:SingleParticle_Stark_Hamiltonian}
\hat{H} = J\sum_{i=1}^{L-1} \left( \vert i\rangle\langle i+1\vert + \vert i+1\rangle\langle i\vert  \right) + \theta \sum_{i=1}^{L}  i \vert i\rangle\langle i\vert,
\end{equation}
where $J$ is the tunneling rate. In this setup, the gradient-field strength $\theta$ is the unknown parameter and aimed to be estimated.  
It is well known that in the thermodynamic limit (i.e.~$L{\rightarrow}\infty$), all energy levels of the Hamiltonian Eq.(\ref{Eq:SingleParticle_Stark_Hamiltonian}) go through a phase transition from an extended to a localized phase at $\theta_c{=}0$~\cite{van2019bloch,kolovsky2008interplay,
schulz2019stark,chanda2020coexistence,yao2020many}. 
Focusing on the ground state, Fig.~\ref{fig1:Stark_probe} (a) presents QFI as a function of $h$ for different sizes of the probe. 
The initial plateaus represent the extended phase in which the QFI takes its maximum for $\theta{\leq}\theta_{\max}$. Clearly, $h_{\max}$ skews to zero and tends to the critical point $\theta_{c}{=}0$, as the length of the probe increases.  
The maximum values of the QFI, denoted by $I_{Q}(\theta_{\max})$, increase dramatically by enlarging the probe, evidencing divergence of the QFI in the thermodynamic limit. 
In the localized phase, namely $\theta{>}\theta_{\max}$, the QFI is size-independent and decays algebraically according to $I_{Q}{\propto}|\theta-\theta_{\max}|^{-\alpha}$ with $\alpha{=}2$ for the ground state. 
The scaling of the QFI with the system size at the critical point has been plotted in Fig.~\ref{fig1:Stark_probe} (b). 
The QFI shows strong quantum-enhanced sensitivity, i.e.~$I_{Q}(\theta_{\max}){\propto}L^{\beta}$ with $\beta{\simeq}6$.
The inset of Fig.~\ref{fig1:Stark_probe} (b) shows how $\beta$ changes by getting distance from the critical point, i.e.~$\theta_{\max}$.
As the Stark localization transition happens in the entire energy spectrum of Eq.~(\ref{Eq:SingleParticle_Stark_Hamiltonian}), one can inquire about higher energy eigenstates' scaling.
Interestingly, one can obtain the quantum-enhanced sensitivity with $\beta{\simeq}4$ for all excited eigenstates.
Following section~\ref{sec:2nd-order-pt}, establishing finite-size scaling analysis results in critical exponents $(\alpha,\nu){=}(2.00,0.33)$ and $(\alpha,\nu){=}(4.00,1.00)$ for the ground state and a typical eigenstate from the mid-spectrum, respectively.
Indeed, three exponents $\beta,\alpha$ and $\nu$ obey $\beta{=}\alpha/\nu$ in Eq.~(\ref{eq:exponents}), regardless of the energy level.
\\

\begin{figure}[t]
\includegraphics[width=\linewidth]{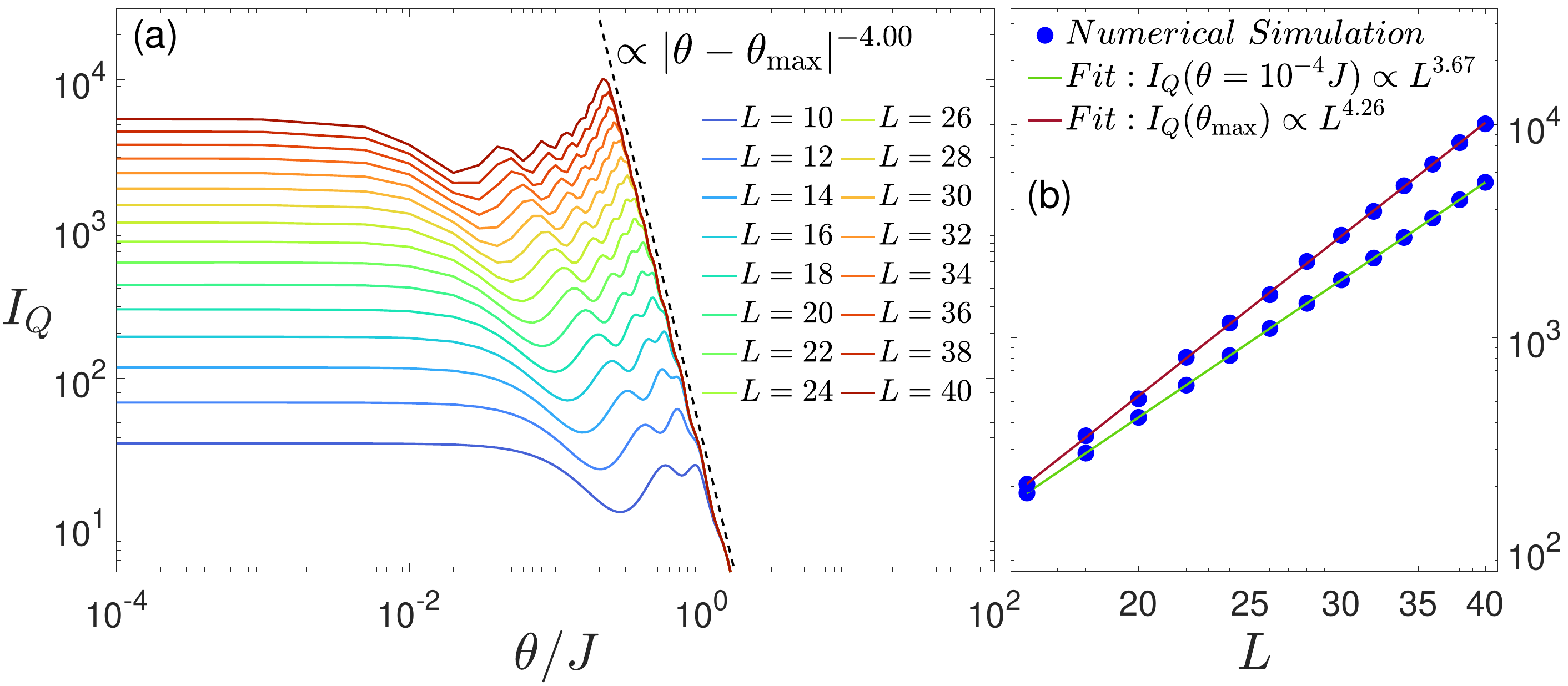} 
\includegraphics[width=\linewidth]{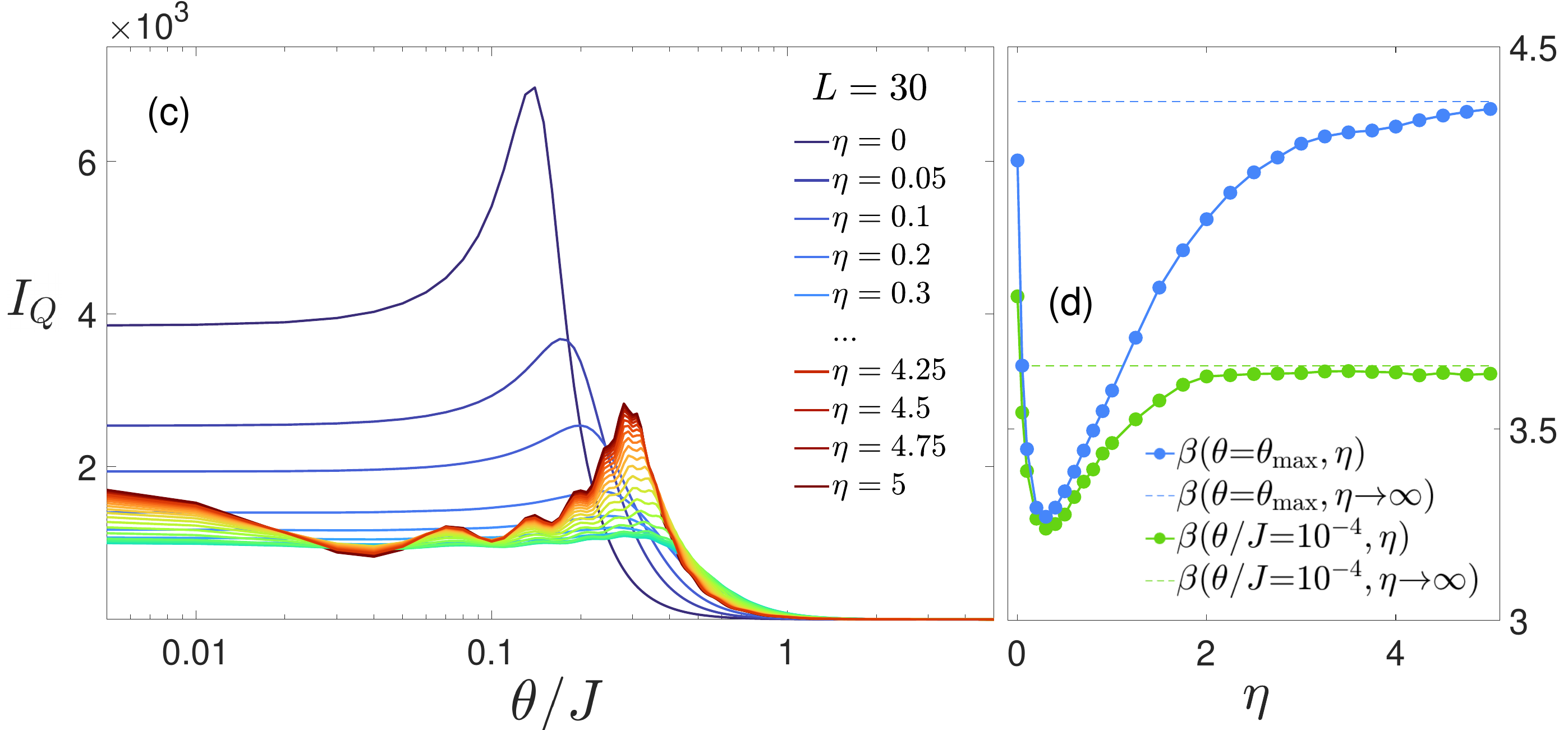} 
\caption{\textbf{Many-body Stark probe;} (a) QFI versus $\theta$ for many-body Stark probe with nearest-neighbor interaction prepared in the ground state of Eq.~(\ref{Eq:Manybody_Stark_Hamiltonian}) with $\eta{\rightarrow}\infty$ and different sizes. The dashed line which is the best-fitting function shows the asymptotic behavior of the QFI in the localized phase.
(b) the QFI versus $L$ in extended phase, i.e.~$\theta{=}10^{-4}J$, and critical point, i.e.~$\theta{=}\theta_{\max}$. The numerical results (markers) are well-described by fitting function $I_{Q}{\propto}L^{\beta}$ (line) with $\beta{=}3.67$ and $\beta{=}4.26$ for the extended phase and critical point $\theta_{\max}$, respectively. 
(c) QFI versus $\theta$ for many-body Stark probe prepared in the ground state of Eq.~(\ref{Eq:Manybody_Stark_Hamiltonian}) with different $\eta$ and $L{=}30$. 
(d) The obtained scales of the QFI, namely $\beta(\theta,\eta)$ 
 versus range of the interaction $\eta$, in both extended phase ($\theta{=}10^{-4}$) and critical points ($\theta{=}\theta_{\max}$). Figure is adapted from~\cite{he2023stark,yousefjani2023longrange}.}\label{fig2:Stark_probe}
\end{figure}

The offered quantum-enhanced sensitivity by Stark localization is not limited to the non-interacting probe. 
To study the effect of the interaction, one can consider a one-dimensional probe of size $L$ containing $N_{\rm ex}{=}L{/}2$ excitations with tunneling to their next neighbor and interaction range that decay algebraically depends on the exponent $\eta{>}0$. 
The Hamiltonian reads
\begin{equation}\label{Eq:Manybody_Stark_Hamiltonian}
\hat{H} = J\sum_{i=1}^{L-1} (\hat{\sigma}^{x}_{i}\hat{\sigma}^{x}_{i+1}+\hat{\sigma}^{y}_{i}\hat{\sigma}^{y}_{i+1}) + \sum_{i<j}\frac{J}{|i-j|^\eta}\hat{\sigma}^{z}_{i}\hat{\sigma}^{z}_{j} + \theta \sum_{i=1}^{L}  i\hat{\sigma}_{i}^{z},
\end{equation}
with $J$ and $\theta$ as the coupling strength and the magnitude of the gradient field, respectively.
By changing the exponent $\eta$, one can smoothly interpolate between a fully connected graph for $\eta{=}0$ and a conventional nearest-neighbor one-dimensional chain for $\eta{\rightarrow}\infty$.
In the following, after presenting the results for the probe with short-range interaction (i.e.~$\eta{\rightarrow}\infty$), we investigate the impact of long-range interaction (i.e.~$\eta{<}\infty$) to address this question that whether longer range of interaction can provide more benefits for sensing tasks.
\\

\textbf{Short-range interaction}.$-$ Fig.~\ref{fig2:Stark_probe}(a) represents the QFI as a function of $\theta$ for various $L$ obtained for the ground state ~\cite{he2023stark}. 
By increasing the Stark field, the system goes through a phase transition from the extended phase to the localized one.
Clearly, the QFI peaks at some $\theta_{\max}$ which gradually moves to zero, indicating that the Stark localization transition in the many-body interacting probe takes place at an infinitesimal gradient field.
While finite-size effects are evident in the extended phase ($\theta{<}\theta_{\max}$), in the localized phase ($\theta{>}\theta_{\max}$), the QFI is size-independent and decays algebraically as $I_{Q}{\propto}|\theta-\theta_{\max}|^{-\alpha}$ with $\alpha{=}4.00$.
The scaling behavior of the Stark probe has been presented in Fig.~\ref{fig2:Stark_probe}(b) for both extended phase ($\theta{=}10^{-4}J$) and transition point ($\theta{=}\theta_{\max}$). 
The best fitting function to describe the numerical results (markers) obtained as $I_{Q}{\propto}L^{\beta}$ (lines) with $\beta{=}3.67$ and $\beta{=}4.26$, for extended phase and transition point, respectively.
Obviously, similar to the single particle probe, the many-body interacting probe also results in strong quantum-enhanced sensitivity. 
\\

\textbf{Long-range interaction}.$-$
To explore the advantages offered by long-range interactions in the Stark probe, Ref.~\cite{yousefjani2023longrange} presents the QFI for a Stark probe of size $L{=}30$ prepared in the ground state of Hamiltonian Eq.~(\ref{Eq:Manybody_Stark_Hamiltonian}) with different $\eta$'s.
The results are shown in Fig.~\ref{fig2:Stark_probe}(c).
Regardless of the range of the interaction, increasing $\theta$ results in transition from extended phase to localize one, highlighted by a peak at $\theta_{\max}(\eta)$.
Various $\eta$'s leaves different imprints on the QFI.
By decreasing the range of the interaction the induced Zeeman energy splitting on each site changes from a uniform to nonuniform (resembling disorder) which causes diminution in QFI.
The effect of nonuniform energy splitting gradually vanishes as the interaction tends to short-range.
By considering a fitting function of the form $I_Q\sim L^\beta$, one can estimate the exponent $\beta$ for various choices of $(\theta,\eta)$. The results are presented in  Fig.~\ref{fig2:Stark_probe}(d) which clearly shows the possibility of quantum-enhanced sensitivity over a wide range of interactions.
\\

Capturing the ultimate precision limit provided by the QFI, and equally saturating the Cram\'er-Rao bound in Eq.~(\ref{eq:cramer-rao-classical-quantum}), demands an experimental-friendly measurement setup. 
In Ref.~\cite{he2023stark} it has been shown that a simple position measurement described by local projectors $\{\hat{\Pi}_{i}=|i\rangle\langle i|\}_{i=1}^{L}$ results in CFI closely matches with the QFI in single particle probe. 
For many-body interacting case, the best observable to closely catch the QFI is obtained as $\hat{\mathcal{O}}{=}\hat{\Pi}_{i=1}^{L}\hat{\sigma}_{i}^{z}$~\cite{yousefjani2023longrange,yousefjani2025nonlinearity}.

\section{Equilibrium quantum sensing: Topological Phase Transitions}\label{sec:topo-phase}

Second order quantum phase transitions are known for various features including spontaneous breaking of a continuous symmetry, local order parameters, gap closing and scale invariance behavior. While such critical phenomena is usually good for quantum-enhanced sensitivity, it is not clear which of these features are responsible for achieving quantum-enhanced sensitivity. Moreover, although these models are not as noise-sensitive as interferometric schemes, detailed analysis shows reduction in the QFI scaling with increasing noise~\cite{chen2021effects}. Thus, building quantum many-body sensors based on beyond second-order phase transitions is crucial. There also exists phase transitions where the corresponding order parameter is global. These are called topological phase transitions  and are characterized by discrete jump in topological indices of the ground state of the physical system such as the Chern number~\citep{zhang2008topological,stanescu2016introduction,goldman2016topological}. In addition, topological systems are expected to be naturally more robust against local noise and thus their performance as a sensor be less noise-affected. In this section, we consider two different topological sensors, namely  Symmetry Protected Topological (SPT) systems and non-Hermitian topological probes. 

\subsection{Symmetry protected topological sensors}

Topological phase transitions, exemplified through SPT transitions in free-fermionic systems, are fundamentally different from conventional second-order quantum criticality. In fact, SPT transitions are characterized by: (i) the emergence of robust edge/surface states which are protected against symmetry-preserving local perturbations~\cite{alldridge2020bulk,alase2019boundary}; (ii) an integer-valued nonlocal quantity called a topological invariant~\cite{bernevig2013topological}; and (iii) the absence of long-range correlations and entanglement~\cite{chen2013symmetry}.  

One of the simplest and earliest examples of free-fermionic systems with SPT transition is the Su-Schrieffer-Heeger (SSH) model~\citep{su1979solitons} which consists of one excitation in a 1D tight-binding lattice with $N$ sites consisting of two different sublattices characterized by bonds of alternating strengths. The Hamiltonian is given by
\begin{equation}
    \hat{H}_{\rm SSH} = -\sum_j \left( t_1 \hat{b}_{j}^{\dagger} \hat{a}_{j} + t_2 \hat{a}_{j+1}^{\dagger} \hat{b}_{j} \right).
\end{equation}
where $t_1$ and $t_2$ are intra- and inter-site hopping rates and $\hat{a}_j$, $\hat{b}_j$ are the annihilation operators for the two sublattices. 
The parameter of interest is $\theta{=}t_1/t_2$.  
For 2D and higher dimensional systems, the quintessential examples of topological systems are the so called Chern insulators and generalizations thereof (see Ref.~\citep{qi2011topological} for a detailed review). For a simple system that has been realized in optical lattices with cold atoms~\citep{wu2016realization}, the spin-orbit coupled Hamiltonian is given by 
\begin{align}
\label{Chham}
    \hat{H}^{\rm Ch} = \displaystyle\sum_{\boldsymbol{k}} \begin{bmatrix} \hat{c}^{\dag}_{\boldsymbol{k},\uparrow} & \hat{c}^{\dag}_{\boldsymbol{k},\downarrow}\end{bmatrix} \enspace \hat{H}_{\boldsymbol{k}}^{\rm Ch} \enspace
    \begin{bmatrix} \hat{c}_{\boldsymbol{k},\uparrow} & \hat{c}_{\boldsymbol{k},\downarrow}\end{bmatrix} ^T \,,
\end{align}
where $\hat{H}_{\boldsymbol{k}}^{\rm Ch} = \boldsymbol{B} \cdot \boldsymbol{\hat{\sigma}}$ is the Bloch Hamlitonian with $\boldsymbol{B} = \left(2 t_1 \cos{k_x}, \, 2 t_1 \cos{k_y}, \, m_z + 2 t_2 (\sin{k_x} + \sin{k_y})\right)$ and $\boldsymbol{\hat{\sigma}}$ the vector of Pauli matrices. 
Here $\uparrow,\downarrow$ denote spin-$1/2$ up and down states, and $m_z, t_1, t_2$ are lattice parameters of a $N \times N$ square lattice. {\color{black} The parameter to be estimated can be considered as the ratio between the two parameters of the Hamiltonian, namely $\theta{=}m_z/t_2$, which is a dimensionless parameter. The eigenvectors form two bands that touch at phase transition at the Dirac points $(k_x, k_y) {=} \pm(\pi/2, \pi/2)$ for nonzero $\theta$. By varying $\theta$ the system goes through a phase transition at $\theta{=}\theta_{\rm c} {=} \mp 4$~\citep{qi2011topological}}. These SPT models are generically short-range entangled and their topological order are protected only under transformations with respect to some discrete symmetry. 

\begin{figure}[t]
\centering
  \begin{tabular}{cc}
    \includegraphics[width=0.45\linewidth]{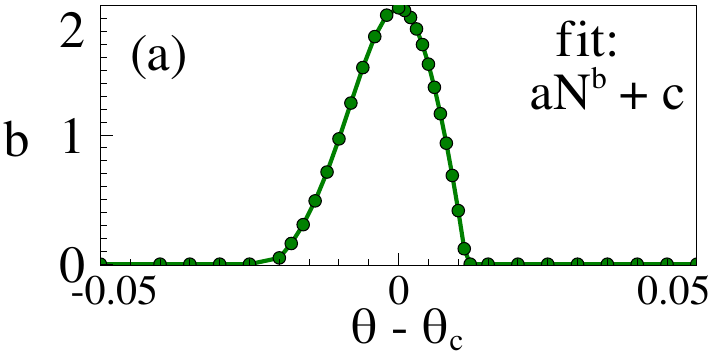} 
    \includegraphics[width=0.45\linewidth]{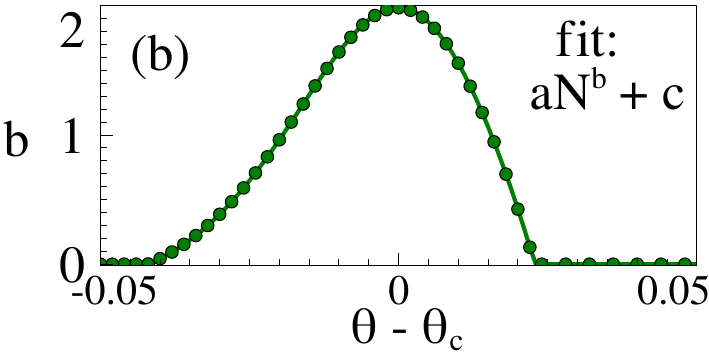} \\
    \includegraphics[width=0.45\linewidth]{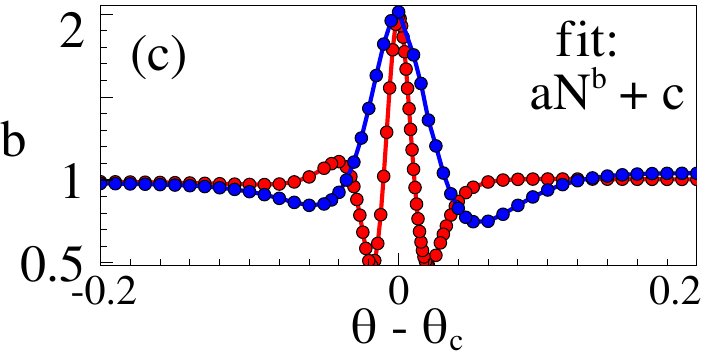} 
    \includegraphics[width=0.45\linewidth]{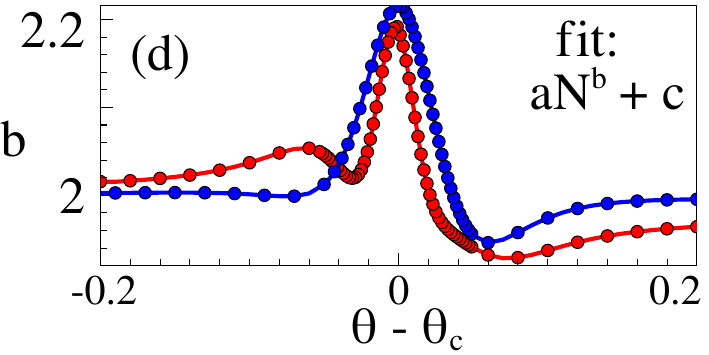} \\
  \end{tabular}
\caption{\textbf{QFI scaling near transition}. Top panels : Scaling exponent of QFI of edge state (for $\theta < \theta_{\rm c}$) and corresponding bulk state (for $\theta > \theta_{\rm c}$) as a function of $\theta$ for (a) SSH model, and (b) Chern insulator model. Bottom panels : Scaling exponent of QFI of many-body ground state as a function of $\theta$ subject to PBC (blue) and OBC (red) for (c) SSH model, and (d) Chern insulator model. Figures adapted from Ref.~\citep{sarkar2022free}.}
\label{fig_edge}
\end{figure}

For topological systems, Ginzburg-Landau theory fails since local order parameters do not exist and no continuous symmetry is broken. Hence the fidelity susceptibility or QFI based approach to detect topological phase transitions holds special relevance for such models, in addition to the practical task of building many-body quantum sensors. Extensive studies in this regard has been made on lower dimensional systems, including 1D Kitaev chain~\citep{mondal2024multicritical},  and Kitaev's deformed toric code models~\citep{abasto2008fidelity} and honeycomb models~\citep{yang2008fidelity,zhao2009singularities} in 2D. Gu and Lin~\citep{gu2009scaling} also pointed out that the critical exponent of fidelity for the honeycomb model is super-extensive and different for two sides of the critical point. Thus, fidelity susceptibility is indeed an useful tool for detecting topological phase transitions. In~\citep{sarkar2022free}, the possibility of quantum enhancement of sensing in edge and ground states of topologically non-trivial single-excitation subspaces of generic free-Fermion Hamiltonians was investigated. For edge states, $N^\beta$ scaling, with $\beta{=}2 $ of QFI was obtained at the transition point and constant scaling away from criticality, as shown in Figs.~\ref{fig_edge}(a) and (b). Significantly, the optimal measurement for saturating the Cram\'{e}r-Rao bound was revealed as a constant site-location measurement on the lattice for every value of the parameter $\theta$. For many-body ground states, quadratic scaling of QFI was again obtained around the topological phase transition point, with classical linear scaling away from criticality, as shown in Figs.~\ref{fig_edge}(c) and (d). 

For multiparameter estimation, it has been suggested recently that the Berry curvature and related topological invariants (e.g., the Chern Number) act as physically relevant bounds for the QFI matrix used for multiparameter estimation~\citep{mera2022relating}. Cai and his collaborators~\citep{yu2022experimental} have succeeded in experimentally demonstrating these bounds for a 2D Chern insulator system synthesized in an NV-center based setup. 

\subsection{Non-Hermitian quantum sensors}

Non-Hermitian Hamiltonians can generate non-Unitary dynamics which is closely connected with open system dynamics~\cite{ashida2020non}.
For example, in the Lindbladian description, the evolution of the density operator can be expressed in terms of the properties of the non-Hermitian Liouvillian super-operator.
Also, for the dynamics during which no Lindbladian jump operator acts (the probability of which decreases exponentially in time), an effective non-Hermitian Hamiltonian can describe a pure state evolution.
With recent experimental progress, it is possible to design specific non-Hermitian Hamiltonians that determine the evolution of the system.
In particular, the topological systems bear many surprising features due to the unique properties of the eigenstates and complex eigenvalues~\cite{bergholtz2021exceptional}.
In the absence of degeneracy, the eigenstates are linearly independent and form a complete basis, but they are not orthogonal.
This stems from the fact that the right- and left-eigenstates are not the same.
For a non-Hermitian Hamiltonian $\hat{H}_{\rm NH}$, they are defined as
\begin{align}
    \hat{H}_{\rm NH} \ket{\psi^{\rm R}_n} &= E_n \ket{\psi^{\rm R}_n}, \nonumber \\
    \bra{\psi^{\rm L}_n} \hat{H}_{\rm NH} &= E_n \bra{\psi^{\rm L}_n} \implies \hat{H}_{\rm NH}^{\dag} \ket{\psi^{\rm L}_n} = E^*_n \ket{\psi^{\rm L}_n} ,
\end{align} 
where $E_n$ is the corresponding eigenvalue.
However, these states are biorthogonal~\cite{brody2014biorthogonal}, and upon normalization gives, $\braket{\tilde{\psi}^{\rm L}_m | \tilde{\psi}^{\rm R}_n} {=} \delta_{mn}$.
Here, $\ket{\tilde{\psi}^{\rm R}_n} {=} \ket{\psi^{\rm R}_n} / \sqrt{\braket{\psi^{\rm L}_n | \psi^{\rm R}_n}}$, and $\ket{\tilde{\psi}^{\rm L}_n} {=} \ket{\psi^{\rm L}_n} / \sqrt{\braket{\psi^{\rm L}_n | \psi^{\rm R}_n}} \, ^*$.
This helps in writing down the completeness relation $\sum_n \ket{\tilde{\psi}^{\rm R}_n} \bra{\tilde{\psi}^{\rm L}_n} {=} \mathds{1}$ and the spectral decomposition $\hat{H}_{\rm NH} {=} \sum_n E_n \ket{\tilde{\psi}^{\rm R}_n} \bra{\tilde{\psi}^{\rm L}_n}$.
A crucial point to note while considering sensing with non-Hermitian systems is that the norm of the state is not conserved during evolution.
Therefore, to obtain a normalized probability distribution from the measurement outcomes, the state needs to be divided by the norm calculated with standard inner products. 
In this context, the valid probe corresponding to a right-eigenstate is $\ket{\tilde{\psi}^{\rm R}_n}\bra{\tilde{\psi}^{\rm R}_n} / \text{Tr}(\ket{\tilde{\psi}^{\rm R}_n}\bra{\tilde{\psi}^{\rm R}_n})$.
Under the action of a non-Unitary operator $\hat{U} {=} e^{-i \hat{H}_{\rm NH} t / \hbar}$ on an initial state $\ket{\psi_0}$, the dynamical probe is given by the normalized density operator $\hat{U} \ket{\psi_0} \bra{\psi_0} \hat{U}^{\dag} / \text{Tr}(\hat{U} \ket{\psi_0} \bra{\psi_0} \hat{U}^{\dag})$.
This approach has been standardized through both theoretical works~\cite{alipor2014quantum, yu2023quantum} and experimental realizations~\cite{Xiao2024Non, Yu2024Toward}.
It ensures that the probe is a valid density operator with which the standard QFI definition~\cite{paris2009quantum, braunstein1994statistical} in Eq.~\eqref{eq:qfi-pure-states} can be applied.

Non-Hermitian Hamiltonians can have degeneracies analogous to Hermitian systems, i.e.~identical eigenvalues with distinct eigenstates.
The points in the Hamiltonian's parameter space where this happens are known as diabolic points.
There are also parameter values where more than one eigenvalues and the corresponding eigenstates can become identical.
These are known as Exceptional Points (EP) and have no Hermitian counterpart.
The right- and left-eigenstates corresponding to the EP are orthogonal to each other and the biorthogonal framework breaks down as the Hamiltonian is not digonalizable anymore.
If the Hamiltonian has parity-time (PT) symmetry, then it commutes with the antilinear and antiunitary $\mathcal{PT}$ operator. 
For example, in spin-$1{/}2$ systems on a lattice, the parity operator is $\hat{\mathcal{P}}{=}\sum_{j} \ket{-j} \bra{j}\otimes \hat{\sigma}^{x}_j$, and the time-reversal operator is $\hat{\mathcal{T}}{=} \sum_{j} \ket{j} \bra{j} \otimes \hat{\sigma}^{y}_j \hat{K}_j$.
Here the term $|{-j}\rangle \langle j |$ swaps the position of spins at sites $j$ and $-j$. 
The action of $\hat{K}_j$ is complex conjugation, and $\hat{\sigma}^{(x,y,z)}_j$ are the Pauli spin operators on $j$-th site that affect the spin degree of freedom~\cite{krasnok2021paritytime, mochizuki2016explicit}.
In an appropriate parameter regime, a PT-symmetric Hamiltonian supports a real energy spectrum, and it is known as the PT-unbroken phase.
In this case, the Hamiltonian and the PT operator share the same eigenstates, known as PT-symmetric eigenstates.
Upon variation of the Hamiltonian parameters, a PT-broken phase can be entered where some or all eigenvalues become complex that appear in complex conjugate pairs.  
This happens through a phase transition associated with spontaneous PT-symmetry breaking.
The PT-symmetry breaking point is an EP of the system.

\begin{figure}[t]
    \centering
    \includegraphics[width=0.99\linewidth]{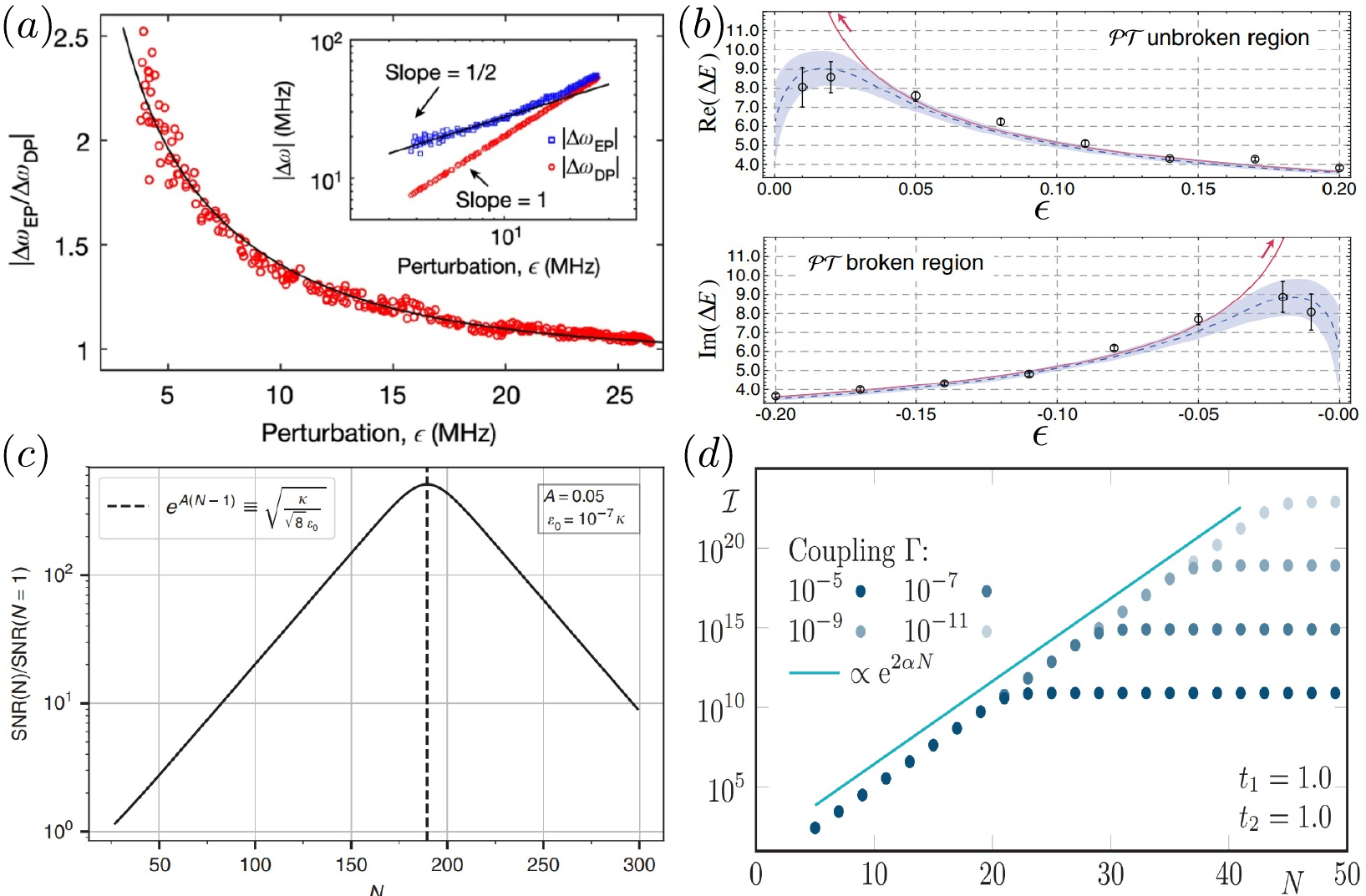} 
    \caption{\textbf{EP based and non-reciprocity based sensors}. 
    (a) Energy slitting versus perturbation strength $\epsilon$ near EP shows enhancement over standard splitting near a diabolic point (DP). 
    For EP of order 2 in Ref.~\cite{chen2017exceptional}, the black line shows the expected $\epsilon^{1/2}$ behavior. 
    The inset shows scaling exponent of EP based sensor (blue squares) increasing form $1/2$ (black line) as $\epsilon$ increases. 
    The exponent for DP based sensor is $1$ (red circles). 
    Figure adapted from Ref.~\cite{chen2017exceptional}. 
    (b) For a PT-symmetric system in Ref.~\cite{yu2020experimental}, EP of order 2 appears at the symmetry breaking point.
    The black circles are the experimental data for the energy splitting with perturbation $\epsilon$.
    The red line shows the expected $\sqrt{|\epsilon|}$ behavior. 
    In the PT-unbroken regime, the energies are real and the upper panel shows the splitting.
    In the PT-broken regime of the system considered , the real part of energy stays constant while the splitting is observed in the imaginary part (lower panel).
    As the EP is approached, deviation from the expected value occurs due to excessive noise.
    Figure adapted from Ref.~\cite{yu2020experimental}. 
    (c) The signal-to-noise ratio (SNR) shows the exponentially enhanced sensitivity with system size $N$ in the non-reciprocity based sensing protocol in Ref.~\cite{mcdonald2020exponentially}.
    It is based on the Hatano-Nelson model in Eq.~\eqref{HN_ham} and $A = \log{(J_R/J_L)}/2$.
    The analysis is based on linear response theory which breaks down for very large $N$, due to amplification of noise.
    The expected value of $N$ for this to happen with fixed values of other parameters is shown with dashed vertical line.  
    Figure adapted from Ref.~\cite{mcdonald2020exponentially}. 
    (d) CFI with respect to perturbation coupling the boundaries ($\Gamma$) as a function of system size $N$ for the non-Hermitian SSH model based proposal in Ref.~\cite{koch2022quantum}.
    Here $t_1$ and $t_2$ are the intra- and inter-site hopping terms.
    The CFI increases exponentially with coefficient $\alpha$ (solid line) until it saturates to a value that increases with decreasing $\Gamma$.
    Figure adapted from Ref.~\cite{koch2022quantum}.}
\label{Fig_EP}
\end{figure}

EPs are branch point singularities in the complex energy manifold and generate non-trivial topology, quantified by winding of eigenvalues and eigenvectors~\cite{leykam2017edge}.
Sensing based on EP was proposed first in Ref.~\cite{wiersig2014enhancing} for coupled cavity systems -- with photon loss and gain terms -- that have resonant modes at EP.
A non-Hermitian perturbation with strength $\epsilon$ would move the system away from the EP.
For an EP of degree $n$ (i.e.~$n$ eigenstates coalescing), the resulting energy splitting ${\sim} \epsilon^{1/n}$. 
In contrast, for Hermitian degeneracies or diabolic points, the leading order splitting is ${\sim} \epsilon$.
The advantage of the EP based sensors can be quantified by the susceptibility $d_{\epsilon} \epsilon^{1/n} \propto \epsilon^{-(1-1/n)}$, which diverges as $\epsilon \to 0$. 
This was experimentally demonstrated with optical microring resonators~\cite{chen2017exceptional, hodaei2017enhanced} and also with further imposition of PT symmetry~\cite{liu2016metrology, yu2020experimental} (see Fig.~\ref{Fig_EP}).
However, these schemes do not reveal the precision of the sensor which is the smallest change in $\epsilon$ that can be measured.
The standard measurement is done by first implementing an excitation, followed by analyzing the resonant scattering output.
The signal-to-noise ratio of the output is a measure of the precision, which corresponds to the QFI if the associated observable can be chosen to be optimal.
As the eigenstates participating in the excitation-deexcitation process become more and more non-orthogonal while approaching EP, it was debated that the resulting quantum noise would overwhelm the advantage of EP based sensors~\cite{langbein2018no, chen2019sensitivity}.
Experimental observation also confirmed this~\cite{wang2020petermann}.
Signal enhancement was nevertheless predicted at specific frequencies~\cite{zhang2019quantum} or with additional gain terms~\cite{lau2018fundamental} and was also observed experimentally for systems slightly detuned from EP~\cite{kononchuk2022exceptional}. 
The effectiveness of EP based sensors is still a topic of active research~\cite{wiersig2020review}.

In Ref.~\cite{lau2018fundamental}, another resource for enhanced signal was shown to be the non-reciprocity.
For tight-binding non-Hermitian Hamiltonians, a concrete proposal based on the Hatano-Nelson model~\cite{hatano1996localization} was put forward in Ref.~\cite{mcdonald2020exponentially}.
Here the non-reciprocity is provided by different hopping rates to left and right, namely $J_L$ and $J_R$, in the 1D Hamiltonain
\begin{align}
    \hat{H}_{\rm HN} = \sum_{j} \left( J_L \ket{j} \bra{j{+}1} + J_R \ket{j{+}1} \bra{j}\right) .
    \label{HN_ham}
\end{align}
The imbalanced hopping terms generate an unique non-Hermitian topology even in the single-band structure, unlike Hermitian cases. 
Considering a $N$-site chain of coupled bosonic modes with a classical resonant drive at site 1, it was shown that the dynamics of the two quadrature modes are governed by a `doubled' Hatano-Nelson Hamiltonian, with opposite reciprocities for the two modes.
This imposes a $\mathbb{Z}_2$ symmetry which is broken by a perturbation on site $N$ that mixes the two modes with strength $\epsilon$.
An input drive in terms of one mode would generate a large signal as the output in the other mode in the steady state.
With judicious choice of the optimal homodyne measurement~\cite{banchi2015quantum}, the signal-to-noise ratio gives the QFI with respect to $\epsilon$ which scales exponentially with system size, namely QFI $\sim |\epsilon| e^{(N-1) \log{(J_R/J_L)}/2}$.
This kind of exponential sensitivity was also reported in Ref.~\cite{koch2022quantum} with a similar coupled cavity system with input drive but here the perturbation term -- with strength $\epsilon$ -- connects the boundary modes (sites 1 and $N$).
Here, the dynamics is governed by a non-Hermitian extension of the SSH model with balanced loss and gain terms on alternating sites.
This breaks the sublattice symmetry, but makes the systems PT-symmetric.
In the topologically non-trivial regime, the sensitivity of the steady state output field scales exponentially with system size, i.e.~${\sim} e^{\alpha N}$, where $\alpha$ is a positive coefficient.
This behavior is closely connected to the acute spectral sensitivity of non-Hermitian topological systems~\cite{koch2020bulk, kunst2018biorthogonal, edvardsson2022sensitivity}. 
In particular, exponential sensitiveness of the edge state energy $E_0$ to the coupling at the boundary $\epsilon$ (i.e.~$\partial_{\epsilon} E_0 \sim e^{\alpha N}$), has also led to sensing proposal in Ref.~\cite{budich2020non}.
These proposals function anywhere in the topologically non-trivial regime and does not require fine-tuning to an EP.
An experimental confirmation of edge state sensitivity at the EP was also performed recently with an optical resonator setup~\cite{guo2021sensitivity}.  
Even in the absence of EP, superiority of sensing has been shown experimentally~\cite{Xiao2024Non} and Heisenberg scaling in time has also been demonstrated with non-Unitary dynamics~\cite{Yu2024Toward}. 
Such scaling behaviors are quite promising regarding sensing in open systems, although recent theoretical arguments cast doubts about actual advantages of non-Hermitian sensors over their Hermitian counterparts~\cite{ding2023fundamental}. 

\begin{figure}[t]
    \centering
    \includegraphics[width=0.95\linewidth]{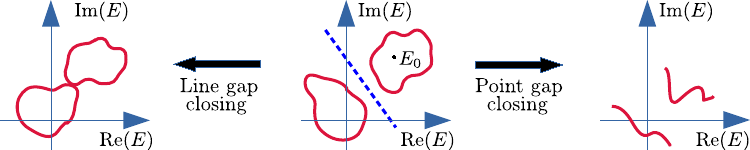}
    \caption{\textbf{Non-Hermitian energy gaps}. 
    Two types of gaps can occur in the complex energy plane (middle panel). 
    Line gap (blue dashed line) separates two bands (red loops).     
    Point gap resides inside a spectral loop (e.g.~$E_0$). 
    Line gap closes when the loops merge (left panel).
    Point gap closes when the loop shrinks to an arc (right panel). Figure taken from Ref.~\cite{Sarkar2024Critical}.}
    \label{Fig_NH_gap}
\end{figure}

\begin{figure}[t]
    \centering
    \includegraphics[width=0.95\linewidth]{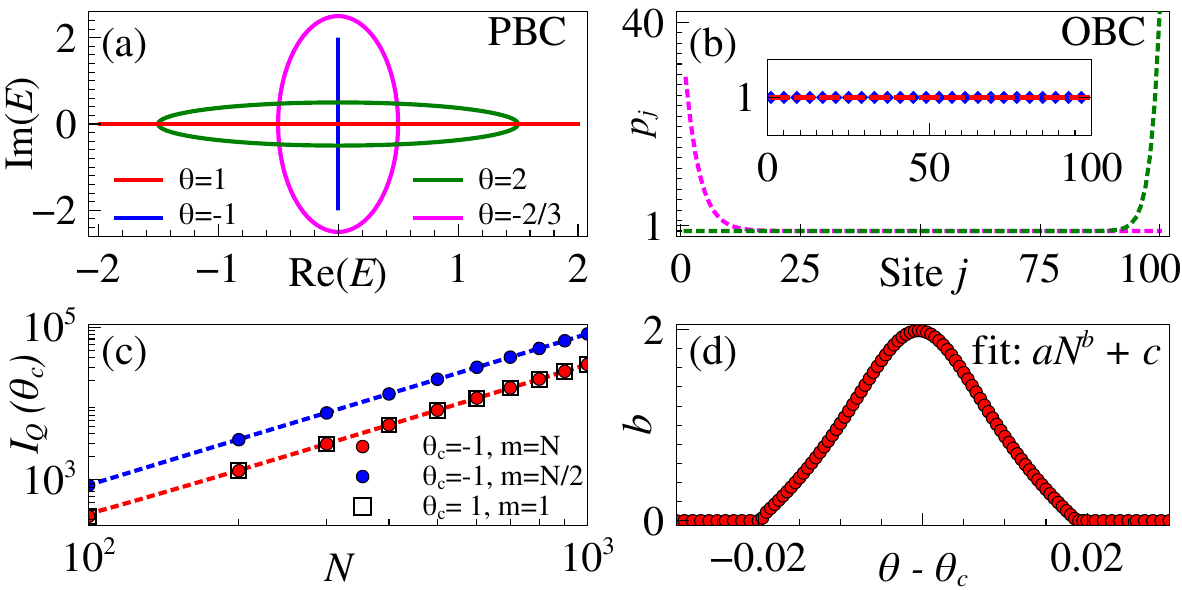} 
    \caption{\textbf{Bulk Hamiltonian parameter estimation with Hatano-Nelson model} with $\theta = J_R/J_L$ in Eq.~\eqref{HN_ham}. 
    (a) Bulk spectrum with periodic boundary conditions (PBC). 
    (b) Cumulative site-population for open boundary conditions (OBC) eigenstates shows the non-Hermitian skin effect and its absence (inset). 
    (c) Quadratic scaling of QFI with system size $N$ at the transition $\theta = \theta_c = \pm 1$. 
    (d) QFI Scaling exponent near transition. 
    Figures adapted from Ref.~\cite{Sarkar2024Critical}.}
    \label{Fig_HN}
\end{figure}

Beyond sensing boundary perturbations, non-Hermitian topology also can be used for estimating bulk Hamiltonian parameters.
The notion of band gaps are extended in the non-Hermitian regime to two different gap structures: line gap and point gap (see Fig.~\ref{Fig_NH_gap})~\cite{kawabata2019symmetry}.
A spectral loop (for 1D systems) or finite spectral area (in higher dimensions) correspond to the presence of non-Hermitian skin effect in finite systems~\cite{alvarez2018non, yao2018edge}, where most of the eigenstates are localized at a boundary~\cite{borgnia2020non, zhang2020correspondence, zhang2022universal, Li2020Critical}.
Changes in the bulk Hamiltonian parameters can make the spectral structure contract to a curve with zero spectral area.
This signals a closure of the point gap, which makes the skin effect vanish.
The corresponding localization transition in the eigenstates can be used for sensing.
As was shown in Ref.~\cite{Sarkar2024Critical}, precision with Heisenberg scaling can be achieved using such probes.
This extends the role of gap closing for quantum-enhanced sensitivity to a truly non-Hermitian domain.
For the aforementioned Hatano-Nelson model, this scaling can be obtained analytically (see Fig.~\ref{Fig_HN}).
The robustness against local disorders is also shown by mapping it to a non-Hermitian Aubrey-Andr\'{e}-Harper model~\cite{longhi2019topological}.
Apart form extension to prototypical multi-band models~\cite{yao2018edge} and 2D systems~\cite{yao2018non}, experimental implementation is also possible in discrete-time non-Unitary quantum walk setups~\cite{xiao2020non, xiao2021observation}.

\section{Resource Analysis for Critical Quantum Sensing}\label{sec:ultimate-limits}

As discussed above, various forms of criticality can achieve quantum-enhanced sensitivity. 
The resource for achieving such enhanced precision is the system size $N$ and the figure of merit is the QFI.  
However, in some cases the time $t$ needed for accomplishing the sensing task also has to be considered as a resource. 
This might be due to practical issues, such as limited coherence time, or size dependent $t$ which can make the sensing time extremely long for large system sizes. 
To account for such complexity, one can use normalized QFI as a figure of merit, namely $I_Q/t$~\cite{albarelli2018restoring,rossi2020noisy,chin2012quantum,chaves2013noisy,brask2015improved,smirne2016ultimate}. 
This can be understood by considering the total time $t_{\mathrm{tot}}$ that is needed for collecting the data through probe preparation and measurement with $M$ repetitions.
The number of measurements that are possible within the available total time is $M{=}t_{\mathrm{tot}}/t$. 
Now from Eq.~\eqref{eq:cramer-rao-classical-quantum} one can see that the upper bound of achievable precision is given by $1/\sqrt{t_{\mathrm{tot}} I^Q/t}$. 
This clearly suggests that the rescaled QFI $I^Q/t$ is the suitable figure of merit. 
For instance, in the case of many-body ground state sensing, one may consider an adiabatic approach for generating the complex ground state. 
This can be achieved by slowly evolving the system initialized in the ground state of a simpler Hamiltonian which is transformed into the desired Hamiltonian over the time interval $t$. If the transformation is slow enough, namely $t$ is large, the quantum state remains in the ground state of the instant Hamiltonian throughout the evolution. This guarantees that at the end of evolution when the Hamiltonian takes the final form and, hence, the system is in the desired ground state. When the system is adiabatically driven towards a second-order quantum phase transition, the required time $t$ scales as $t{\sim} N^z$~\cite{rams2018limits}, where  $z$ is the dynamical critical exponent showing how the energy gap $\Delta E$ vanishes at the critical point by increasing the system size $\Delta E_{\rm min}\sim 1/N^z$. Incorporating this into the scaling of the QFI across second-order quantum phase transitions in Eq.~(\ref{eq:QFI_finit_size}) leads to
\begin{equation} \label{eq:normalized_QFI}
    I_Q(\theta{=}\theta_c) /t \sim N^{\beta-z}.
\end{equation}
This clearly shows that the scaling diminishes as time is included in the resource analysis. 
Note that the issue of time as a resource can arise in interferometric setups as well, for example in the adiabatic sweeping through quantum phase transitions for entanglement generation and detection~\cite{lee2006adiabatic, huang2018non, zhuang2020symmetry}.

Similar results can be found for different quantum sensing strategies where the required time depends on the system size. For instance, in Ref.~\cite{montenegro2023quantum} quantum-enhanced sensitivity, i.e.~$\beta{>}1$, is achieved for the steady state at the boundary time crystal transition when the figure of merit is the QFI $I_Q$. However, if one changes their figure of merit to normalized QFI, i.e.~$I_Q/t$, with $t$ being the time that the system reaches its steady state, the scaling with respect to system size is reduced to standard quantum limit. 
{\color{black} On the other hand, in the case of Stark systems, presented in section~\ref{Section:Stark_Probe}, it has been shown that regardless of the number of particles and the range of the interaction, one always has $\beta-z{>}1$ ensuring that the quantum-enhanced sensitivity is always achievable for the Stark probes~\cite{he2023stark,yousefjani2023longrange}.}

{\color{black} Valuable insights on the normalized QFI can be obtained from the fundamental bounds on QFI.
When the probe is the ground state of the Hamiltonian in the form $\hat{H}_{\theta} {=} \hat{H}_0(\theta) {+} \hat{H}_1$, with a control term $\hat{H}_1$ and the parameter dependent term $\hat{H}_0(\theta)$, the upper bound of the QFI is given by~\cite{abiuso2025fundamental}, 
\begin{equation}
    I_Q(\theta) \le \frac{|| \partial_{\theta} \hat{H}_0(\theta) ||^2}{\Delta^2},
    \label{eq:gap-bound}
\end{equation}
where the operator seminorm $||.||$ is the difference between the maximum and minimum eigenvalues of the operator, and $\Delta$ is the energy gap between the ground state and the first excited state.
On the other hand, when the probe state is prepared by evolving an appropriate initial state with a Hamiltonian with a time-dependent control term in the form $\hat{H}_{\theta, t} {=} \hat{H}_0(\theta) {+} \hat{H}_1(t)$, the QFI is upper bounded by the generalized Heisenberg limit~\cite{boixo2007generalized, puig2024dynamical}
\begin{equation}
    I_Q(\theta, t) \le t^2 || \partial_{\theta} \hat{H}_0(\theta) ||^2 .
    \label{eq:t-bound}
\end{equation}
According to the adiabatic theorem, the time needed for adiabatic preparation of the ground state probe is inversely proportional to the square of the critical energy gap.
However, with certain case specific shortcuts to adiabaticity, it can be made inversely proportional to the critical gap, namely $\sim 1/\Delta_c$~\cite{roland2002quantum}.
Therefore, the rescaled critical QFI in this case is bounded as $I_Q/t \le \frac{|| \partial_{\theta} \hat{H}_0(\theta) ||^2}{\Delta_c}$. 
This indicates that the rescaled QFI can benefit from the scaling of both the energy gap as well as the seminorm factor.
The fundamental bounds of the QFI in the context of different probes is provided in Table \ref{tab:comp}, which is adapted from Ref.~\cite{abiuso2025fundamental}.}

\begin{table}[t]
    \centering
    \begin{tabular}{c|c}
 \sf{Encoding} & \sf{Maximum} $I_Q(\theta)$  
 \\ \hline \rule{0pt}{4ex} 
  \text{Dynamical}  $\rho_\theta = e^{-i \hat{H}_\theta t/\hbar}\rho  e^{i \hat{H}_\theta t/\hbar}$ & \hspace{1mm} $
  \dfrac{t^2}{\hbar^2} \|\partial_\theta \hat{H}_\theta\|^2$  
  \rule[-2.5ex]{0pt}{0pt}  \\ \hline \rule{0pt}{4ex}    
 \text{Thermal} $ \rho_\theta= \dfrac{e^{-\beta \hat{H}_\theta}}{\Tr{e^{-\beta \hat{H}_\theta}}} $ & \hspace{1mm}  $\dfrac{\beta^2}{4}  \|\partial_\theta \hat{H}_\theta\|^2$ \hspace{1.7mm}  
 \rule[-2.5ex]{0pt}{0pt}\\ \hline \rule{0pt}{4ex} 
 Ground state  
 of $\hat{H}_\theta$ with spectral gap $\Delta$ & \hspace{1.7mm}  $\dfrac{1}{\Delta^2} \|\partial_\theta \hat{H}_\theta\|^2$  \hspace{1mm} \rule[-2.5ex]{0pt}{0pt} \\ \hline
\end{tabular}
    \caption{\textbf{QFI upper bounds} for probes prepared dynamically, at thermal equilibrium (with inverse temperature $\beta$), and as ground states. Table adapted from Ref.~\cite{abiuso2025fundamental}.}   
    \label{tab:comp}
\end{table}

The time needed for accomplishing the sensing task, is directly related to the strategy that one exploits. In Ref.~\cite{Gietka2021adiabaticcritical}, the authors show that by adiabatic evolution, one cannot reach Heisenberg scaling for quantum sensing. However, the proof relies on a specific choice of the adiabatic evolution. In their formalism the Hamiltonian is written as $\hat{H} {=} \theta \hat{H}_0 {+} \hat{H}_1(t)$, where $\hat{H}_0$ and $\hat{H}_1$ do not depend on the target parameter $\theta$. In this scenario, $\hat{H}_0$ is always on and the time dependent Hamiltonian $\hat{H}(t)$ varies slowly. The resulted dynamics encodes the information about $\theta$ into the state of the system whose initial state does not depend on $\theta$.  In this case, as proved in Ref.~\cite{Gietka2021adiabaticcritical}, Heisenberg scaling cannot be reached. Nonetheless, one may consider to prepare the probe in the ground state of $\hat{H}_1$ and as $\hat{H}_1$ is turned off, the probe is put under the action of a general $\hat{H}_0(\theta)$ adiabatically to prepare the probe in the ground state of a general $\hat{H}_0(\theta)$ instead of $\theta \hat{H}_0$. In this situation, obtaining Heisenberg scaling is possible.

Note that a true resource analysis depends on the sensing scheme. If the time is not a restricting issue QFI can be used as the figure of merit. Equivalently, for those sensing schemes where time is a key limiting factor, a better figure of merit is normalized QFI, namely $I_Q/t$.

\section{Non-Equilibrium Quantum Sensing: Dynamical Quantum Phase Transitions} \label{section:Dynamical_QPT}

\begin{figure*}
   \centering
    \includegraphics[width=0.4\linewidth]{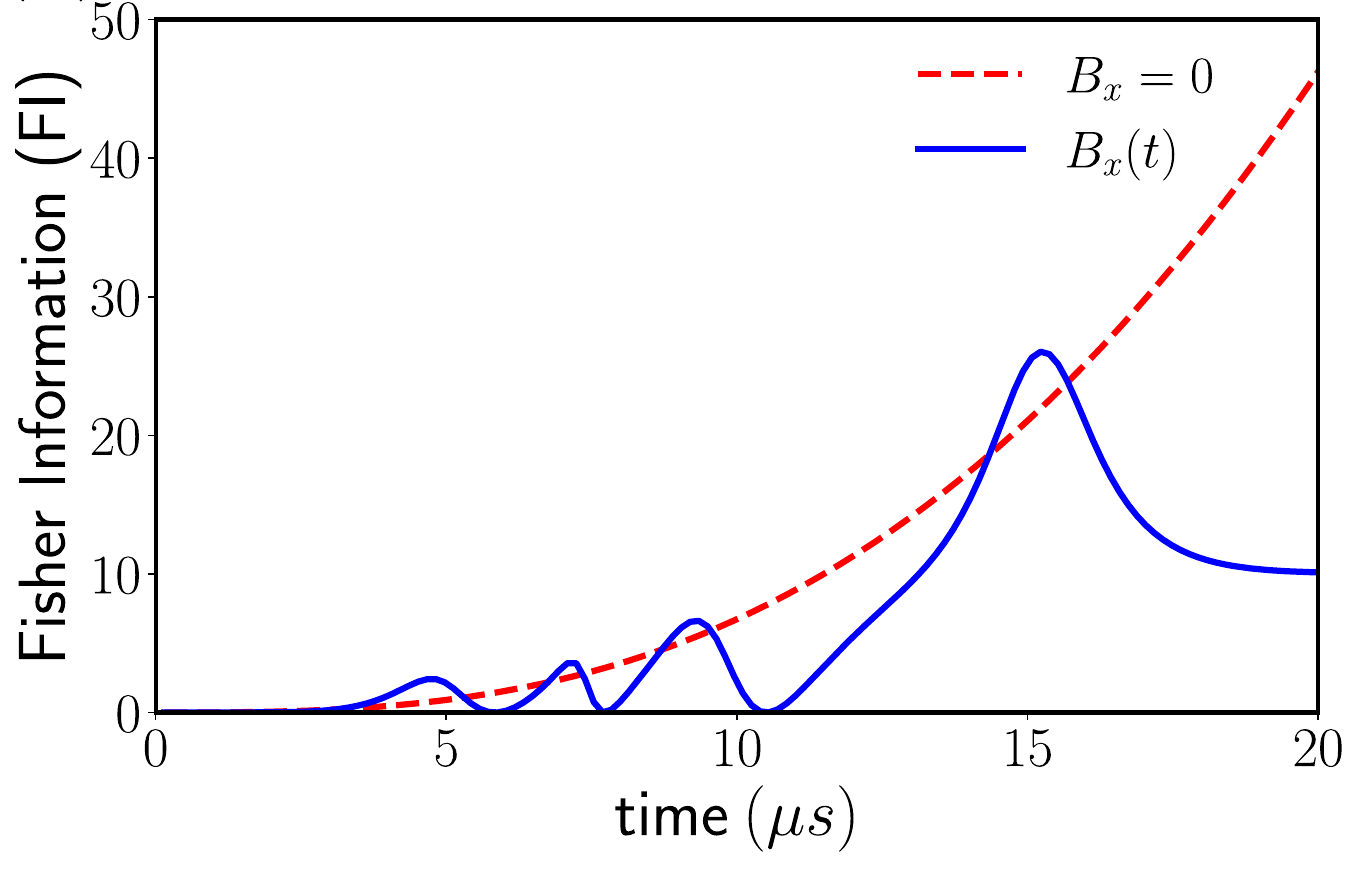}
    \includegraphics[width=0.45\linewidth]{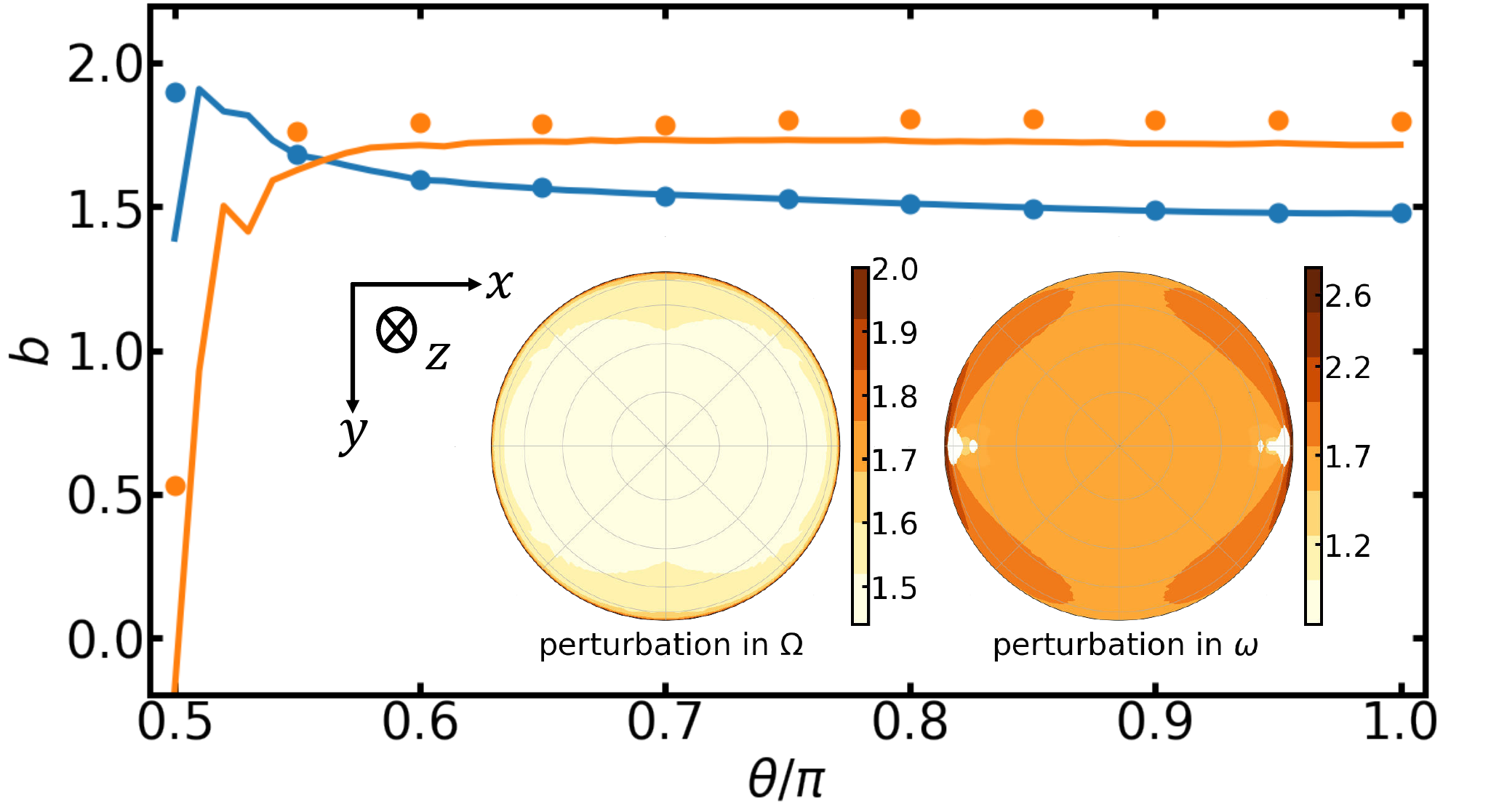}
    \caption{{\color{black}\textbf{DQPT quantum sensor}. (left) A Gaussian longitudinal magnetic field of the form $ B_ {x}(t) =B \exp\{-(t-\tau)^2/(2\sigma^2)\} $ improves the Fisher information for measuring the  coupling at $\tau=3t_{c_{1}}$. On the other hand for other $\tau$ and $B_x=0$ does not provide improvement in the Fisher information. The other parameters  are $B=200$ G, $\sigma = \tau/2$, and $B_{z}=50 G$  is fixed for both cases. Figure is adapted from~\cite{gonzalez2022dynamical}.
    (right) The scaling exponent of quantum Fisher information $I^{t\to \infty}_Q\sim N^{b}$ corresponding to $\Omega$ at the critical points $\Omega=\Omega_{c}$ of non-equilibrium phase transition. Abscissa represents the initial state for various choices of $\theta$ and $\phi=0$ of the spin-coherent state, $\omega/\chi=10^{-4}$, and $N\in[100,200]$.  The lines are obtained numerically and dots are obtained using long-time secular contribution~\cite{pang2014quantum}. Figure is adapted from~\cite{guan2021identifying}}}
    \label{fig:Figure_DQPT}
\end{figure*}

The footprints of the ground state quantum criticality in equilibrium can also be observed in non-equilibrium dynamics. For a closed system described by a time-independent Hamiltonian $\hat{H}$, its time evolution which is described by Schr\"odinger equation as
\begin{equation}
    |\Psi(t)\rangle  = e^{-i\hat{H} t}|\Psi(0)\rangle,
\end{equation}
where $|\Psi(0)\rangle$ is the initial state. There are two different notions of dynamical quantum phase transitions, which we will discuss separately in the following subsections. 

\subsection{Dynamical quantum phase transitions quantified by Loschmidt echo }

Considering the projection of the state with the initial state that changed in time, one can define the Loschmidt echo, ${\cal L}(t)=|\langle \Psi(0)|\Psi(t)\rangle|^2$, interpreted as the probability of return. One can define the rate of return probability ${\cal G}$, as
\begin{equation}
    {\cal G}(t) = -\lim_{N\to \infty}\frac{1}{N}\log {\cal} \mathcal{L}(t).
\end{equation}
It is discovered that if $|\Psi(0)\rangle$ and $|\Psi(t)\rangle$ belong to two different equilibrium phases, then $ {\cal G}(t) $ can be non-analytic at a certain time $t{=}t_c$. The occurrence of non-analytic behavior has been rigorously established as the dynamical quantum phase transition (DQPT)~\cite{heyl2018dynamical}.  On the other hand, if $|\Psi(0)\rangle$ and $|\Psi(t)\rangle$ belong to the same equilibrium phase, then ${\cal G}(t)$ is a regular function of time. There are a few peculiarities to these cases~\cite{jafari2017loschmidt} as well as their extension in non-Hermitian systems~\cite{Zhou2018Dynamical, Jing2024Biorthogonal}. Such dynamical phase transitions have been experimentally realized in large quantum simulators~\citep{zhang2017observation}. Some recent efforts have been made to explore its relevance as a useful resource for various quantum technologies, including quantum-enhanced sensing~\cite{gonzalez2022dynamical}, where the authors have proposed a negatively charged Nitrogen-Vacancy (NV$^{-}$) center in nearby Carbon-13 nuclear spins as a platform for studying dynamical phase transitions. First, they have shown that the system undergoes a dynamical phase transition by a suitable choice of a Gaussian time-varying magnetic field. Then it is shown that the Fisher information, for estimating the coupling strength between the two carbon-13 nuclear spins, improves for $\tau=3t_{c_{1}}$ as compared to the other choices of $\tau$ at regular times. This is shown in the left panel of Fig.~\ref{fig:Figure_DQPT}.

\subsection{Dynamical quantum phase transitions quantified by a time-averaged order parameter }

A second category of dynamical phase transitions in many-body systems is detected by the time-averaged order parameter ${\cal O}(t)$.  The time average of the order parameter in the evolved state $|\Psi(t)\rangle$  is given by 
\begin{equation}
  \overline{\langle  {\cal O}(t)\rangle} = \lim_{\tau \to \infty}\frac{1}{\tau}\int_{0}^{\tau}dt \langle  {\cal O}(t)\rangle. 
\end{equation}
The order parameter distinguishes an ordered phase when $ \overline{\langle  {\cal O}(t)\rangle} \neq 0$ from a disordered phase when $ \overline{\langle  {\cal O}(t)\rangle}=0$. {\color{black} It has been shown that Fisher information can be a tool for identifying such steady state phase transitions~\cite{cite-key,PhysRevE.84.041116,Huang_2016,paz2024entanglement,guan2021identifying}}. {\color{black}Moreover, using a suitable collective spin model, steady-state phase transitions have been identified for quantum-enhanced sensing of single-parameter estimation~\cite{guan2021identifying}}. In this paper, the authors have studied a fully connected spin model under sudden quenching. They focused on the LMG model which is given by 
\begin{eqnarray}
    \hat{H}_{LMG}{=}-\frac{\chi}{N}\hat{S}^{z \, 2}-\Omega \hat{S}^x -\omega \hat{S}^z
\end{eqnarray}
where  $\hat{S}^{x,y,z}{=}\sum_{i=1}^{N}\hat{\sigma}^{x,y,z}_{i}/2$  are defined as collective spin operators and $\hat{\sigma}^{x,y,z}_{i}$ are the Pauli matrices for the $i$th spin-1/2  particle. The criticality of the ground state of the model occurs at $\Omega_c/\chi{=}0.5$ and $\omega_c/\chi{=}0$.  The Fisher information in long-time state behaves as an order parameter and shows a peak at the critical point. Moreover, the scaling of the long-time  QFI at the critical point as a function of system size $N$ reaches a sub-Heisenberg limit for the transverse field, $I^{t\to \infty}_{Q,\Omega}\sim N^{1.5}$ and $I^{t\to \infty}_{Q,\omega}\sim N^{1.75}$ for the longitudinal field. For various choices of initial state, the scaling exponent $b$, obtained by fitting the numerical data in the function $I_{Q}^{t\to \infty}\sim N^{b}$, is shown in the right panel of Fig.~\ref{fig:Figure_DQPT}.


\section{Non-Equilibrium Quantum Sensing: Discrete Time Crystal Phase Transitions}\label{sec:discrete_time_crystal}

\begin{figure}
    \centering
    \includegraphics[width=0.49\linewidth]{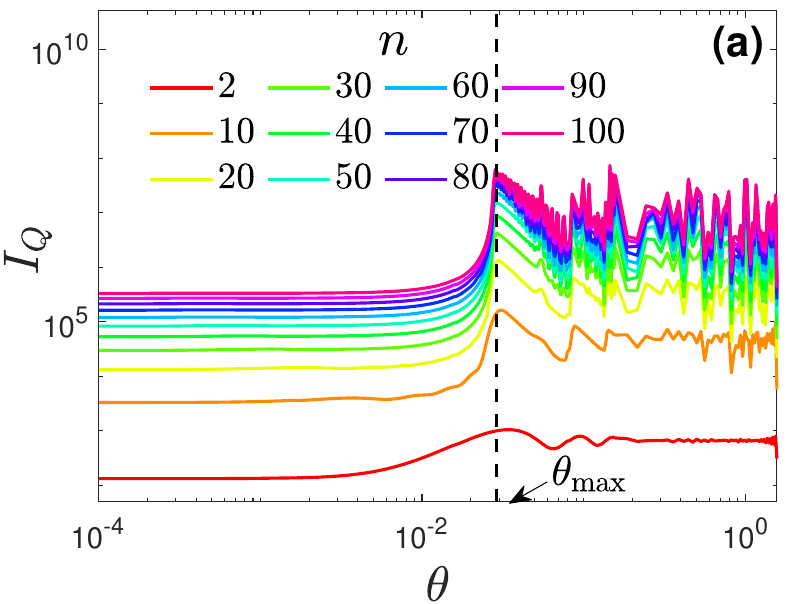}
    \includegraphics[width=0.49\linewidth]{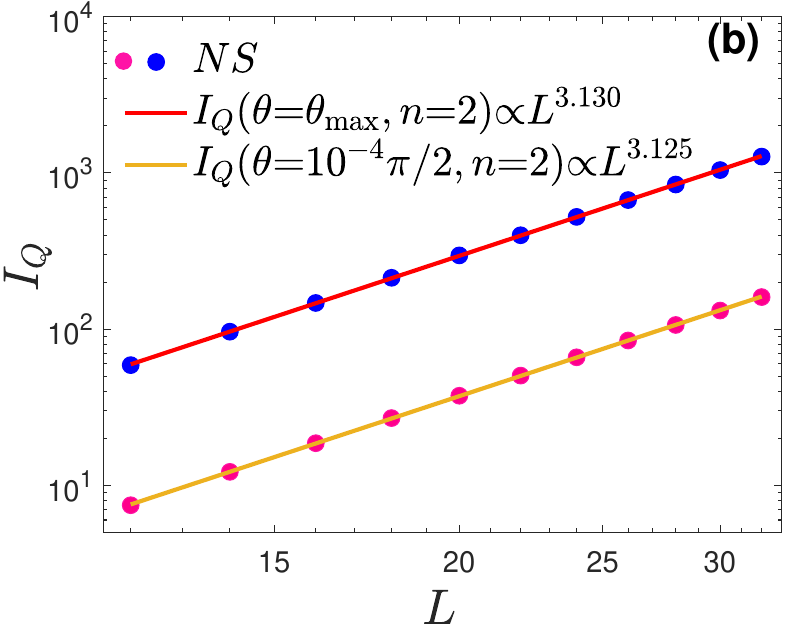}
    \caption{\textbf{DTC quantum sensor;} (a) The QFI versus $\theta$ in stroboscopic times in a system of size $L{=}30$. The onset of the phase transition is determined by  $\theta{=}\theta_{\max}$ (dashed line), the point where QFI peaks in different period cycles $n$'s. (b) The values of the QFI after $n=2$ in DTC phase (for $\theta{=}10^{-4}\pi/2$) and at transition points ($\theta{=}\theta_{\max}$) versus $L$. The numerical simulation (NS) is well-mapped by a function as $I_{Q}{\propto}L^{\beta}$ (solid lines) with $\beta{>}3$. Results are obtained for $\varepsilon{=}0.01$. Figure is adapted from~\cite{yousefjani2024discrete}.}
    \label{fig:FigDTC}
\end{figure}

Apart from dynamical phase transitions, discussed in Sec.~\ref{section:Dynamical_QPT}, there are other categories of phase transitions which occur through non-equilibrium dynamics of many-body systems. An important class of such systems are discrete-time crystals (DTCs) which are defined in an isolated non-equilibrium quantum many-body system that undergoes periodic drives. 
Breaking the transnational symmetry of time results in the emergence of a new phase of matter known as the DTC phase with persistent indefinite oscillations~\cite{sacha2015modeling,khemani2016phase,else2016floquet,yoshinaga2022quantummetrologyprotectedhilbert}. 
This concept has also been studied in the non-Hermitian systems~\cite{yousefjani2025non}.
Recently, long-range spatial and time-ordered dynamics in the DTC phase
have been harnessed to measure an AC field~\cite{iemini2024floquet,gribben2024quantum,yousefjani2025discrete}. 
The optimal performance for long periods along with intrinsic robustness to the imperfections in the sensing protocol based on DTC phase allows one to hit the standard quantum limit in estimating the AC field. 
However, obtaining a true sensing enhancement based on a resource such as system size requires a wised mechanism for establishing DTC. 
In Ref.~\cite{yousefjani2024discrete} the proposed mechanism results 
in a stable DTC with period-doubling oscillations that persist indefinitely even in finite-size systems. The time-dependent Hamiltonian that governs the dynamic in the considered spin-$1/2$ chain is
\begin{align}\label{Eq.DTCHamiltonian}
\hat{H}(t) = J \hat{H}_{I} + \sum_n \delta(t-nT)\hat{H}_{P}, 
\cr
\hat{H}_{I} = \sum_{j=1}^{L-1}j\hat{\sigma}^{z}_{j}\hat{\sigma}^{z}_{j+1}, \quad  \hat{H}_{P}  = \Phi \sum_{j=1}^{L}\hat{\sigma}^{x}_{j}.
\end{align}
Here $J$ is the spin exchange coupling, and $\hat{\sigma}^{x,y,z}_{j}$ are the Pauli operators.
For one period of evolution, the Floquet unitary operator reads
\begin{equation}\label{Eq.DTCFloquetUnitary}
\hat{U}_{F}(\Omega,\varepsilon) = e^{-i\hat{H}_{P}} e^{-i\Omega \hat{H}_{I}},    
\end{equation}
here $\Theta{=}JT$, and $\Phi$  is tuned to be $\Phi{=}(1{-}\varepsilon)\frac{\pi}{2}$, with $\varepsilon$ as deviation from a $\pi/2$ $x$-rotation.
While setting $\Theta{=}\pi/2$ results in a stable period doubling  DTC that is robust against arbitrary imperfection $\varepsilon$ in the rotating pulse,  the system goes through a sharp second-order phase transition as the spin exchange coupling, namely $\Theta$ varies from $\frac{\pi}{2}$, denoting this variation as $\theta{=}|\Theta-\frac{\pi}{2}|$. 
Relying on this transition, a DTC quantum sensor has been devised to sense $\theta$ with benefits from multiple features.
As it is clear from Fig.~\ref{fig:FigDTC}(a), in the DTC phase the QFI provides a plateau whose value depends on period cycles $n$, and in the non-DTC region it shows nontrivial and fast oscillations. 
By approaching the transition point, denoted by $\theta_{\max}$ (dashed line), the QFI shows a clear peak at all stroboscopic times.
The extreme sensitivity to the exchange coupling across the whole DTC phase ($\theta{<}\theta_{\max}$) as well as at the transition point ($\theta_{\max}$) providing quantum-enhanced sensitivity as $I_{Q}{\propto}L^{\beta}$ with $\beta{>}3$ (see Fig.\ref{fig:FigDTC} (b)). 
The obtained quantum enhancement is independent of the initial state and can be boosted by increasing imperfection in the pulse to a certain value. 

\section{Non-Equilibrium Quantum Sensing: Floquet Phase Transition}

Quantum metrologically advantageous states can also be generated using the application of external time-periodic driving to the system. {\color{black} In fact, in interferometry-based quantum sensing one can overcome the detrimental effect of decoherence by driving the system in a proper way to restore Heisenberg scaling~\cite{bai2023floquet}.} 
In periodically driven systems, the frequency of the external driving can resonate with the lowest energies of the unperturbed systems. This resonance can be harnessed for quantum-enhanced sensing~\cite{lang2015dynamical,wang2022sensing,jiang2022floquet,pang2017optimal,PhysRevX.10.031003,mishra2021driving,mishra2022integrable,yu2022quantum,ivanov2021enhanced}. This resonance phenomena has also been observed in gaped many-body quantum systems when the frequency of external driving matches with the lowest energy gap of the unperturbed Hamiltonian.~\cite{russomanno2016entanglement,mishra2021driving,mishra2022integrable}.    
Various important physical features of systems with time-periodic influence can be extracted by employing Floquet formalism. 
For a  time-periodic Hamiltonian, $\hat{H}(t+\tau)=\hat{H}(t)$, with periodicity $\tau=2\pi/\omega$, the solution of the Schr\"{o}dinger  equation follows from the Floquet theorem. The Floquet theorem gives an ansatz of the form $|\Psi(t)\rangle=\sum_{\ell} e^{-i\mu_{\ell}t}|\Phi_{\ell}(t)\rangle$. Here $\mu_{\ell}'s$ are quasienergies (the Floquet phase), $|\Phi_{\ell}(t)\rangle$ the Floquet modes, and $\ell \in \mathbb{Z}$. 
Substituting the ansatz to the Schr\"{o}dinger equation, $ i \frac{\partial |\Psi(t)\rangle}{\partial t} = \hat{H}(t)|\Psi(t)\rangle$, the Floquet modes satisfy
\begin{equation}
\Big(\hat{H}(t)-i \frac{\partial }{\partial t}\Big)|\Phi_{\ell}(t)\rangle=\mu_{\ell}|\Phi_{\ell}(t)\rangle.
\end{equation} 
It can be noted that $|\Phi_{\ell}(t+\tau)\rangle$ is also a solution of the above equation with Floquet quasienergies $\mu_{\ell}$, therefore,  one can write $|\Phi_{\ell}(t+\tau)\rangle = |\Phi_{\ell}(t)\rangle$. Furthermore, the time evolution propagator, describing the dynamics of the system, is given by 
\begin{equation}
\label{eq:U_t}
    \hat{U}(t,t_0)={\cal T}_{t}\exp\big(-i\int_{t_0}^{t}\hat{H}(t')dt'\big),
\end{equation}
where ${\cal T}_{t}$ represent time ordering operator. 
It is noted that $\hat{U}(t+\tau)|\Psi(t)\rangle=|\Psi(t+\tau)\rangle$. Using the ansatz for $|\Psi(t)\rangle$ in the above equation, one gets $\hat{U}(t+\tau)e^{-i\mu_{\ell}t}|\Phi_{\ell}(t)\rangle = e^{-i\mu_{\ell}t}|\Phi_{\ell}(t+\tau)\rangle$. This shows that the Floquet modes are eigenvectors of the one time-period propagator $\hat{U}(\tau,0)$ with $\{e^{-i\mu_{\ell}\tau}\}$ the corresponding eigenvalues. 

Properties of Floquet modes $\{|\Phi_{\ell}(t)\}$ and Floquet phases $\{\mu_{\ell }\}$ have been explored for sensing periodic magnetic fields due to clusters of few qubits~\cite{lang2015dynamical}. Here, the effective magnetic field due to the cluster of spins is detected using a single qubit as a sensor. The initial state of the system is $|\psi(0)\rangle{=}\frac{1}{\sqrt{2}}(\ket{\uparrow}{+}\ket{\downarrow})\otimes |\mathcal{B}(0)\rangle$, where $\ket{\mathcal{B}(0)}$ is the initial state of the detected spin cluster at time $t=0$ and $\ket{\uparrow}$ , $\ket{\downarrow}$ are the computational basis states of the electronic spin, respectively. At later times, due to the interaction with the sensor (the qubit state), the spin cluster state becomes entangled with the sensor. The magnetic field signal is detected from the data of temporal coherence of the time-evolved state of the detected spin cluster. The coherence can be expressed in terms of the eigenvalues and eigenstates of the one-period unitary evolution operator or Floquet operator. The coherence displays a dip which occurs at the avoided crossings of the Floquet eigenstates where $e^{i\mu_{\ell}}\approx e^{-i\mu_{k}}$ or when the Floquet gap, $\Delta_F = \mu_{\ell}-\mu_{k}$, vanishes. Thus, by measuring the location of the coherence dip, the energy-difference between Floquet states can be observed with precision higher than the one predicted by the average Hamiltonian theory.   
However, this method can only be applicable for a sensing around the resonance frequencies. Sensing of arbitrary frequency fields has been proposed using  a single qubit as a sensor in Floquet systems~\cite{wang2022sensing}. In addition, periodically driven systems and the transition of Floquet quasienergies have been used for magnetic-field signal amplification~\cite{jiang2022floquet}. In Ref.~\cite{pang2017optimal}, an optimal control method is developed to estimate the amplitude, $B$, of a time-periodic magnetic field $B(t)=B(\cos(\omega t)\hat{\sigma}_x+\sin(\omega t)\hat{\sigma}_z)$. In Ref.~\cite{yang2024variational},  a systematic variational  approach for establishing optimal control was set up, where it was shown that for magnetometry, restricted controls in the form of one and two-body interactions are already sufficient to approximately achieve Heisenberg scaling.
It is shown here that the QFI with respect to $B$ can achieve quantum-enhanced sensitivity with interrogation time, $t$, by applying an optimal control pulse as compared to the one without control.  A solid state sensor with an ensemble of high density Nitrogen Vacancy (NV) centers in diamonds has been investigated both theoretically and experimentally by Floquet formalism ~\cite{PhysRevX.10.031003}. The system, here, is a dense ensemble of electronic spins. Each electronic spin act as an independent probe for the external signal. Thus, one would expect that, by increasing the number of particles in the system, the sensitivity can be improved. The improvement, however, is severely hindered by interaction between the spins, since a large spins are confined in a small volume, the spin-spin interaction is unavoidable.  By applying a  series of dynamical decoupling pulses, the interaction between the spins has been suppressed, which results in a larger coherence time. In this way the standard quantum limit is shown to be restored which is lost due to spin-spin interaction.

\begin{figure*}[t]
\centering \includegraphics[width=\linewidth]{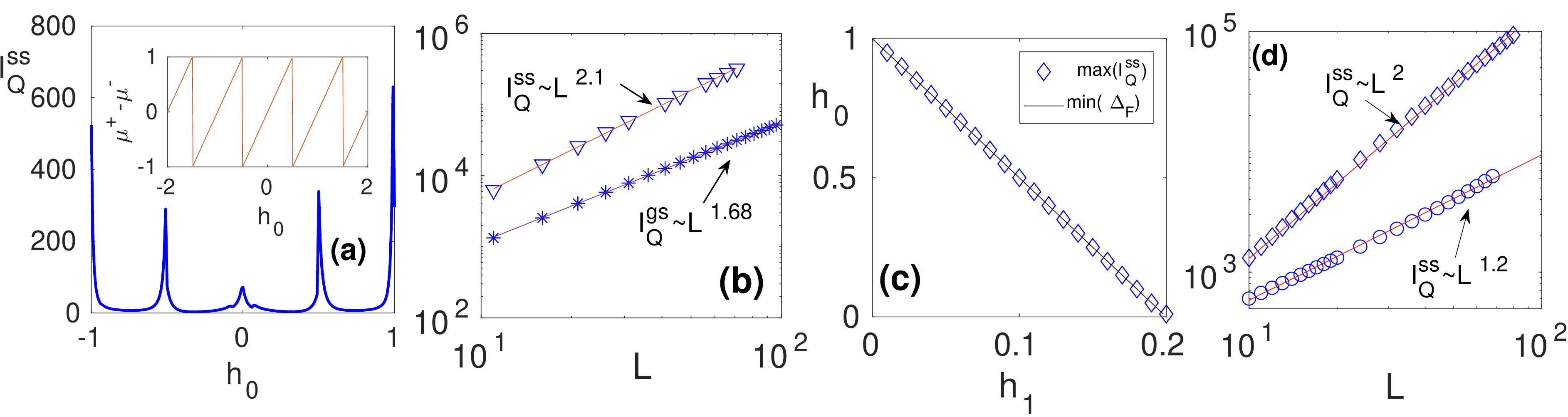}
\caption{\textbf{Driven enhanced sensing}. (a-b) DC field sensing for $h(t)=h_0+h_1\sin(\omega t)$. (a) Steady-state QFI $I^{ss}_Q$ as a function of $h_0$ for $\gamma=1$, $\omega=1$, $h_1=1.5$, and the total system size $N=6000$.  (b) The scaling of QFI as a function of $L$ in the steady-state ($I^{ss}_Q$) and the ground state ($I^{gs}_Q$). The total system size here is $N=6000$ and the other parameters for the $I^{ss}_Q$ and $I^{gs}_Q$ are same as in panel (a). (c-d) AC field sensing of the form $h(t)=h_0+h_1\sum_{n=0}^{\infty}\delta(t-n\tau)$.  (c) The line of maximum of $I^{ss}_{Q}$ and minimum of Floquet gap. (d) Scaling of $I^{ss}_Q$ as a function of $L$ for points ($h_0,h_1$) such that $\Delta_F=0$ in blue diamonds and for points $\Delta \neq 0$ in the blue circles. For (c-d), $\tau=0.2$. Figure adapted from Refs.~\cite{mishra2021driving} and~\cite{mishra2022integrable}.}
\label{fig5:floquet_DC}
\end{figure*}

It is important to note that the above sensing protocols either do not employ interaction between the particles directly or use schemes to suppress the particle-particle interaction. Thus, the above schemes are equivalent to the single-particle probe quantum metrology schemes with several rounds of repetition of the protocol. It is natural to extend the formalism to many-body systems with Floquet dynamics without applying any dynamical decoupling pulses and ask if still one can harness any property for improving the precision well beyond the standard quantum limit. Sensing protocols have been developed in many-body systems using Floquet formalism for detecting  both static~\cite{mishra2021driving} and periodic fields~\cite{mishra2022integrable}. The Hamiltonian considered in those studies is
\begin{eqnarray}
    \hat{H}(t)&=& -\frac{J}{2}\sum_{i=1}^{N}  \Big[\Big(\frac{1+\gamma}{2}\Big)\hat{\sigma}^{x}_{i}\hat{\sigma}^{x}_{i+1}+\Big(\frac{1-\gamma}{2}\Big)\hat{\sigma}^{y}_{i}\hat{\sigma}^{y}_{i+1}\Big]\nonumber\\
    &-& \frac{(h_0+h(t))}{2} \sum_{i}\hat{\sigma}^{z}_{i},
\label{eq:model}
\end{eqnarray}
where, $\hat{\sigma}^{x,y,z}$ are the Pauli matrices, $J$ is the exchange coupling,   $-1{\leq} \gamma {\leq} 1$ is the anisotropic parameter, and the periodic-boundary conditions, i.e., $\hat{\sigma}^{x,y,z}_{N+1} {\equiv} \hat{\sigma}^{x,y,z}_{1}$, is imposed. Furthermore, the time-dependent field $h(t)$ is periodic with a period $\tau=\frac{2\pi}{\omega}$. Thus, the Hamiltonian is a time-periodic $\hat{H}(t+n\tau)=\hat{H}(t)$, where $n\in \mathbb{Z}$. In the absence of the periodic field, i.e.~$h(t)=0$, the Hamiltonian (\ref{eq:model}) shows quantum criticality at $h_0{=}h_c{=}J$ for all values of $\gamma$~\cite{sachdev1999quantum}. Using Floquet formalism for solving the dynamics driven by the Hamiltonian in Eq.~(\ref{eq:model}), one can estimate both the static field  $h_0$~\cite{mishra2021driving} and the amplitude of the periodic field $h(t)$~\cite{mishra2022integrable} even with  partial accessibility to a block of spins. Note that partial accessibility implies that one only has access to a certain block of spins to perform measurement and gains information about the parameter of interest.  It is to be emphasis here that the sensitivity to the parameter of interest naturally deceases as the accessibility reduces to smaller blocks. For example, it is shown that the QFI with respect to $h_0$ in the ground state of Eq.~(\ref{eq:model}), when $h(t)=0$, decreases as compared to the full system~\cite{mishra2021driving}.  By driving the system periodically using an external drive field $h(t)$ the QFI can be made to exhibits peaks not only at the critical field, but at the points where the Floquet resonances occur. Thus, even with partial accessibility, the probe state becomes more sensitive with respect to the magnetic field $h_0$ when a suitable controlled driving magnetic field is applied such that the floquet gap closes. 

By utilizing the time translation invariance of the Hamiltonian, the unitary dynamics in the integer multiple of time-period $\tau$ can be given by 
\begin{equation}
    \hat{U}(\tau,0) = e^{-i \hat{H}_{F}\tau},
\end{equation}
where $\hat{H}_F$ is called the Floquet Hamiltonian and it describes the evolution of the system at the stroboscopic times $t=n \tau$.
The knowledge of the state of the system at the later times $t=n\tau$ can be fully obtained  by successive application of $\hat{U}(\tau,0)$. The  $\hat{U}(\tau,0)$  admits spectral decomposition in terms of eigenvectors $\{|\mu\rangle\}$ and eigenvalues $\{\mu(\tau)\}$ of $\hat{U}(\tau,0)$ which depends on unknown parameter to be estimated. The eigenvalues are periodic and are given by $\mu_{\ell}=\mu_{\ell}+n \omega$. Starting from a initial state, $|\Psi_0\rangle$, the time-evolved state can be written  as $|\Psi(n\tau)\rangle{=}\sum_{i}e^{-{\it i}\mu_{i}n\tau}|\mu_{i}\rangle|\langle\mu_{i}|\Psi_0\rangle|^2$. 
 
Since the system is continuously absorbing heat from the driving, it may heat up to an infinite temperature state which does not carry any information about the unknown parameter $h_0$~\cite{Ishii2018heating}.
This time is exponentially large for most of the generic integrable finite-size and disorder-free many-body systems. Thus, a lot of information about the unknown parameter is gathered by performing measurements at the intermediate time, especially at the transient time.  For a system with partial accessibility, one can only perform measurement on a block of size $L$. This practical limitation makes the sensing even more challenging as due to entanglement between the spins in the block and the rest of the system some information might get lost. To investigate the sensing with partial accessibility, the QFI has to be computed for the reduced density matrix $\rho_L  = \mbox{Tr}_{N-L}\Big(|\Psi(t)\rangle\langle \Psi(t)|\Big)$. This  reaches a non-trivial steady-state which serves as a potential probe state. 
 
For static field sensing, i.e., for estimating $h_0$, the time-dependent external driving field  $h(t)=h_1\sin(\omega\ t)$  is used as a control field to enhance sensitivity. The QFI of the reduced density matrix $\rho_L$ with respect to $h_0$, denoted as $I^{ss}_{Q}$,  tends to a steady-state~\cite{mishra2021driving}. It is shown that $I^{ss}_{Q}$ exhibits Floquet resonances at $h_0=h_c+\ell \omega$, $\ell \in \mathbb{Z}$. Thus, at these points of $h_0$, the $I^{ss}_{Q}$ displays peaks, as can be seen in Fig.~\ref{fig5:floquet_DC}(a). Moreover, the driving of the system with certain frequency $\omega$ recovers the Heisenberg scaling of $I^{ss}_{Q}\sim L^{\beta(\omega)}$ as a function of $L$  which is absent in the ground state. In the ground state the QFI $I^{gs}_{Q}$ scales with $L$ as $I^{gs}_{Q} \sim L^{1.68}$ as shown in Fig.~\ref{fig5:floquet_DC}(b). While for the steady-state, the exponent  $\beta(\omega=2)=2.06$, obtained in Fig.~\ref{fig5:floquet_DC}(b).  

Now for periodic field sensing, the Hamiltonian in Eq.~(\ref{eq:model}) is taken to be a a transverse -field Ising model ($\gamma=1$) with a time-dependent magnetic field of unknown amplitude $h_1$ added to the Hamiltonian~\cite{mishra2022integrable}. For simplicity, the form is taken as Dirac-delta pulse $h(t)=h_1\sum_{n=0}^{\infty}\delta(t-n\tau)$. The aim is to estimate the unknown amplitude $h_1$. The $I^{ss}_{Q}$ is maximum and has a peak along a line in the $(h_0-h_1)$ plane. Interestingly, this line corresponds to the line of vanishing Floquet gap. Moreover,  in the steady-state, the Fisher information $I^{ss}_Q$ with respect to $h_1$ exhibits the Heisenberg type scaling $I^{ss}_{Q}\sim L^{\beta}$. An important finding is that at the location of the vanishing Floquet gap, the exponent $\beta \approx 1.96 $ while for points where the Floquet phases do not cross, the exponent decreases towards the standard quantum limit. 

\section{Non-Equilibrium Quantum Sensing: Dissipative Phase Transitions}

Physical systems unavoidably interact with their environment, often referred to as the bath or reservoir, which possesses a larger number of degrees of freedom than the system of interest. The study of open quantum systems, which involves understanding the interaction between our system and the environment, has led to remarkable theoretical findings and practical applications~\cite{breuer2002theory}. However, the presence of noisy environments causes the systems to lose their coherence (decohere) and, consequently, results in unavoidable detrimental effects on their dynamics. This issue is particularly relevant for upcoming quantum technologies, as fighting against decoherence is essential for achieving more efficient quantum probes~\cite{goldstein2011environment}, developing algorithms for specific quantum tasks~\cite{plenio1997quantum}, and quantum simulators with significantly longer coherence times~\cite{kjaergaard2020superconducting}. \vm{See Ref.~\cite{jiao2023quantum} for a recent and comprehensive review on quantum metrology in the noisy intermediate-scale quantum (NISQ) era. Moreover, various sensing strategies have been proposed to estimate unknown parameters in open quantum systems. A few notable examples include recent works~\cite{wang2017quantum, bai2019retrieving, wei2021nonmarkovian, wei2021threshold, wu2021gaussian}---see also Ref.~\cite{mehboudi2019thermometry} for a thorough and comprehensive review on thermometry. Since the interaction between the system and its environment is unavoidable, a natural question arises: Can the detrimental effects of decoherence be mitigated or even exploited to enhance metrological performance? In the following, we review these possibilities in detail.}

\subsection{Profiting from dissipation}\label{sec:metrology-decoherence-A}
 
Since non-classical states are crucial for countless applications and theoretical advances, it would be highly desirable to have a mechanism where instead of avoiding dissipation, one could harness the very process of dissipation to one's advantage. 
The seminal work by Alfred Kastler in 1950 pointed in that direction~\cite{Kastler}. The process, known as \textit{optical pumping}, prepares any initial quantum state into a well-defined target state via selection rules of spontaneous emission. Building on Kastler's work, \textit{coherent population trapping} was experimentally demonstrated in the 1970s by Arimondo and Orriols~\cite{alzetta1976experimental, arimondo1976nonabsorbing}. In a nutshell, a three-level energy configuration with transitions levels $|a\rangle\longleftrightarrow |b\rangle$ and $|a\rangle\longleftrightarrow |c\rangle$ and Hamiltonian $\hat{H}=\hbar\Omega(|a\rangle\langle b| - |a\rangle\langle c| + h.c)$ allows for a quantum state with a zero eigenvalue (a \textit{dark state}). Remarkably, such a dark state, denoted as $|\psi_d\rangle \sim |b\rangle + |c\rangle$, decouples from the external pumping field $\Omega$, thus allowing excitations that reach that energy level to remain there. In contrast, the \textit{bright state}, denoted as $|\psi_b\rangle \sim |b\rangle - |c\rangle$, oscillates with the driving field $\Omega$ and the state $|a\rangle$. Indeed, this idea lies at the heart of laser cooling~\cite{PhysRevLett.61.826} as the ground state represents another specific quantum state. The paradigm shift of harnessing dissipation as a resource rather than a drawback has led to systematic studies of decoherence itself~\cite{myatt2000decoherence}, the ability to prepare a quantum state from any arbitrary initial state~\cite{PhysRevLett.77.4728}, the use of atomic engineered reservoirs to create two-mode Einstein-Podolsky-Rosen-entangled states~\cite{pielawa2007generation, pielawa2010engineering}, protection against noise through the engineering of pointer states~\cite{carvalho2001decoherence}, dissipative quantum computation~\cite{Verstraete2009}, the creation of many-body states and non-equilibrium quantum phases~\cite{diehl2008quantum},  thermometry of cold Fermi gases with dephasing-impurities~\cite{mitchison2020in, bouton2020single, brattegard2024thermometry}, quantum memories using dissipative mechanisms~\cite{pastawski2011quantum}, the generation of Bell states on an open-system quantum simulator~\cite{barreiro2011open}, and the possibility to enhance sensing precision nearly to the Heisenberg limit~\cite{goldstein2011environment}.

Other proposals include decoherence-free subspaces~\cite{lidar2014review, lidar1998decoherence, palma1996quantum, PhysRevLett.79.1953, zanardi1997dissipative}, dynamical decoupling~\cite{lidar2014review, viola1999dynamical, viola2003robust, khodjasteh2005fault}, and reservoir engineering techniques~\cite{myatt2000decoherence, PhysRevLett.77.4728, pielawa2010engineering, pielawa2007generation, carvalho2001decoherence, Verstraete2009, diehl2008quantum}. The latter technique has been utilized to prepare maximally entangled states of two qubits~\cite{PhysRevLett.106.090502, PhysRevA.82.054103}, W-states~\cite{PhysRevA.87.042323, li2018dissipation}, and many-body entangled states~\cite{PhysRevLett.106.020504, stannigel2012driven}. In the continuous variable scenario, this technique allows the study of multi-mode entanglement~\cite{PhysRevA.76.022312, li2010multi} and the preparation of steady entanglement in bosonic dissipative networks~\cite{PhysRevA.90.062322}. Notably, it also enables the generation of quantum many-body entangled states which remains robust in the presence of dephasing, thermal effects, and exhibits scalability and robustness against disorder in the model parameters~\cite{de2017steady}.

These approaches could provide efficient solutions for the fields of quantum computation, quantum communication, and quantum metrology where long-lived quantum states and quantum correlations are essential.

\subsection{Restoring quantum sensing advantage in noisy environments}\label{sec:noisy_metrology}

The previous section demonstrates the ability to deterministically prepare strongly correlated multi-mode and many-body quantum states by harnessing dissipation. Generating such quantum states is particularly advantageous for advancing quantum metrology. For instance, in lossless optical interferometry, the use of N00N states $|\Psi_\mathrm{N00N}\rangle \sim |N,0\rangle + |0,N\rangle$ in Mach-Zender interferometers enables achieving the Heisenberg limit of precision for phase estimation. Therefore, the deterministic production of N00N states is highly desirable. However, even if the N00N state is faithfully prepared, the loss of a single excitation destroys all the information about the unknown phase. This poses a significant challenge for quantum state preparation, as unavoidable noise can severely undermine sensing precision, even if states are efficiently prepared. In fact, in the asymptotic limit of the number of photons $N\gg 1$, it has been proven that phase estimation in the presence of losses can only improve the phase sensitivity up to a certain factor~\cite{kolodynski2010phase, demkowicz2012elusive}. Similarly, S. Knysh et al.~\cite{knysh2011scaling} demonstrated that in dissipative systems and in the asymptotic limit $N\gg 1$, the scaling with respect to the number of photons $N$ transitions from the Heisenberg limit towards the shot-noise limit. Interestingly, the crossover between the Heisenberg to standard quantum limit scaling is solely a function of dissipation.

One solution to overcome photon losses was experimentally demonstrated by M. Kacprowicz et al.~\cite{kacprowicz2010experimental}. They employed a modified N00N state of the form $|\psi\rangle = x_1|2,0\rangle + x_2|0,2\rangle + x_3|1,1\rangle$ with non-negative weights $x_i$, which enhances the robustness of phase estimation in the presence of single excitation loss. Remarkably, even if one excitation is lost, the modified N00N state remains maximally entangled. However, it is essential to note that the above approach still falls under the category of quantum probe preparation.

A more fundamental route to overcome noisy metrology was proposed by B. M. Escher et al.~\cite{escher2011general}. Their proposal involves treating the system in addition to the environment as a whole unitary process, aiming to restore quantum-enhanced sensitivity. This approach entails an upper bound given by
\begin{equation}
C_Q[\rho_{\mathrm{S}+\mathrm{E}}(\theta)]\equiv I_Q[\rho_{\mathrm{S}+\mathrm{E}}(\theta)] \geq I_Q[\rho_{\mathrm{S}}(\theta)]
,\end{equation}
where $I_Q[\rho_{\mathrm{S}+\mathrm{E}}(\theta)]$ represents the QFI of the joint system plus environment ($S + E$), and $I_Q[\rho_{\mathrm{S}}(\theta)]$ denotes the QFI of the system ($S$) only, i.e.~with the environment traced out. This framework enables the retrieval of information lost in the environment. Such a general approach has found applications in lossy optical interferometry and atomic spectroscopy in the presence of dephasing~\cite{escher2011general}, force and displacement estimation using a noisy quantum-mechanical oscillator probe~\cite{PhysRevA.88.042112}, and in variational approaches~\cite{escher2012quantum}. The proposed general protocol in the extended space entails a purified evolution and, remarkably, demonstrates that there is always a purification channel such that one can restore the quantum-enhanced limit of precision. However, observing the joint system plus environment might be practically very challenging. 

\vm{Another strategy in noisy quantum metrology is the use of time-continuous monitoring~\cite{plenio2016sensing}. Indeed, continuous monitoring of the environment has been proposed as a method to restore Heisenberg-limited precision~\cite{albarelli2018restoring}. For a pedagogical introduction, see Albarelli et al.~\cite{albarelli2024apedagogical}. Within this framework, the following chain of inequalities has been established~\cite{albarelli2018restoring} (for the case of continuous measurements in linear Gaussian quantum systems, see M. G. Genoni~\cite{genoni2017cramer}):  
\begin{equation}  
I_Q[\rho_\mathrm{unc}] \leq \tilde{I}_{Q_{\mathrm{unr}, \eta_j}} \leq \overline{I_Q}_{\mathcal{L}_\theta},  
\end{equation}  
where $I_Q[\rho_\mathrm{unc}]$, $\tilde{I}_{Q_{\mathrm{unr}, \eta_j}}$, and $\overline{I_Q}_{\mathcal{L}_\theta}$ denote the QFI for the unconditional state, the unraveling process, and the joint system plus environment, respectively~\cite{albarelli2018restoring}. This continuous sensing methodology has been applied in various contexts, including quantum frequency estimation under independent dephasing, where the recovered sensing precision matches the noiseless case~\cite{albarelli2020quantum}; noisy quantum metrology enhanced by continuous nondemolition measurement~\cite{rossi2020noisy}; Heisenberg-limited quantum magnetometry~\cite{albarelli2017ultimate}; characterization of spin-squeezing via continuous monitoring of the output field in a cavity-QED setup~\cite{caprotti20242024}; fundamental tests of physics, such as discriminating wave function collapse models~\cite{genoni2016unravelling}; and applications in quantum batteries~\cite{morrone2023daemonic}.}

\subsection{Dissipative phase transition as a sensing resource}

As opposed to zero temperature (zero entropy) phase transitions lead by quantum fluctuations, dissipative phase transitions emerge due to the competition between the unitary (Hamiltonian) and the non-unitary (Lindbladian) system's parameters~\cite{lee2013unconventional}. Leading to a phase transition in the steady-state as this competition varies. Indeed, in the thermodynamic limit, the competition between external drivings, the Hamiltonian evolution, and the dissipation mechanisms can trigger a non-analytical change in the steady-state of the system~\cite{minganti2018spectral}. One of the first studies on general properties of the dissipative phase transitions was addressed by Kessler et al., ~\cite{kessler2012dissipative} by investigating the steady-state properties of the central spin model
\begin{equation}
\dot{\rho} = -i[\hat{H}_\mathrm{CSS}, \rho]{+}J\gamma\left(\hat{S}^-\rho \hat{S}^+ - \frac{1}{2}\{\hat{S}^+\hat{S}^-,\rho\} \right),
\end{equation}
where the central spin system (CSS) Hamiltonian $\hat{H}_\mathrm{CSS} = \hat{H}_S + \hat{H}_I + \hat{H}_{SI}$ decomposes in $\hat{H}_S=J\Omega(\hat{S}^+ + \hat{S}^-)$, $\hat{H}_I=\delta\omega \hat{I}_z$, and $\hat{H}_{SI}=a/2(\hat{S}^+\hat{I}^-+\hat{S}^-\hat{I}^+) + a\hat{S}^+\hat{S}^-\hat{I}_z$. In the above, $\hat{S}^{\pm}$ and $\hat{I}_{z}$ are the collective electron and nuclear spin operators, respectively. $J\Omega$ is the Rabi frequency, $\delta\omega$ is the difference of hyperfyne detuning $\omega$, $a$ is the individual hyperfine coupling strength, and $\gamma$ the dissipation rate~\cite{kessler2012dissipative}. {\color{black} Interestingly, the character of the dissipative phase transitions, which may entail pure or mixed quantum states, was describe by having one or more steady states. This  is conjectured to happen when the Liouvillian spectral gap closes for both the real and imaginary parts~\cite{kessler2012dissipative}, see Table.~\ref{fig_table_dissipative}.}
\begin{table*}[ht]
\begin{center}
\begin{tabular}{ |m{3cm}|m{4.5cm}|m{4.5cm}|m{4.5cm}|  }
 \hline
 \textbf{} & \textbf{TPT} & \textbf{QPT} & \textbf{DPT} \\ \hline\hline
 \textbf{System operator}   & Hamiltonian \par $\hat{H} = \hat{H}^\dagger$ & Hamiltonian \par $\hat{H}=\hat{H}^\dagger$ & Liouvillian \par $\hat{\mathcal{L}}$ - Lindblad \\ \hline
 \textbf{Relevant quantity}   & Free energy \newline $F(\rho)=\langle \hat{H} \rangle_\rho - T\langle S\rangle_\rho$    & Energy eigenvalues \newline $E_\psi:\hat{H}\ket{\psi}=E_\psi\ket{\psi}$ & ``Complex energy" eigenvalues \newline $\lambda_\rho : \hat{\mathcal{L}}\rho = \lambda_\rho\rho$ \\ \hline
 \textbf{State} & Gibbs state \newline $\rho_T=\underset{\rho\geq 0, \mathrm{Tr}(\rho)=1}{\mathrm{argmin}}[F_{\rho}]$ \newline $\rho_T \propto \mathrm{exp}[-\hat{H}/k_BT]$ & Ground state \newline $\ket{\psi_0}=\underset{||\psi||=1}{\mathrm{argmin}}[\bra{\psi}\hat{H}\ket{\psi}]$ \newline $[\hat{H}-E_{\psi_0}]\ket{\psi_0}=0$ & Steady state \newline $\rho_0=\underset{||\rho||_{\mathrm{tr}}=1}{\mathrm{argmin}}[||\hat{\mathcal{L}}\rho||_\mathrm{tr}]$ \newline $\hat{\mathcal{L}}\rho_0=0$ \\ \hline
 \textbf{Phase transition}       & Non-analyticity in $F(\rho_T)$ & $\Delta = E_{\psi_1}-E_{\psi_0}$ vanishes & ADR $= \max[\mathrm{Re}(\lambda_{\rho})]$ vanishes \\ \hline
 \end{tabular}
 \end{center}
\caption{\textbf{Non-exhaustive comparison of thermal phase transitions (TPT), quantum phase transitions (QPT) and dissipative phase transitions (DPT):} The concepts for DPT parallel in many respects the considerations for QPT and TPT. $||\cdot ||_\mathrm{tr}$ denotes the trace norm and $S$ the entropy. Note that if the steady state is not unique, additional steady states may come with a non-zero imaginary part of the eigenvalue and then appear in pairs: $\mathcal{L}\rho=\pm iy\rho$, $(y \in \mathbb{R})$. Table and caption taken from~\cite{kessler2012dissipative}.}
\label{fig_table_dissipative}
\end{table*}

\begin{figure}[b]
    \centering
    \includegraphics[width=\linewidth]{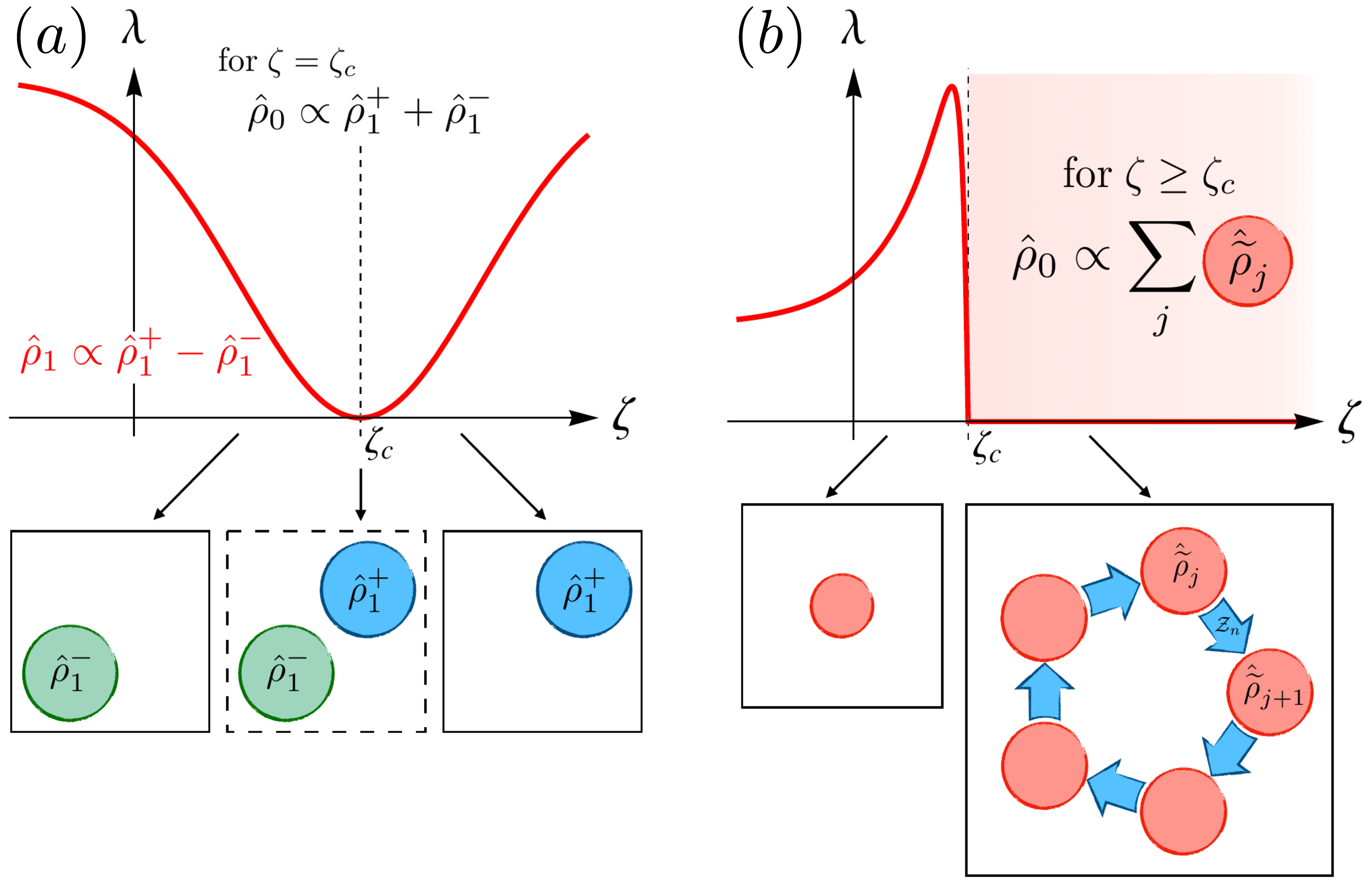}
    \caption{\textbf{Sketches illustrating the paradigms of first-order and second-order dissipative phase transitions:} \boldsymbol{$(a)$} \textbf{First-order case.} In the thermodynamic limit, both the real part $\mathrm{Re}[\lambda_1]$ and the imaginary part $\mathrm{Im}[\lambda_1]$ of the Liouvillian gap close when the parameter $\zeta$ (that triggers the transition) of the Liouvillian reaches its critical value $\zeta_c$, i.e., $\zeta \simeq \zeta_c$. At the critical point $\zeta = \zeta_c$, the steady state $\rho_{ss}$ is bimodal. Specifically, $\rho_{ss}$ is a statistical mixture of $\rho_1^+$ and $\rho_1^-$, which represent two distinct phases of the system. \boldsymbol{$(b)$} \textbf{Second-order case.} In the sketch, it is considered the breaking of a $Z_n$ symmetry with $n = 5$. In the thermodynamic limit, the Liouvillian gap closes over the entire region $\zeta \geq \zeta_c$. Moreover, for $\zeta \geq \zeta_c$, all eigenvalues $\lambda_0, \lambda_1, \ldots, \lambda_{n-1}$ of the Liouvillian are zero. When $\lambda \neq 0$ (here $\zeta < \zeta_c$), the steady-state density matrix $\rho_{\text{ss}}$ is mono-modal. In the symmetry-broken phase ($\lambda = 0$ and $\zeta \geq \zeta_c$), $\rho_{\text{ss}}$ is an $n$-modal statistical mixture of density matrices $\hat{\tilde{\rho}}_j$, which are mapped one onto the other under the action of the symmetry superoperator $Z_n$. Figure taken from~\cite{minganti2018spectral}.}     
    \label{fig_spectral_dissipative}
\end{figure}

A more rigorous (general) theory of Liouvillian spectra analysis was indeed addressed by Minganti et al.,~\cite{minganti2018spectral}, confirming that the Liouvillian gap, i.e., $\mathrm{Re}[\lambda_1]$ such that the Liouvillian superoperator $\hat{\mathcal{L}}$ satisfies $\hat{\mathcal{L}}\rho_i=\lambda_i\rho_i$, is $\mathrm{Re}[\lambda_1]=0$ only at the critical point if the transition is of the first-order [see Fig.~\ref{fig_spectral_dissipative}(a)], whereas for the second-order type, $\mathrm{Re}[\lambda_1]=0$ in the whole region of broken symmetry~\cite{minganti2018spectral} [see Fig.~\ref{fig_spectral_dissipative}(b)]. Consequently, the steady state of the system exhibits a divergent susceptibility with respect to parameters of the system~\cite{kessler2012dissipative}. Other analyses include the use of the Keldysh formalism~\cite{dalla2012dynamics, marino2016driven} and numerical approaches~\cite{vicentini2018critical, rota2018dynamical}.

Intense theoretical work has been pursued to understand the critical behavior at the dissipative phase transition~\cite{rota2017critical, vicentini2018critical, letscher2016bistability, overbeck2017multicritical, capriotti2005dissipation, eisert2010noisedriven, sieberer2014nonequilibrium, garbe2020critical, ivanov2020enhanced}, including photonic systems~\cite{carmichael2015breakdown, benito2016degenerate, arenas2016beyond, casteels2016power, bartolo2016exact, casteels2017critical, casteels2017quantum, fossfeig2017emergent, biondi2017nonequilibrium, biella2017phase, savona2017spontaneous, munoz2018hybrid}, lossy polariton condensates~\cite{sieberer2013dynamical, sieberer2014nonequilibrium}, and spin systems~\cite{kessler2012dissipative, lee2013unconventional, jin2016cluster, lee2011antiferromagnetic, chan2015limit, maghrebi2016nonequilibrium, overbeck2017multicritical}. In addition, fundamental limitations on criticality and driven-dissipative phase transitions in \textit{quadratic} open systems, namely bilinear Hamiltonian, linear Lindblad operators, and bilinear Hermitian Lindblad operators, have been recently studied~\cite{zhang2022criticality}. Example of the above have been investigated for driven-dissipative Bose-Einstein condensate~\cite{zhang2024driven}. Exact steady-state expressions~\cite{lawande1981effects, puri1980dispersion, bartolo2016exact} and approximate methods for their evaluation based on variational approaches have been pursued~\cite{weimer2015variational, raghunandan2018high}. The presence of phase symmetry breaking in a single qubit-laser system~\cite{fernandez2017quantum}, semiclassical first-order dissipative phase transition in Kerr parametric oscillators with parity symmetry breaking~\cite{heugel2019quantum}, and second-order transitions in the two-photon Kerr resonator~\cite{di2021critical}. The universality class of driven-dissipative systems has also been extensively addressed~\cite{marcuzzi2014universal, carollo2019critical, sieberer2013dynamical, marino2016driven, maghrebi2016nonequilibrium} with in-depth analysis of the phase diagram and critical exponents of a dissipative transverse Ising spin chain~\cite{werner2005phase}. From a geometric perspective, the Uhlmann curvature has been proposed to shed light upon the nature of criticality~\cite{carollo2018uhlmann}, and a recent method based on the coherent anomaly approach has been proposed for determining critical exponents~\cite{jin2021determination}.

In addition to the above extensive theoretical findings, several experiments have already been realized in various systems, including trapped ions~\cite{cai2022probing}, ultracold atoms~\cite{ferioli2022observation, benary2022experimental, brennecke2013real, ferri2021emerging, baumann2010dicke}, cavity-polariton~\cite{fink2018signatures, ohadi2015spontaneous}, superconducting circuits~\cite{fitzpatrick2017observation, collodo2019observation}, Rydberg atom ensembles~\cite{ding2016phase, letscher2016bistability}, to name a few.

Dissipative phase transitions have been extensively studied in quantum optics, particularly in the context of cooperative resonance fluorescence~\cite{Agarwal1977, Carmichael1977, Walls1978, Walls1980, Carmichael1980, morrison2008dissipation}. Notably, the quantum optical model used to explain cooperative resonance fluorescence has also been employed to investigate \textit{Boundary Time Crystals} (BTCs), which are open quantum many-body systems situated at the boundary of a large bulk undergoing everlasting oscillations of a certain observable in the thermodynamic limit~\cite{iemini2018boundary, carollo2022exact, lledo2020dissipative}. Time crystals are intriguing phenomena resulting from the breaking of time-translational symmetry~\cite{wilczek2012quantum,shapere2012classical,li2012space}. They have been studied for both discrete and continuous temporal symmetry breakings. 

\begin{figure}
    \centering
    \includegraphics[width=\linewidth]{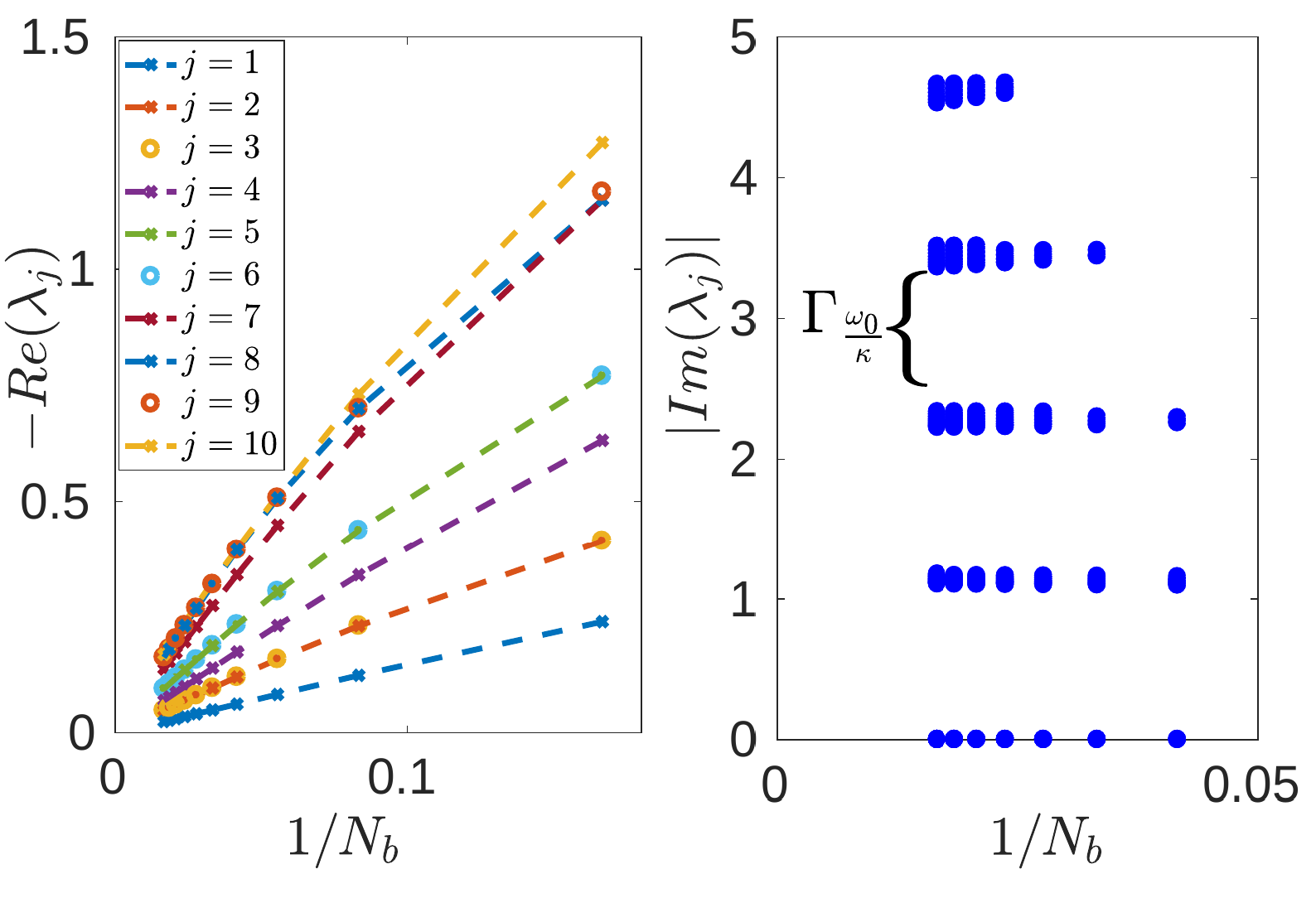}
        \caption{\textbf{BTC Liouvillian spectra analysis: (Left)} Finite size scaling for the real part of the Liouvillian eigenvalues in the BTC phase. The index $j$ labels the eigenvalues. The Liouvillian eigenvalues $\lambda_j$ are ordered as a function of their real part ($|\text{Re}(\lambda_j)| \leq |\text{Re}(\lambda_{j+1})|$, and $j = 0$ has zero real part). In the $\omega_0/\kappa > 1$ phase, they scale to zero as a power-law of the inverse system size. \textbf{(Right)} The imaginary parts of the eigenvalues show a band structure, with a fundamental frequency separation $\Gamma_{\omega_0/\kappa}$. For fixed excitation thresholds ($\lambda_j$ is selected such that $\nu = \frac{j^2}{N_b} \leq \epsilon$) [$N_b$ being the system size of the boundary], the width of the bands remains finite in the thermodynamic limit (here $\nu < 0.025$). The widths of the bands tend to decrease as lower excitation thresholds are considered. The eigenvalues are plotted in units of $\kappa$. Figure taken from~\cite{iemini2018boundary}.}
    \label{fig_btc_gap}
\end{figure}

Discrete time crystals and their potential for serving as quantum sensors in closed quantum systems were discussed in Sec.~\ref{sec:discrete_time_crystal}. Here, the focus is on the case of  continuous symmetry breaking, which is responsible for the emergence of BTC, the Liouvillian spectral gap closes only for the real part, while the imaginary part forms band gaps~\cite{minganti2018spectral, iemini2018boundary, minganti2020correspondence}, see Fig.~\ref{fig_btc_gap}. This is the most distinctive feature of BTC as the presence of bands give rise to everlasting oscillations in their stationary dynamics~\cite{sacha2017time, iemini2018boundary}. BTCs have been studied using mean-field analysis~\cite{carollo2022exact} and through continuous monitoring~\cite{cabot2022quantum}. Recently, a correspondence between second-order dissipative phase transitions and dissipative time crystals in the thermodynamic limit~\cite{minganti2020correspondence} and a thorough analysis of genuine multipartite correlations in a BTC have been put forward~\cite{lourencco2022genuine}.

BTCs serve as valuable probes for addressing key issues in quantum sensors, namely: (i) they avoid the need for specific quantum many-body probe preparation through open dynamics and inherently benefit from dissipation, and (ii) an undemanding measurement basis can extract a fair fraction of the ultimate sensing. In concrete terms, the BTC probe is governed by the following non-unitary dynamics~\cite{Agarwal1977, iemini2018boundary, carollo2022exact}
\begin{equation}
\frac{d}{dt}\rho = -i\omega[\hat{S}^x,\rho]+\frac{\kappa}{S}\left(\hat{S}^-\rho \hat{S}^+-\frac{1}{2}\left\{\hat{S}^+\hat{S}^-,\rho\right\} \right)=\hat{\mathcal{L}}[\rho],\label{eq:master-equation-BTC}
\end{equation}
where $N$ non-interacting spin-1/2 particles are described by a (pseudo-)spin of length $S{=}N/2$. The collective angular momentum operators are given by $\hat{S}^{x,y,z}{=}1{/}2\sum_j\hat{\sigma}^{x,y,z}_{j}$, where $\hat{\sigma}^{x,y,x}_{j}$ is the Pauli matrix at site $j$. In Eq.~\eqref{eq:master-equation-BTC} $\omega$ is the single particle coherent splitting, $\hat{\mathcal{L}}[\rho]$ is the Liouvillian operator, and $\kappa$ is the effective collective emission rate. Deviations of Eq.~\eqref{eq:master-equation-BTC}, including local pumping and anisotropies in the coherent splitting, shows the robustness of such BTC probes even in this noisy case~\cite{Tucker_2018}. Such modified master equation faithfully represents state-of-the-art experiments~\cite{Norcia_2018, Shankar_2017}. The steady-state $\rho_\mathrm{SS}=\rho(t\rightarrow\infty)$ of the boundary undergoes a phase transition from an unbroken symmetry phase (determined by $\omega<\kappa$) to a boundary time crystal phase with everlasting total spin oscillations (determined by $\omega>\kappa$).  

\begin{figure}[t]
\centering \includegraphics[width=\linewidth]{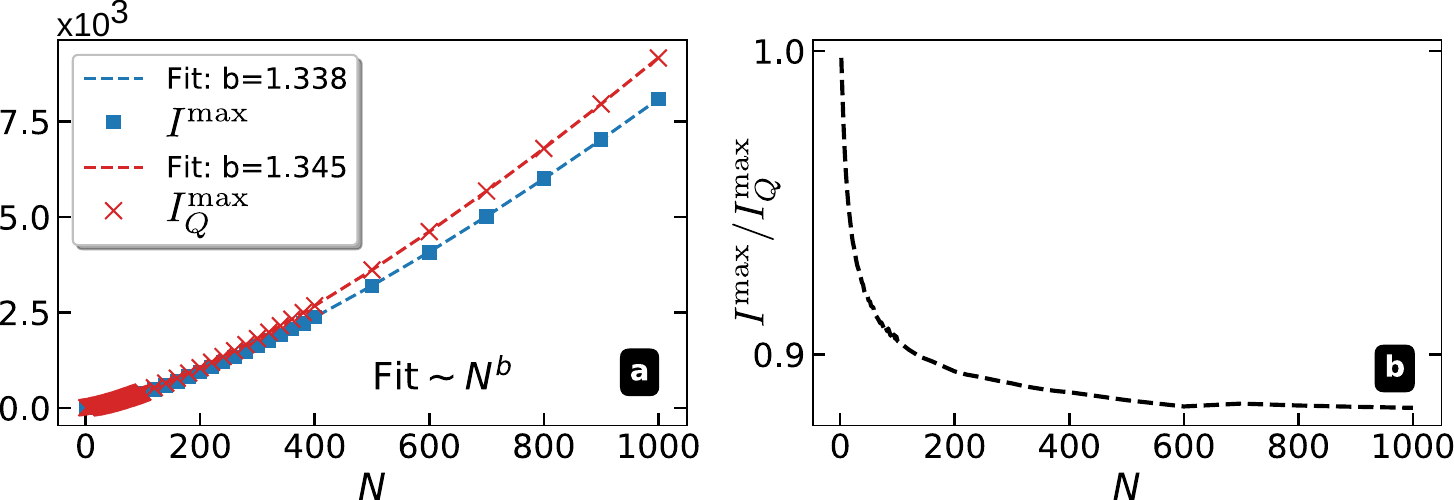}
\caption{\textbf{BTC quantum-enhanced sensing:} \boldsymbol{$(a)$} Maximal quantum (classical) Fisher information $I_Q^\mathrm{max}$ ($I^\mathrm{max}$) as a function of the system size $N$. Fitting functions with super-linear coefficient $b>1$, evidencing quantum-enhanced sensing using a suboptimal observable. \boldsymbol{$(b)$} Efficiency ratio $I^\mathrm{max}/I^\mathrm{max}_Q$ as a function of the system size $N$. Figure from~\cite{montenegro2023quantum}.}\label{fig5:CFI}
\end{figure}

Quantum-enhanced sensitivity using the BTC probe of Eq.~\eqref{eq:master-equation-BTC} can be evidenced in Fig.~\ref{fig5:CFI}(a) via QFI analysis~\cite{montenegro2023quantum}. As seen from the figure, the maximum of the QFI $I_Q^\mathrm{max}$ grows polynomially with the system size $N$. A fitting function of the form $I_Q^\mathrm{max}=aN^b + c$ ($c\rightarrow 0$) reveals a coefficient $b=1.345$. Interestingly, in Fig.~\ref{fig5:CFI}(a), a simple measurement basis of the total magnetization also scales super-linearly with the system size $N$. A fitting function of the same form as above exhibits quantum-enhanced sensitivity with a coefficient $b=1.338$. To show the performance between the QFI and the CFI, in Fig.~\ref{fig5:CFI}(b), the ratio between their maximum values is plotted, showing an asymptotic performance of 90\% ---a fair fraction achieved with a sub-optimal measurement basis.

\section{Non-Equilibrium Quantum Sensing: Quantum Many-Body Scars and Sensing}

Quantum many-body scars (QMBSs)~\cite{shiraishi2017systematic, turner2018weak, serbyn2021quantum}, observed in isolated interacting many-body systems, are known for violation of the eigenstate thermalization hypothesis~\cite{deutsch1991quantum,srednicki1994chaos,rigol2008thermalization} and, hence, exhibiting non-equilibrium dynamics. Recently, the potential application of QMBSs in quantum sensing,  using either their long-lived coherence time~\cite{dooley2021robust} or their strong multipartite entanglement~\cite{dooley2023entanglement,desaules2022extensive}, has been identified. 
In Ref.~\cite{dooley2021robust}, the authors consider $N$ spin-$1$ particles in a 1D lattice with Dzyaloshinskii-Moriya interaction (DMI).
The aim is to estimate the strength of an unknown magnetic field $\theta$ which acts uniformly on particles. The Hamiltonian reads
\begin{eqnarray}\label{Eq:DMI_Hamiltonian}
\hat{H}(\phi)&=&\sum_{i<j}\dfrac{J}{(i-j)^2}[\cos\phi(\hat{S}_i^{x}\hat{S}_j^{x} + \hat{S}_i^{y}\hat{S}_j^{y}) \cr &+& \sin\phi(\hat{S}_i^{x}\hat{S}_j^{y} - \hat{S}_i^{y}\hat{S}_j^{x})] + \dfrac{\theta}{2}\sum_{i=1}^{N}\hat{S}_i^{z}, 
\end{eqnarray}
where $\hat{S}_i^{x,y,z}$ are the spin-1 operators, and $J$ as the coupling strength.
The interaction part of the Hamiltonian rotates between a pure XX interaction for $\phi{=}0,\pm\pi$ and DMI for $\phi{=}\pm\pi/2$.  
Spins that are prepared in the initial product state
$\vert\psi_{t=0}\rangle{=}\otimes_{i=1}^{L}(\vert {+1}\rangle{+}\vert {-1}\rangle){/}\sqrt{2}$
are allowed to evolve under the action of Hamiltonian $\hat{H}$ for a sensing time $t$.
Then the local observable $\hat{\mathcal{O}}_{\zeta}{=}e^{-i\zeta}\otimes_{i=1}^{L}(\vert {+1}\rangle\langle {-1}\vert + \text{h.c.})$, with properly tuned $\theta$, will be measured.
In the absence of interaction between spins, i.e.~$J{=}0$,
the time evolved state can be obtained as $\vert\psi_{t}\rangle{=}\otimes_{i=1}^{L}(e^{-it\theta/2}\vert {+1}\rangle{+}e^{it\theta/2}\vert {-1}\rangle){/}\sqrt{2}$
which results in the estimation error  $\delta \theta=1/\sqrt{tL  t_\mathrm{tot}}$ without restriction on $t$.
Generally, the interactions between the spins through generating entanglement and thermalization scramble quantum information irreversibly and, hence, deteriorate the sensing performance. 
In this case, one needs to complete the sensing task in an optimal time $t^{*}$ which is much smaller than the termalization time, resulting in the minimal value of the estimation error $\delta \theta^{*}{=}\min_{t}\delta \theta$. 
\begin{figure}
    \centering
\includegraphics{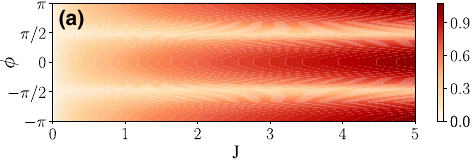}   \includegraphics[width=\linewidth]{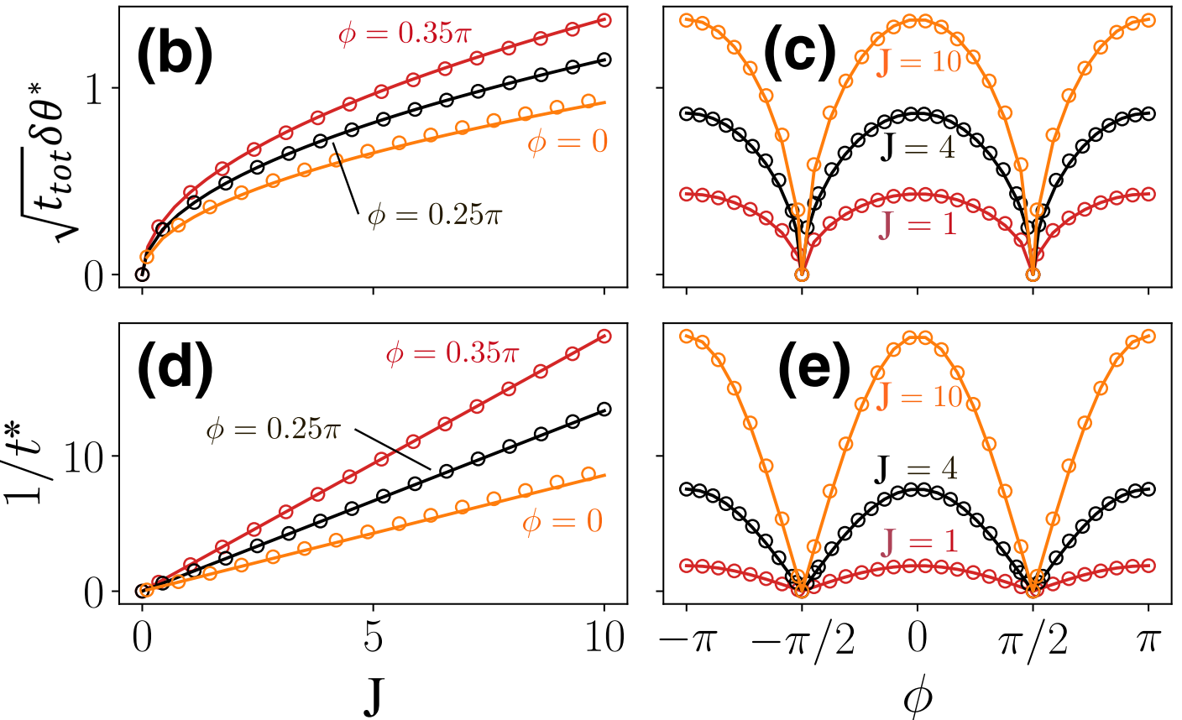}
    \caption{\textbf{Quantum Scar based sensor;} (a) the minimal error $\sqrt{t_\mathrm{tot}}\delta \theta^{*}$ obtained numerically for optimal sensing time $t^{*}$, in a system of size $L{=}10$, versus $J$ and $\phi$. (b) and (c), some cross-sections of panel (a). The numerical simulation (markers) is well fitted by the function $\delta \theta^{*}{=}1.09{/}\sqrt{t_\mathrm{tot}Lt^{*}}$ (solid line). (d) and (e) the optimal sensing time $t^{*}$ and its behavior for different values of the coupling strength $J$ and $\phi$. The numerical simulation (markers) can be closely described by the fitting function $t^{*}{=}0.53{/}|J\cos\phi|$ (solid line). This figure is adapted from Ref.~\cite{dooley2021robust}. }
    \label{fig:Fig_Scar}
\end{figure}
In Fig.~\ref{fig:Fig_Scar}(a), the minimal error $\delta h^{*}$, for optimal time $t^{*}$ and $L{=}10$, as a function of $\phi$ and the coupling strength $J$ is presented. Clearly, regardless of the interaction strength, the minimal error can be obtained for $\phi{=}{\pm} \pi/2$.
For clarifying the relationship between all the parameters, in Figs.~\ref{fig:Fig_Scar}(b) and (c) for selected values of $J$ and $\phi$, the minimal error $\sqrt{t_\mathrm{tot}}\delta \theta^{*}$ is plotted. The numerical simulation (markers) are well fitted by the function $\delta \theta^{*}{=}1.09{/}\sqrt{t^{*}Lt_\mathrm{tot}}$ (solid line). The same approach can be followed to formulate the optimal time of sensing $t^{*}$, see Figs.~\ref{fig:Fig_Scar}(d) and (e).
The numerical simulation is well fitted by $t^{*}{=}0.53{/}|J\cos\phi|$ (solid line).
Clearly, for $\phi{=}\pm\pi{/}2$, i.e.~when the interaction between spins is purely DMI, one has a diverging optimal sensing time $t^{*}{\rightarrow}\infty$ and its associated vanishing error, even for strong interactions.
The origin of this quite surprising result is QMBSs, as it has been shown in Ref.~\cite{dooley2021robust}.   
For proving this connection, the authors start by introducing the
spin-$1/2$ operators for two local basis states $\vert{\pm 1}\rangle$ of each spin-$1$ particle.
The corresponding operators are $\hat{S}^{\pm}_{i}{=}\vert{\pm 1}\rangle\langle{\mp 1}\vert$, and $\hat{S}^{z}_{i}{=}[\hat{S}^{+}_{i},\hat{S}^{-}_{i}]$ with the associated collective operators as $\hat{\mathcal{J}}^{\pm}{=}\sum_{i}\hat{S}^{\pm}_{i}$, and  $\hat{\mathcal{J}}^{z}{=}(1/2)\sum_{i}\tilde{S}^{z}_{i}$. 
It has been shown that the symmetric Dicke states $\{\vert\psi(s)\rangle\}_{s=0}^{L}$ defined as 
\begin{equation}
\vert \psi(s)\rangle = \sqrt{\dfrac{(L-s)!}{s!L!}}(\hat{\mathcal{J}}^{+})^s\big(\otimes_{i=1}^{L}\vert{-1}\rangle\big)
\end{equation}
are the scar states of Hamiltonian Eq.~(\ref{Eq:DMI_Hamiltonian}) with $\phi{=}\pm\pi/2$~\cite{mark2020eta}.
In other words, the eigenstates $\vert \psi(s)\rangle$, defined by $\hat{H}(\phi{=}{\pm}\pi/2)\vert \psi(s)\rangle{=}\theta(s{-}L/2)\vert \psi(s)\rangle$, violate the eigenstate thermalization hypothesis  as they show a subvolume-law entanglement growth.
Rewriting the initial product state $\vert\psi_{t=0}\rangle{=}\otimes_{i=1}^{L}(\vert {+1}\rangle{+}\vert {-1}\rangle){/}\sqrt{2}$ in terms of the elements of the Dicke-scar subspace results in  
$\vert\psi_{t{=}0}\rangle{=}\sum_{s=0}^{L}
\sqrt{\dfrac{L!}{2^{L}s!(L-s)!}}\vert\psi(s)\rangle$.
The time-evolved state in this subspace is obtained as 
$\vert\psi_t^{*}\rangle{=}\otimes_{i=1}^{L}(e^{-it^{*}\theta/2}\vert{+1}\rangle+e^{it^{*}\theta/2}\vert{-1}\rangle){/}\sqrt{2}$ results in $\delta\theta{=}1/\sqrt{t^{*}Lt_\mathrm{tot}}$, despite the presence of strong interaction.
Therefore, the initial state which is prepared in the Dicke-scar subspace of Hamiltonian $\hat{H}(\phi{=}\pm\pi/2)$ can result in a vanishing estimation error via its long-lived coherence property $t^{*}{\rightarrow}{\infty}$.
More discussion on the advanced role of quantum many-body scar in quantum-enhanced sensing can be found in Refs.~\cite{dooley2021robust,dooley2023entanglement,desaules2022extensive}. 

\section{Non-Equilibrium Quantum Sensing: Sequential Measurements Metrology}\label{sec_seq_measurements}

Previous sections have demonstrated that placing the probe at the critical point of a general phase transition or a special highly entangled state, such as GHZ-type and N00N-type states, leads to quantum-enhanced sensitivity---see Secs.~\ref{sec:2nd-order-pt}-~\ref{sec:ultimate-limits} for more details. These sensing advantages primarily exploit quantum superposition to achieve enhanced sensitivity. However, not all quantum probes experience a phase transition, and experimental limitations prevent us from generating arbitrary superpositions of many-body states. Therefore, it becomes essential to explore other features of quantum mechanics to attain quantum sensing advantages.

Measurement represents another distinct feature that sets apart the classical from the quantum world. Indeed, the concept of wave-function collapse due to projective measurements on the quantum system has sparked several discussions within quantum theory itself~\cite{machida1980theory1, machida1980theory2, furry1936note, zeh1970interpretation}. In quantum many-body probes with partial accessibility, measurements can only be conducted locally on a subsystem. However, in these scenarios, even though the measurement is local, the wave function collapse is global and affects the entire system's wave function. This intriguing phenomenon has been the subject of intensive studies~\cite{Busch1990, Schmidt_2020, BAN2021127383, PhysRevX.9.031009, PhysRevLett.128.010604, benoist2023limit, Haapasalo_2016, benoist2019invariant, burgarth2014exponential, PhysRevA.92.042315}, such as a unique feature for quenching many-body systems~\cite{bayat2017scaling, pouyandeh2014measurement, ma2018phase, bayat2018measurement}, leading to potential new types of phase transitions~\cite{choi2020quantum, feng2022measurement, minato2022fate, skinner2019measurement, gullans2020dynamical, block2022measurement}. These \emph{measurement-induced phase transitions} have also been recently linked with quantum-enhanced sensitivity~\citep{paviglianiti2023multipartite, di2024metrology}. 

\begin{figure}
    \centering
    \includegraphics[width=\linewidth]{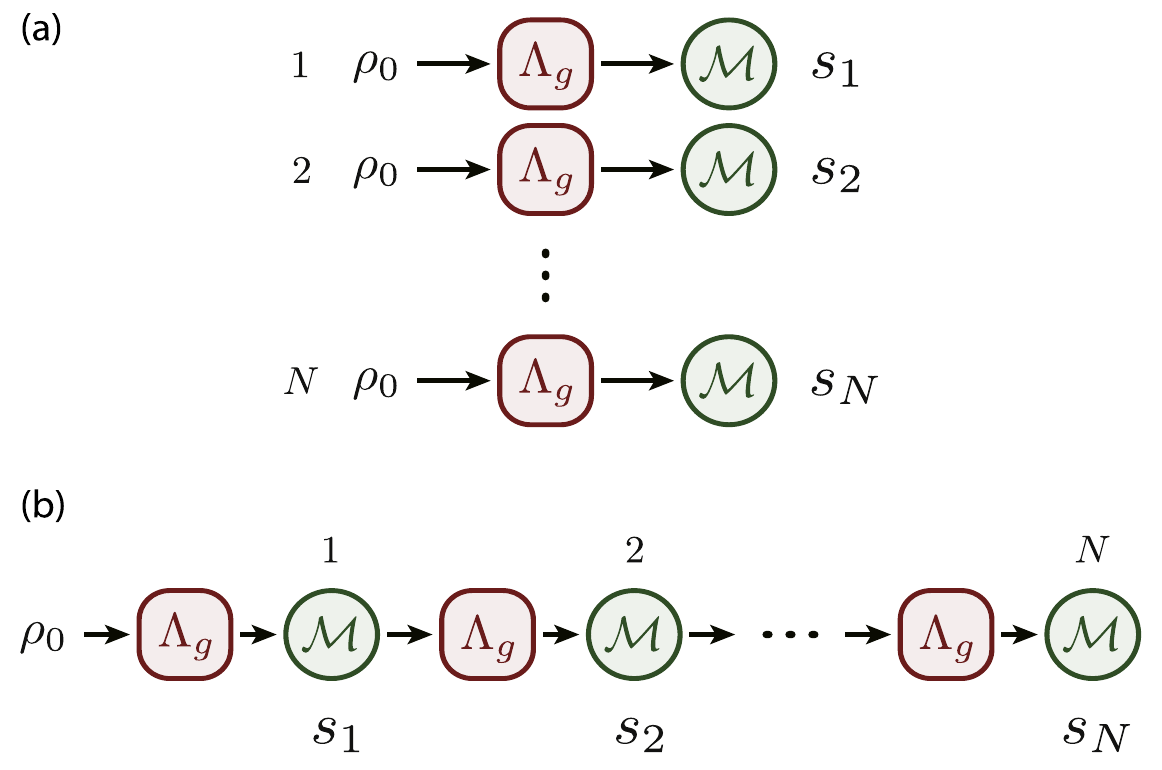}
    \caption{\textbf{Comparison between conventional and sequential measurements sensing strategies:} \boldsymbol{$(a)$} Conventional strategy for estimating a parameter $g$ of a quantum system, where measurement data $\{s_1, \ldots, s_N\}$ are collected by \textit{independent and identical} experiments. Every time the experiment is performed, the system is reset to some specific known initial state $\rho_0$. \boldsymbol{$(b)$} Sequential scheme for estimating a parameter $g$ of a quantum system, where the measurements are performed sequentially to collect data $\{s_1, \ldots, s_N\}$ \textit{without resetting the state of the system every after the measurement} and the initial state $\rho_0$ can be arbitrary. Figure taken from~\cite{burgarth2015quantum}.}
    \label{fig_seq_measurements_protocol}
\end{figure}

In conventional sensing strategies, data is collected through independent and identical (IID) probability distributions. This means that after measuring the probe, one needs to reset it to its original quantum state for another round of experiments or, equivalently, use identical copies. Note that the formulation of the Cram\'{e}r-Rao inequality (see Eq.~\eqref{eq:cramer-rao-bound-single}) assumes the resetting of the probe after each measurement or equivalently using $M$ identical probes at once. A simple yet versatile sensing strategy has been devised using sequential measurements on the probe at regular time intervals~\cite{burgarth2015quantum, guta2017information, catana2015fisher, montenegro2022sequential, HMabuchi_1996, van2015sanov}. This innovative sequential sensing scheme is of significant importance in quantum metrology for several reasons: (i) it avoids extremely time-consuming resetting times, (ii) it avoids the need for highly correlated or complex initial states, and (iii) available experimental measurements suffice for evidencing quantum-enhanced sensitivity. The sequential measurement sensing scheme, see Fig.~\ref{fig_seq_measurements_protocol}, is an iterative sensing protocol:
\begin{enumerate}
    \item[(i)] A quantum probe $\rho^{(i)}(0)$ freely evolves to $\rho^{(i)}(\tau_i)$,
    
    \item[(ii)] At time $\tau_i$ a positive operator-valued measure (POVM) $\hat{\Upsilon}_{\gamma_i}=\hat{\Pi}^\dagger_{\gamma_i}\hat{\Pi}_{\gamma_i}$, where $\hat{\Upsilon}_{\gamma_i}\geq0$ and $\{\hat{\Upsilon}_{\gamma_i}\}$ are the elements of the POVM with random outcome $\gamma_i$, is performed on the probe, updating the quantum state into 
    \begin{equation}
        \rho^{(i{+}1)}(0){=}\frac{\hat{\Pi}_{\gamma_i}\rho^{(i)}(\tau_i)\hat{\Pi}_{\gamma_i}^\dagger}{p(\gamma_i)},
    \end{equation}
    where 
    \begin{equation}p(\gamma_i){=}\text{Tr}[\hat{\Pi}_{\gamma_i}\rho^{(i)}(\tau_{i})\hat{\Pi}_{\gamma_i}^\dagger],
    \end{equation}
    is the probability associated to $\gamma_i$ at step $i$,
    
    \item[(iii)] The outcome $\gamma_i$ is recorded and the new initial state $\rho^{(i{+}1)}(0)$ is replaced in (i),
    
    \item[(iv)] The above steps are repeated until $n_\mathrm{seq}$ measurements outcomes are consecutively obtained,
    
    \item[(v)] After gathering a data sequence $\pmb{\gamma}{=}(\gamma_1,{\cdots},\gamma_{n_\mathrm{seq}})$, the probe is reset to $\rho_0$ and the process is repeated to generate a new trajectory.
\end{enumerate}
In general, the data collected from sequential measurements are non-IID, meaning that the measurement data are not independent of each other---i.e., one deals with effectively different probes at each measurement step. It is essential to emphasize that, unlike conventional sensing schemes, there is no need for specific maximally entangled probe, the need for supporting quantum phases, feedback mechanisms, or quantum control. Naturally, standard sensing scheme reduces to the particular case of $n_\mathrm{seq}{=}1$. 

By harnessing the sequential measurements followed by free evolution at regular time intervals, the sequential sensing scheme has led to significant advancements, including Hamiltonian identification~\cite{HMabuchi_1996}, sequential measurements sensing schemes~\cite{li2023fisher, oconnor2024fisher, PhysRevA.99.022102, yang2024sequential, de2017estimating, clark2019quantum, rasbi2022jumps, burgarth2015quantum, catana2015fisher, guctua2011fisher, PhysRevA.96.012316, Ritboon_2022, bompais2023asymptotic, bompais2022parameter, proceedings2019012011, PhysRevA.64.042105, PhysRevA.94.042322, nagali2012testing, PhysRevA.92.032124}. 

Quantum-enhanced magnetometry powered by sequential measurements sensing in quantum many-body probes has been investigated for short number of sequential measurement sequences. An unknown local magnetic field $\bm{B}{=}(B_x{,}0{,}B_z)$ is aimed to be estimated using $N$ interacting spin$-1/2$ particles with Heisenberg Hamiltonian~\cite{montenegro2022sequential}
\begin{equation}
\hat{H}=-J\sum_{j=1}^{N-1}\hat{\boldsymbol{\sigma}}_j \cdot \hat{\boldsymbol{\sigma}}_{j+1} + B_x \hat{\sigma}_1^x + B_z \hat{\sigma}_1^z,\label{eq:hamiltonian-single}
\end{equation}
where $\hat{\boldsymbol{\sigma}}_j{=}(\hat{\sigma}_j^x{,}\hat{\sigma}_j^y{,}\hat{\sigma}_j^z)$ represents a vector composed of Pauli matrices acting on qubit at site $j$, $J$ denotes the exchange interaction, and $\bm{B}{=}(B_x{,}0{,}B_z)$ stands for the local magnetic field to be estimated located at first site. For the sake of simplicity, a ferromagnetic state $|\psi(0)\rangle = |\downarrow\downarrow\downarrow\ldots\rangle$ is probed at site $N$ in the $\hat{\sigma}_z$ basis at regular time intervals $J\tau = N$. The latter allows enough time for the quantum correlations to spread throughout the spin chain. It has been shown~\cite{montenegro2022sequential} that a \textit{synchronization-like} magnetization dynamics occurs between the first and last sites. This opens up the possibility of extracting information about the local magnetic field (located at the first site) by observing the dynamics of the readout qubit (located at the last $N$ site). Remote sensing holds special significance in biology, where non-invasive sampling of organic tissues is desirable to protect the biological material (see quantum illumination techniques~\cite{albarelli2020perspective,Casariego2022bifrequency,reichert2023quantum} and networks~\cite{zhao2025quantum}).

From a practical perspective, both measuring and initializing the probe at each sequential step consume time, which is a crucial resource in actual experiments. Sequential measurements metrology accounts for these technical constraints straightforwardly by considering the total interrogation time as follows:
\begin{equation}
t_\mathrm{tot}=M(t_\mathrm{init}+t_\mathrm{evo}+t_\mathrm{meas}n_\mathrm{seq}),\label{eq:total-time}
\end{equation}
where $t_\mathrm{init}$, $t_\mathrm{evo}$, and $t_\mathrm{meas}$ are the initialization, evolution, and measurement times, respectively. Typically, $t_\mathrm{init} {\gg} t_\mathrm{meas} {\gg} t_\mathrm{evo}$. Eq.~\eqref{eq:total-time} shows that for a fixed total time $t_\mathrm{tot}$, increasing the number of sequences $n_\mathrm{seq}$ reduces the number of sequential trials $M$. However, the total number of measurements $M\times n_\mathrm{seq}$ increases with an increase in $n_\mathrm{seq}$ for a fixed time $t_\mathrm{tot}$. By introducing the dimensionless average squared relative error given by: $\delta B_x^2=\int f(\widetilde{B_x}|\bm{\Gamma})(|\widetilde{B_x}-B_x|/|B_x|)^2d\widetilde{B_x}$, where $f(\widetilde{B_x}|\bm{\Gamma})$ is the posterior distribution of obtaining $B_x$ given the observed data $\bm{\Gamma} = {\pmb{\gamma}_1,\pmb{\gamma}_2,\cdots,\pmb{\gamma}_M}$, where each sequential run $\pmb{\gamma}_k$ contains $n_\mathrm{seq}$ spin outcomes, and $|\widetilde{B_x}-B_x|/|B_x|$ corresponds to the relative error of the estimation. $\delta B_x^2$ accounts for both the uncertainty and the bias in the estimation simultaneously. For short number of sequential measurements, it has been shown that a suitable fitting function of the form $\overline{\delta B_x^2} \sim t_\mathrm{tot}^{-\nu} n_\mathrm{seq}^{-\beta}$ with $\nu=1$ and $\beta > 1$, evidences quantum-enhanced sensitivity for estimating $B_x$ with respect to the number of sequential measurements $n_\mathrm{seq}$ and a given total protocol time $t_\mathrm{tot}$. Hence, one can consistently achieve higher sensitivity by employing sequential projective measurements on a probe subsystem for a finite and experimental-friendly sequence of measurements~\cite{montenegro2022sequential}.

\vm{Nonetheless, sequential measurement sensing schemes rapidly become limited to short measurement sequences due to the exponential growth of measurement outcomes with the number of sequences---$(2^N)^q$ for $N$ measurements on $q$ qubits.} Thus, evaluating all probabilities distributions becomes intractable. For estimating sensing precision with a large number of sequential measurements, indirect approaches have been proposed, such as functional analysis of the measurement outcomes, showing a Fisher information increasing linearly with the number of sequential measurements $n_\mathrm{seq}$~\cite{burgarth2015quantum}, namely $I(\theta) \simeq n_\mathrm{seq}$, or via correlated stochastic processes~\cite{Radaelli_2023}. Moreover, single-trajectory based sensing~\cite{gambetta2001state, HMabuchi_1996} utilizing maximum-likelihood estimators~\cite{7403443} (with convergence proof~\cite{bompais2022parameter}) has recently been proposed. This approach demonstrates that a single sequential run (a trajectory) with $n_\mathrm{seq}{\gg}1$ is enough to achieve estimation with arbitrary precision. Recently, using a Monte Carlo methodology, a comprehensive analysis of sequential measurements metrology for $n_{\mathrm{seq}} \gg 1$ has been presented. This analysis evaluates the increment in Fisher information as~\cite{yang2023extractable} \vm{(for clarity of presentation, we have omitted the unknown parameter to be estimated, that is $I^{(n_\mathrm{seq})}(\theta){:=}I^{(n_\mathrm{seq})}$):}
\begin{equation}
I^{(n)}{=}I^{(n{-}1)}{+}{\Delta}I^{(n)};\hspace{0.5cm}\Delta I^{(n)}{:=}\sum_{\pmb{\gamma}} P_{\pmb{\gamma}} f^{\pmb{\gamma},(n)},\label{eq_joint-cfi-us}
\end{equation}
where $I^{(j)}$ represents the CFI at step $j$, $\Delta I^{(n)}$ is the increment of the CFI after performing one more measurement following the recording of $(n{-}1)$ measurements, and $f^{\pmb{\gamma},(n)}(\theta){:=}f^{\pmb{\gamma},(n)}$ is the CFI obtained from the $n$-th measurement $p(\gamma_n)$ in trajectory $\pmb{\gamma}$. Utilizing this approach, one can examine the behavior of the CFI for arbitrarily large numbers of sequential measurements. Conducting local projections on a quantum many-body probe leads to probe's finite memory, attributed to repeated local projections and free evolution, resulting in a rank-1 matrix~\cite{yang2023extractable}. This probe's finite memory is linked to the transition from super-linear to linear scaling of CFI. After losing memory of an early state, CFI scales linearly with $n_\mathrm{seq}$, limiting the extractable information capacity of the quantum probe in sequential measurements. 

\begin{figure}
    \centering
    \includegraphics[width=\linewidth]{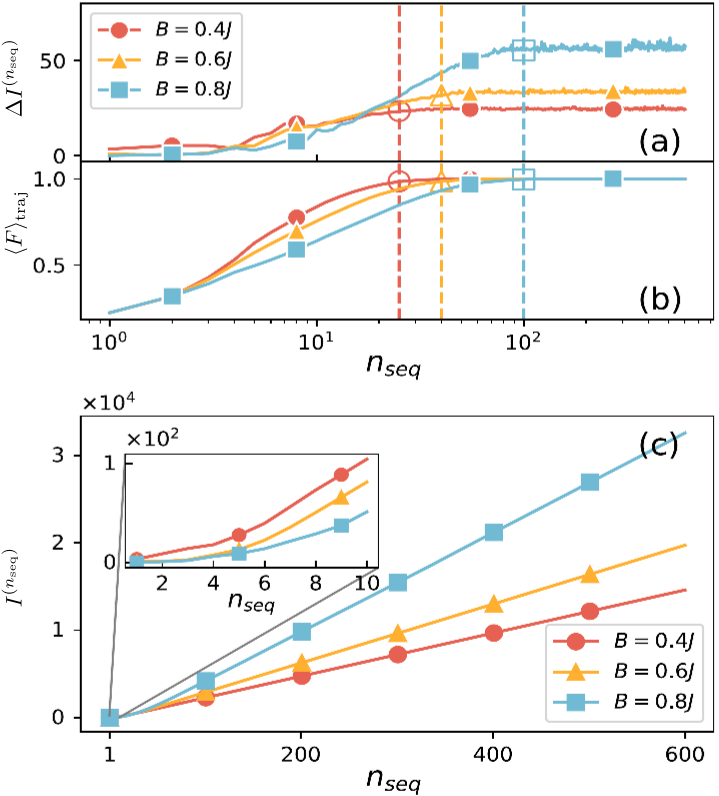}
    \caption{\textbf{Sequential measurements metrology for the quantum-many body probe of Eq.~\eqref{eq:hamiltonian-single} and large number of consecutive measurements \boldsymbol{$n_\mathrm{seq}$}:} \boldsymbol{$(a)$} CFI increment $\Delta I^{(n_\mathrm{seq})}$ as a function of $n_\mathrm{seq}$, \boldsymbol{$(b)$} Fidelity averaged over $10^4$ different trajectories (each trajectory evaluates the fidelity between two distinct initial states following the same trajectory) as a function of $n_\mathrm{seq}$, \boldsymbol{$(c)$} CFI $I^{(n_\mathrm{seq})}$ as a function of $n_{\mathrm{seq}}$. Figure taken from~\cite{yang2023extractable}.}
    \label{fig_seq_large_nseq}
\end{figure}

For the quantum many-body probe described in Eq.~\eqref{eq:hamiltonian-single}, Figs.~\ref{fig_seq_large_nseq}(a)-(b) demonstrate a correspondence between the transition from super-linear to linear scaling of CFI and the probe's finite memory. Specifically, the CFI increment $\Delta I^{(n_\mathrm{seq})}$ becomes nearly constant at around the same $n_{\mathrm{seq}}$ where the averaged fidelity $\langle F \rangle_{\mathrm{traj}}$ also becomes nearly constant. This averaged fidelity measures the similarity between two distinct initial states following the same trajectory~\cite{yang2023extractable} (a single trajectory is composed of $n_\mathrm{seq}$ measurement steps). In Fig.~\ref{fig_seq_large_nseq}(c), the CFI $I^{(n_\mathrm{seq})}$ is plotted as a function of $n_{\mathrm{seq}}$. This plot clearly shows a transition from non-linear to linear behavior, in agreement with the quantum-enhanced sensing shown for short sequential measurements~\cite{montenegro2022sequential} and the linear scaling with indirect methods for large $n_\mathrm{seq}$~\cite{burgarth2014exponential}. The Monte Carlo-based methodology discussed above for arbitrarily large $n_{\mathrm{seq}}$ it has been addressed for both quantum many-body probes and light-matter probes~\cite{yang2023extractable}.

\vm{In yet another sensing application, sequential measurement sensing directly addresses a fundamental challenge in quantum sensing and metrology: it overcomes the singularity of the QFIM, thereby enabling the simultaneous estimation of multiple parameters~\cite{yang2025overcoming}. As discussed in Sec.~\ref{sec:elements-multi}, in multi-parameter quantum sensing, achieving optimal estimation for multiple parameters simultaneously is generally not possible due to the inherent non-commutativity of quantum mechanics~\cite{lu2021incorporating, carollo2019quantumness, di2022multiparameter}. As a result, the matrix inequality $\text{Cov}[\tilde{\pmb{\Theta}}] \geq I(\pmb{\theta})^{-1} \geq I_Q(\pmb{\theta})^{-1}$ from Eq.~\eqref{eq:quantum-cramer-rao-multi} is not always saturable~\cite{szczykulska2016multi, ragy2016compatibility, lu2021incorporating, albarelli2020perspective, albarelli2023fisher, carollo2019quantumness, paris2009quantum, demkowicz2020multiparameter, di2022multiparameter, belliardo2021incompatibility}. Furthermore, the Fisher information matrices often become singular, meaning their inverses do not exist~\cite{namkung2024unifiedcramerraoboundquantum, liu2019quantum, mihailescu2024multiparameter, candeloro2024dimension}.}

\vm{While existing approaches address QFIM singularity using the Moore-Penrose pseudoinverse of the CFIM~\cite{namkung2024unifiedcramerraoboundquantum} or by formulating the Cram\'{e}r-Rao bound as an unconstrained quadratic optimization problem~\cite{stoica2001parameter}, sequential measurement strategies provide a direct and scalable solution. Indeed, by implementing a sequential measurement protocol with minimal system control~\cite{burgarth2015quantum, montenegro2022sequential, yang2023extractable, bompais2022parameter, clark2019quantum, ma2018phase, benoist2019invariant, benoist2023limit, block2022measurement, skinner2019measurement}, it has been shown that one can always ensure the invertibility of the CFI matrix simply by increasing the length of the measurement sequence~\cite{yang2025overcoming}. This approach is highly efficient, as the invertibility condition of the CFI matrix improves exponentially with the sequence length.}

\vm{In summary, as presented throughout this section, sequential measurement metrology addresses several key challenges in quantum sensing and metrology. Notably, (i) it eliminates the need for preparing highly entangled quantum states~\cite{burgarth2015quantum}; (ii) allows for system evolution (whether unitary or governed by open quantum dynamics~\cite{yang2024sequential}) with minimal control~\cite{montenegro2022sequential}; (iii) relies solely on local, experimentally feasible measurements~\cite{montenegro2022sequential}; and (iv) it enables the simultaneous estimation of multiple parameters by overcoming the QFIM singularity~\cite{yang2025overcoming}.}

\vm{One challenge, however, is the practical difficulty of implementing a large number of sequential measurements in an experiment. Yet, as discussed, excessively long measurement sequences would inevitably degrade the encoded information in the quantum probe, making such an approach impractical~\cite{yang2023extractable, burgarth2015quantum}. Despite the above, experimental progress in sequential measurement techniques has been already pursued. For instance, in 2012, Nagali et al.~\cite{nagali2012testing} experimentally demonstrated a sequential measurement protocol involving up to two consecutive measurements by exploiting different degrees of freedom of a single photon. Their experimental setup used a Sagnac interferometer with a polarizing beam splitter. Furthermore, using an adaptive strategy, they investigated the optimal trade-off between the maximum extractable information and the disturbance introduced to the system~\cite{nagali2012testing}.}

\vm{As a final note, sequential measurement sensing may align well with the framework of weak measurements~\cite{aharonov1988how, duck1989thesense}, which has already been shown to enhance metrological capabilities~\cite{zhang2015precision, yin2021improving, alves2015weak} as well as play a significant role in various other contexts~\cite{rozema2012violation, mundarain2016quantumness, feizpour2011amplifying, lundeed2009experimental, coto2017thepower}. In fact, Pfender et al.~\cite{pfender2019high} demonstrated that weak measurement-based techniques enable the use of measurement-correlation schemes for detecting extremely weakly coupled single-spin signals. By employing weak measurements, they achieved nuclear magnetic resonance on a $^{13}$C nuclear spin at room temperature with a spectral resolution of 3.8 Hz---with the spin undergoing a quantum dynamical phase transition from coherent trapping to coherent oscillation. Similarly, Wang et al.~\cite{wang2019characterization} characterized correlations in a quantum bath to arbitrary order using a weak measurement protocol. Meinel et al.~\cite{meinel2022quantum}, exploiting weak measurements in a nitrogen-vacancy center in diamond, demonstrated that quantum nonlinear spectroscopy can extract arbitrary types and orders of correlations in a quantum system. As shown above, the potential connection between sequential measurement metrology and weak measurements might create experimental opportunities to enhance metrological capabilities.}

\section{Global Estimation Theory: Parameter Estimation with Minimal Prior Information}\label{sec:global-sensing}

An implicit assumption underlies all preceding discussions: the assumption that sufficient information about both the unknown parameter of interest and the control parameters is available beforehand. This scenario is known as \textit{local estimation theory}~\cite{Casella98,vantrees2004detection,helstrom1976quantum,Hajek1970, LeCam-1986, Janssen2013Convolution, Inagaki1970}, where the unknown parameter varies within a very narrow interval. It is important to note that the limitations of local estimation theory have been identified in a wide range of contexts. This is because the Cram\'{e}r-Rao inequality of Eq.~\eqref{eq:cramer-rao-bound-single} presents some subtleties, including: (i) although the bound performs excellently for unbiased estimators, a typical unbiased estimator, such as the maximum likelihood, is sometimes infeasible due to the large data set needed to saturate the bound asymptotically~\cite{1523369} (see also Rubio et al.~\cite{rubio2019quantum, rubio2020bayesian} for metrology with limited data). Furthermore, in practice, most estimators are biased~\cite{bj/1186078362}; (ii) regularity conditions of the Cram\'{e}r-Rao bound are hard to verify~\cite{Borovkov1984}; and (iii) a limiting variance may not coincide with the variance of a limiting distribution~\cite{bj/1186078362}. 

In contrast to local estimation theory, \textit{\vm{global estimation} theory} applies when no (or some) prior information is available. Recently, the impact of uncertainty in control parameters on the sensitivity of critical sensors has been explored~\cite{mihailescu2024uncertain}. Therefore, in these situations the unknown parameter varies within a broader sensing interval. Consequently, the choice of measurement to be performed on the probe and the analysis of the estimation data should be optimal on average, that is it should be equally effective for any possible value of the parameter~\cite{helstrom1976quantum,holevo2011probabilistic}.

One of the earliest studies on \vm{global estimation} theory, aimed at overcoming the limitations of local estimation theory, was proposed by van Trees~\cite{vantrees1968, bj/1186078362} (a stronger type of inequality is also provided by Klaassen~\cite{KLAASSEN1989267}). The van Trees inequality addresses the comprehensive concept of \vm{global estimation}, where the unknown parameter $\lambda$ varies according to some prior distribution $z(\lambda)$, determining that the average variance is given by the Bayesian Cram\'{e}r-Rao bound or posterior Cram\'{e}r-Rao bound~\cite{vantrees1968, paris2009quantum}:
\begin{equation}
\overline{\mathrm{Var}[\lambda]}=\int z(\lambda) \left[\hat{\Lambda}(x) - \lambda \right]^2 dxd\lambda \geq \frac{1}{Z_F},    
\end{equation}
where $\hat{\Lambda}(x)$ is the estimator of $\lambda$ (mapped from the measurement outcomes $x$), and $Z_F$ can be demonstrated to be
\begin{equation}
    Z_F = \int z(\lambda)I(\lambda)d\lambda + M\int z(\lambda)[\partial_\lambda \mathrm{log}z(\lambda)]^2 d\lambda,
\end{equation}
where the first term accounts for the average of the CFI over the prior distribution $z(\lambda)$ and the second term is the CFI of the prior distribution itself~\cite{vantrees1968}. A quantum bound can also be derived~\cite{paris2009quantum} by substituting $I(\lambda)$ with $I_Q(\lambda)$. A very narrow prior distribution reduces $Z_F$ to the local CFI bound, i.e., $\int z(\lambda)I(\lambda)d\lambda \rightarrow I(\lambda)$. Examples of van Trees \vm{global estimation} are: collisional thermometry (where a series of ancillas is sent sequentially to probe the system’s temperature) utilizing Bayesian inference, and the significance of prior information using the modified Cram\'{e}r-Rao bound associated with van Trees and Sch\'{u}tzenberger~\cite{alves2022bayesian}; Bayesian thermometry approach based on the concept of thermodynamic length, applicable in the regime of non-negligible prior temperature uncertainty and limited measurement data, exemplified using a probe of non-interacting thermal spin-1/2 particles~\cite{jorgensen2022bayesian}; ultimate bounds in Bayesian thermometry approach for arbitrary interactions and measurement schemes (including adaptive protocols) are explored. Notably, a derivation of a no-go theorem for non-adaptive protocols that does not allow for better than linear (shot-noise-like) scaling, even having access to arbitrary many-body interactions, is presented in Ref.~~\cite{mehboudi2022fundamental}. A strict hierarchy of \vm{global estimation} protocols was recently established linking them with local strategies~\cite{zhou2024strict}.

Note that, in general, van Trees bound is not tight and cannot be saturated. To achieve a tighter bound a systematic and general metric evaluating the performance of \vm{global estimation} and its significance for quantum many-body probes with criticality has been examined in Ref.~\cite{montenegro2021global}. The \vm{global estimation} procedure involves considering a modified quantum Cram\'{e}r-Rao bound, which introduces the average uncertainty of the estimation as follows:
\begin{equation}
\overline{\delta \lambda^2} := \int_{\Delta \lambda} \delta \lambda^2 z(\lambda) d\lambda \geq g(\boldsymbol{B}) := \int_{\Delta \lambda} \frac{z(\lambda)}{M I_Q(\lambda|\boldsymbol{B})}d\lambda, \label{eq:g-single}
\end{equation}
where $z(\lambda)$ is the prior distribution of the unknown parameter $\lambda$ to be sensed, $M$ is the number of measurements, $I_Q(\lambda|\boldsymbol{B})$ is the QFI, $\boldsymbol{B}{=}(B_1,B_2,\ldots)$ are external tunable parameters interacting with the probe, and the unknown parameter varies over a sensing interval 
\begin{equation}
\Delta \lambda \in [\lambda^{\mathrm{min}}, \lambda^{\mathrm{max}}], \hspace{0.5cm} \text{with} \hspace{0.5cm} \lambda^{\mathrm{cen}}:=\frac{\lambda^\mathrm{min} + \lambda^\mathrm{max}}{2}.
\end{equation}
The minimization of the right-hand side of Eq.~\eqref{eq:g-single} with respect to the control parameters $\boldsymbol{B}$ defines the figure of merit for determining the optimal probe, namely:
\begin{equation}
g(\boldsymbol{B}^*){:=}\min_{\boldsymbol{B}} \left[ g(\boldsymbol{B}) \right].\label{eq_global_g_montenegro}
\end{equation}
Note that, in general, the optimal measurement basis varies across $\Delta \lambda$, and no measurement setup can saturate the modified global Cram\'{e}r-Rao bound of Eq.~\eqref{eq:g-single} over the entire interval. Note that it has been proven that the van Trees inequality is always smaller than the mean classical Cram\'{e}r-Rao lower bound, that is~\cite{2647}: $\overline{\mathrm{Var}[\lambda]} \geq \int z(\lambda) \frac{1}{I(\lambda)} d\lambda \geq \frac{1}{Z_F}$. This indeed shows that the average uncertainty $g$ is a tighter bound than the one given by the Van Trees.

\begin{figure}[t]
\centering \includegraphics[width=\linewidth]{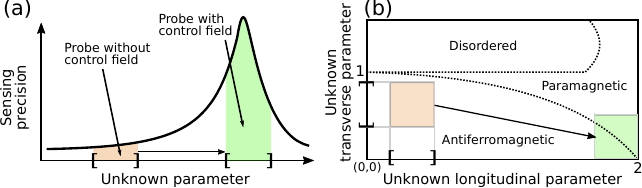}
\caption{\textbf{Sketch of phase diagram and the role of \vm{global estimation}:} \boldsymbol{$(a)$} The \vm{global estimation} scheme tunes the quantum probe to operate optimally on average. This procedure ensures that the control fields will optimize the probe to exploit the critical point. \boldsymbol{$(b)$} As the phase diagram becomes more complex, the probe's offset becomes non-trivial. In general, the minimization of $g(\boldsymbol{B})$ (see Eq.~\eqref{eq:g-single}) systematically optimizes the probe to exploit criticality across the entire phase diagram. Figure taken from~\cite{montenegro2021global}.}\label{fig:model-offset}
\end{figure}

Eq.~\eqref{eq_global_g_montenegro} has been evaluated for the transverse Ising quantum many-body probe:
\begin{equation}
\hat{H} = J\sum_{i=1}^{L} \hat{\sigma}_{i}^{x} \hat{\sigma}_{i+1}^{x} - \sum_{i=1}^{L}(B_z + \lambda_z)\hat{\sigma}_{i}^{z},\label{eq:hamiltonian}
\end{equation}
where $L$ is the system size, $\hat{\sigma}_{i}^{x,y,z}$ represents the Pauli operator at site $i$, $J>0$ denotes the exchange interaction, $B_z$ corresponds to the controllable magnetic field that can be adjusted, $\lambda_z$ stands for the field to be estimated, and periodic boundary conditions are imposed. Assuming that the unknown parameter varies over a wide region $\Delta \lambda_z$, the protocol introduced in Ref.~\cite{montenegro2021global} systematically optimizes the quantum many-body probe to deliver the best sensitivity performance on average. In the case of single-parameter estimation, the control field $B_z$ acts as an offset of the critical point, see Fig.~\ref{fig:model-offset}(a). However, when the phase diagram becomes more complex or in the presence of multi-parameter estimation, such an offset becomes non-trivial, see Fig.~\ref{fig:model-offset}(b).

\begin{figure}[t]
\centering \includegraphics[width=\linewidth]{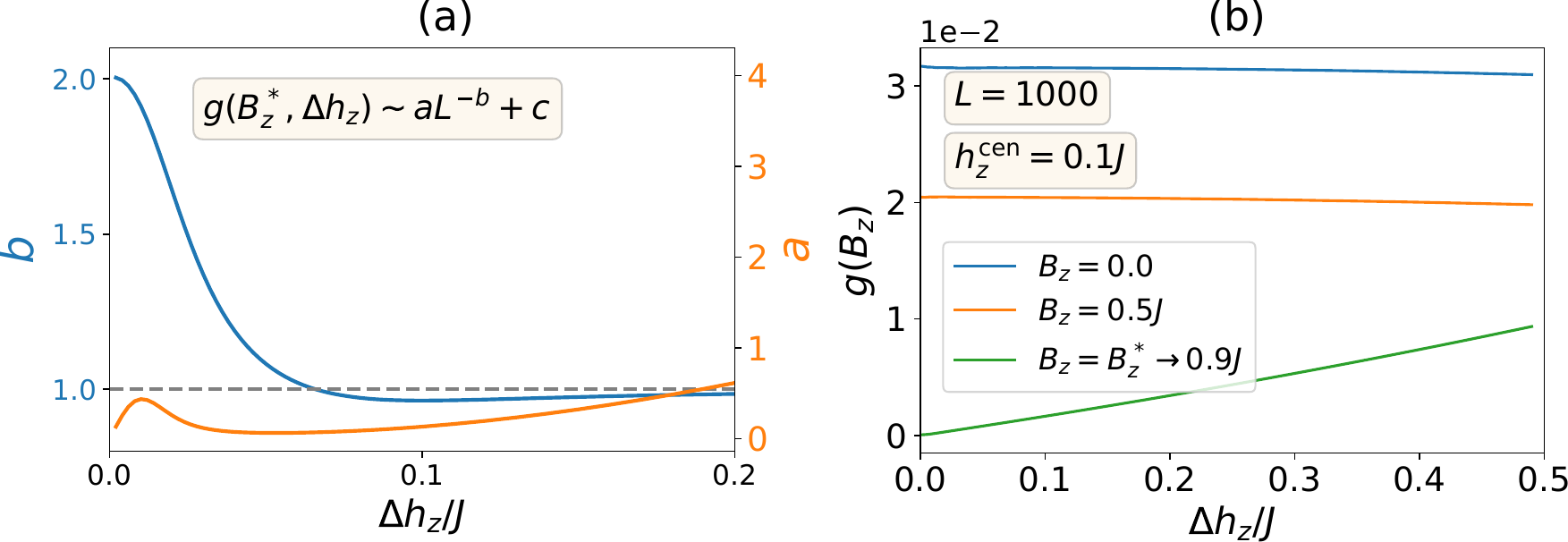}
\caption{\textbf{\vm{Global estimation} protocol for transverse Ising quantum many-body probe} \boldsymbol{$(a)$} Fitting coefficients $a$ and $b$ as a function of $\Delta \lambda_z/J$. A smooth transition from the Heisenberg limit ($b=2$) to the standard quantum limit ($b=1$) is observed. Notably, for a prior distribution as wide as $0\leq \Delta \lambda_z/J \lesssim 0.07$, the global sensor still delivers quantum-enhanced sensitivity ($b>1$). \boldsymbol{$(b)$} Minimizing the average uncertainty $g(B_z)$ always results in better sensitivity when compared to a probe without applying control fields. Figure taken from~\cite{montenegro2021global}.}\label{fig:global-VM-scaling}
\end{figure}

By defining the size of the probe $L$ as the relevant sensing resource, it has been demonstrated that the QFI scales super-linearly with $L$ over a moderate wide interval of $\Delta\lambda_z$~\cite{montenegro2021global}. Fig.~\ref{fig:global-VM-scaling}(a) shows the fitting coefficients $a$ and $b$ obtained from the fitting function $g(B_z^*,\Delta \lambda_z)\sim aL^{-b}+c$, as a function of the interval $\Delta \lambda_z/J$. Here, $\Delta \lambda_z/J {\rightarrow} 0$ represents \vm{the case of local estimation sensing, where much prior information about the unknown parameter is at hand.} As the figure illustrates, the $b$ coefficient decreases from the (expected) Heisenberg limit ($b=2$) to the standard quantum limit ($b=1$) as the interval $\Delta\lambda_z$ increases. The \vm{global estimation} protocol still demonstrates quantum-enhanced sensitivity for sensing intervals $0 \leq \Delta\lambda_z \lesssim 0.07J$. In Fig.~\ref{fig:global-VM-scaling}(b), even under the seemingly standard quantum limit (i.e., $b = 1$), the \vm{global estimation} strategy still delivers superior sensing precision compared to a non-optimized probe. This superiority arises from the \vm{global estimation} strategy, which ensures probe optimization by minimizing $g(\boldsymbol{B})$ over the prior distribution.

Thermometry, the process of estimating the temperature of a sample, plays a crucial role in quantum thermodynamics~\cite{campbell2017global,de2018quantum, zhang2022approaching} and has various applications, such as determining macroscopic objects in their ground state~\cite{chan2011laser, montenegro2020mechanical}. In the context of previously discussed equilibrium quantum sensing, it has been demonstrated that finite temperature can surprisingly enhance the performance of critical quantum metrology protocols~\cite{ostermann2024temperature}. In addition, in certain scenarios, local estimation for thermometry may not be sufficient, making an alternative estimation strategy highly important. Note that the bound in Eq.~(\ref{eq:g-single}) may not be saturated as, in general, the optimal measurement setup vary for different values of the unknown parameter $\lambda$. In some cases that the optimal measurement is not a function of $\lambda$ the bound can indeed be saturated. An example of such scenario is thermometry for which energy measurement is optimal, independent of the temperature of the system. This allows to design an optimal probe for measuring temperature over an arbitrary interval. In Ref.~\cite{correa2015individual}, it is shown that optimization of the probe reduces to optimizing the energy eigenvalues. In Ref.~\cite{campbell2018precision}, this analysis was extended to investigate the spectral features required for sensitivity to multiple temperatures.  Interestingly, for the case of local thermometry, where temperature varies over a narrow interval, the optimal probe contains only two distinct eigenenergies, with a unique ground state and full degeneracy for the rest of the spectrum. The energy gap between the ground state and the  rest of the spectrum depends on the temperature~\cite{Paris_2016}and can be obtained by solving a nonlinear equation~\cite{correa2015individual}. Using the \vm{global estimation} formulation of Eq.~(\ref{eq:g-single}), one can show that the optimal probe requires more distinct energy eigenvalues as the temperature varies over a larger interval~\cite{mok2021optimal}.  In fact, through exploiting an evolutionary algorithm the authors of Ref.~\cite{mok2021optimal} find out that an effective two-level probe for local sensing turns into a three-level system with the two lowest energy eigenvalues being 
unique and then the rest of the spectrum gets fully degenerate as the temperature interval increases. By further enhancing the interval more and more energy levels are separated from the degenerate band. It is worth emphasizing that the bounds, given  by Eq.~(\ref{eq:g-single}) can be saturated for all these optimal thermometers.

\begin{figure}
    \centering
    \includegraphics[width=\linewidth]{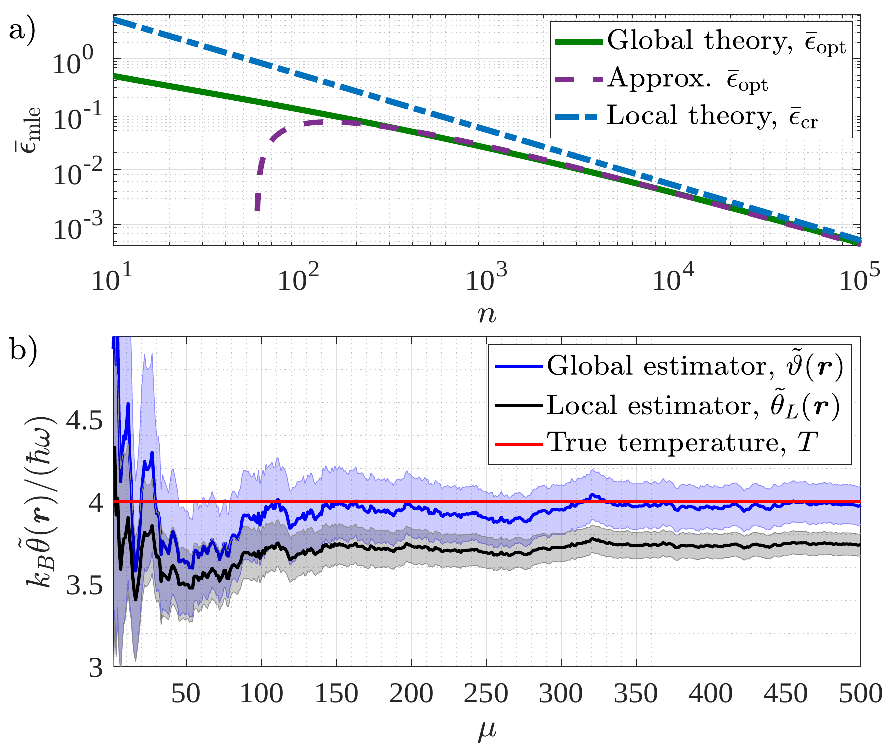}
    \caption{\textbf{\vm{Global estimation} procedure for temperature estimation:} \boldsymbol{$(a)$} Log-log plot of the global optimum $\bar{\epsilon}_\mathrm{opt}$ and local Cram\'{e}r–Rao-like bound (dot-dashed blue) for a gas of $n$ non-interacting spin-1/2 particles in thermal equilibrium. The global optimum is lower than the local bound unless $n\rightarrow\infty$, indicating that the local bound misses information for small $n$. \boldsymbol{$(b)$} Data analysis in global thermometry. The global estimate converges to the true temperature after $\mu \simeq 10^2$ shots ($\mu$ are energy measurements on the $n$-spin gas). In contrast, the local theory leads to a biased estimate even for $\mu \simeq 500$. Figure taken from~\cite{rubio2021global}.}
    \label{fig_global_thermometry}
\end{figure}

While both the van Trees and the average uncertainty approaches focus on minimizing the variance of estimation, a new formalism for global thermometry has been developed based on minimization of the mean logarithmic error. 
By exploiting scale invariance and other well-behaved properties, such as symmetric invariance and monotonicity increment (decrement) from (towards) the actual hypothesis temperature value~\cite{rubio2021global}, the mean logarithmic error $\bar{\epsilon}_\mathrm{mle}$ is defined as the global thermometry figure of merit when the prior information about the temperature is wide
\begin{equation}
\bar{\epsilon}_\mathrm{mle} = \int p(E,\theta)\mathrm{log}^2\left[ \frac{\tilde{\Theta}(E)}{\theta}\right] dE d\theta.
\end{equation}
In the above, $E$ is the energy, $\theta$ is the hypothesis for the true temperature $T$, $p(E,\theta)=p(E|\theta)p(\theta)$ their joint probability, and $\tilde{\Theta}(E)$ the optimal estimator of $T$ given by
\begin{equation}
\frac{k_\mathrm{B}\tilde{\Theta}(E)}{\epsilon_0}=\mathrm{exp}\left[\int p(\theta|E)\log\left(\frac{k_\mathrm{B}\theta}{\epsilon_0}\right) d\theta\right],
\end{equation}
with $k_\mathrm{B}$ the Boltzmann constant, $\epsilon_0$ an arbitrary constant with energy units, and $p(\theta|E)$ the posterior function given by the Bayes rule $p(\theta|E)=p(E|\theta)p(\theta)/p(E)$. Notably, the mean logarithmic error $\bar{\epsilon}_\mathrm{mle}$ applies to both biased and unbiased estimators. Furthermore, to observe the failure of local thermometry and the performance of the global quantum thermometer, refer to the example using the mean logarithmic error for a non-interacting gas of $n$ spin-$1/2$ particles at thermal equilibrium in Ref.~\cite{rubio2021global}, see Fig.~\ref{fig_global_thermometry}.

\section{Optimal Control and Machine Learning Strategies for Quantum Sensing}
 
Let us recall that the ultimate precision attainable through any quantum sensing algorithm depends on four steps (see Fig.~\ref{fig_parameter_estimation_steps}): \textit{(i) initialization} of the probe, \textit{(ii) evolution} of the probe under the influence of the system's (known and unknown) parameters, subjected to external controls, 
\textit{(iii) extraction} of information from the probe's state through a suitably chosen measurement, and \textit{(iv) estimation} of the unknown parameter value from the extracted measurement statistics. We have so far reviewed results where these steps were all assumed to be perfectly achievable. However, for practical sensing tasks, three valuable resources are inevitably limited: \textit{(i) limited probe size} available for sensing, \textit{(ii) finite coherence timescale} of the probe available for initialization and evolution, and \textit{(iii) finite measurement time} leading to degradation of measurement accuracy and precision with each round. Attempts to lift these restrictions have been partially discussed in Secs.~\ref{sec:metrology-decoherence-A}-\ref{sec:noisy_metrology}-\ref{sec_seq_measurements}. 
In general, performing a constrained optimization over all these valuable yet limited resources is analytically intractable and numerically demanding in all but the simplest cases. So far, broadly two approaches have been successfully employed in this direction, which we briefly review here. 

\subsection{Optimal Control Based Approaches} Quantum optimal control theory (QOCT) is now an area of extensive literature. We refer the interested reader to the review in Ref.~\cite{koch2022quantum} for a more detailed treatment of QOCT in other areas of quantum technology. For sensing purposes within QOCT approach, an extra control Hamiltonian is added to the probe's evolution. If the probe evolves through a Hamiltonian $\hat{H}_\theta$ containing information about the unknown parameter $\theta$, then the time-dependent total Hamiltonian is given by \begin{equation}
    \hat{H}_{tot}(t) = \hat{H}_{\theta} + \hat{H}_{c} (\textbf{u}(t), t),
\end{equation}
where the control Hamiltonian $\hat{H}_{c}$ is a function of control parameters $\textbf{u}(t)$. The goal is to design the control Hamiltonian $\hat{H}_c (\textbf{u}(t), t)$ in such a way that maximum precision in estimation of the parameter $\theta$ is achieved. This can be broadly accomplished in two ways: (quasi)analytically and numerically. In terms of control strategies, broadly two distinct strategies are possible, viz., open-loop control and closed-loop, i.e., feedback control. 

\subsubsection{Open-loop QOCT strategies} The simplest and oldest systematic approach to QOCT is to apply Pontryagin's Maximum Principle (PMP), which builds on the calculus of variations to derive a set of necessary conditions for optimal control, see Ref.~\cite{boscain2021introduction} for a detailed introduction to PMP in quantum settings. PMP introduces an auxiliary system described in terms of a costate $|\mu (t)\rangle$ which must satisfy the original evolution equation backwards in time, i.e., $i\hbar \frac{|d\mu (t)\rangle}{dt} = -\hat{H}_{tot}(t) |\mu(t)\rangle$ with the same boundary conditions. Now if the task is to optimize a cost-function $\mathcal{C}$, which can be generally written as a sum of two cost functions, i.e., $\mathcal{C} = \mathcal{C}_1 + \mathcal{C}_2$, with $\mathcal{C}_1$ being a cost function in terms of some suitable path-dependent quantity integrated over the trajectory of evolution (say, the average QFI for sensing tasks) and $\mathcal{C}_2$ being a terminal cost function (say, the fidelity achieved with a target state etc); then PMP puts the necessary condition for the control parameters being optimized, i.e., $ \textbf{u}(t) = \textbf{u}^{\text{opt}}(t)$, as the optimization over the following function for every time $t$ and the quantum state $|\psi (t)\rangle$ along the original evolution equation

\begin{equation}
    \textbf{u}^{\text{opt}}(t) = \arg \max_{\textbf{u}(t)} \left[\langle \mu(t)|\hat{H}_{tot}|\psi(t)\rangle - \mathcal{C} \right] .
    \label{eq:PMP}
\end{equation}

For very simple problems, this optimization is sometimes possible to perform analytically. \textcolor{black}{In Ref.~\cite{lin2021optimal}, the authors consider such a problem with the \emph{twist and turn} angular momentum Hamiltonian $\hat{H} = \chi \hat{J}_{z}^2 + \omega \hat{J}_z + \Omega(t) \hat{J}_x$, where the unknown parameter $\omega$ is augmented with a control Hamiltonian $\Omega(t) \hat{J}_x$, which is also the \emph{turn} term. Ref. \cite{lin2021optimal} then finds the optimal trajectory of $\Omega(t)$ by maximizing the QFI of the final state with respect to $\omega$ as the cost function}. However, for more complicated problems like those involving many-body interactions, analytically solving Eq.~\eqref{eq:PMP} is no longer presumed to be viable. Thus numerical algorithms must be adopted. Popular gradient-based optimization algorithms for quantum control include \textit{gradient ascent pulse engineering} (GRAPE)~\cite{khaneja2005optimal} and Krotov's method~\cite{reich2012monotonically}. In the former, updates are made simultaneously for all times $t$, while in the latter updates are made sequentially. However, for exponentially growing Hilbert space dimensions in generic many-body systems, gradient-based algorithms are highly limited. Moreover, in noisy environments gradient updates are highly unreliable in the first place. Thus gradient-free optimization algorithms are often required. Primary approach along this direction is to Fourier-decompose the control pulse sequence into a smaller number of bases. Randomly sampling a small number of bases via \textit{chopped random basis} (CRAB)~\cite{caneva2011chopped} thus avoids the dimensionality problem for many-body systems. 

\subsubsection{Closed-loop QOCT strategies}

The second approach is the so called \textit{feedback-based} strategies, where measurement results obtained in each round are fed into the quantum control for the next round. Pang and Jordan~\cite{pang2017optimal} considered a phase estimation problem for a parameter $\theta$ imprinted into a time-dependent Hamiltonian $\hat{H}(\theta, t)$. The goal is to optimize the QFI along the entire trajectory. To do so, they noted that QFI  at each instant $t$ is upper bounded by the following integral 
\begin{equation}
    I_Q (\theta) \leq \left[ \int_{0}^{t} \left(\nu_{\max} (\tau) - \nu_{\min} (\tau) \right) d\tau \right]^2, 
    \label{eq:pangjordan}
\end{equation}
where $\nu_{\max} (\tau)$ and $\nu_{\min} (\tau)$ are respectively maximum and minimum eigenvalues of $\partial_{\theta} \hat{H}(\theta,\tau)$ at any time $\tau$. Thus, Pang and Jordan employed an additional control Hamiltonian $\hat{H}_{c}(t)$ to steer the trajectory towards the eigenspace of  $\partial_{\theta} \hat{H}(\theta,t)$ as much as possible. The crucial aspect is that the control Hamiltonian is independent of the actual parameter $\theta$, but written in terms of $\tilde{\theta}$, a known prior estimate of $\theta$  which is updated at each round. As an illustration, they considered a qubit in a uniformly rotating magnetic field $\textbf{B}(t) = B (\cos \omega t, \sin \omega t,0) $, where the goal is to estimate $B$ or the field rotation frequency $\omega$. For estimating $\omega$ from evolution up to time $t$, their approach furnishes a QFI $I_{Q} (\omega) = B^2 t^4$. In contrast, in the absence of the control Hamiltonian, QFI scales only quadratically with time. This result indicates that adaptive control strategies are eminently useful for quantum sensing. Yang \emph{et al} in Ref.~\cite{yang2020probe} addressed a slightly different question, preparation of the optimal probe for maximum precision via adaptive feedback control. They used purity loss instead of QFI for ease of direct measurement. Their approach circumvents the problem of exponentially growing Hilbert spaces by avoiding any computation involving matrix representation of the  time-independent Hamiltonian $\hat{H}$. In particular use for many-body sensing, they considered a spin-chain probe with the control being limited to a sequence of local rotations and even the rotation strength can be adjusted only a fixed number of times. Their results, experimentally implemented on an NMR platform, indicate that Heisenberg scaling is possible even with such a severe limitation on the control. 

Until now, we assumed that the control itself is precisely known beforehand. This assumption can be relaxed by modelling the control with an extra unknown nuisance parameter, say $\theta'$, in contrast to the parameter $\theta$ of interest. Using Eq.~\eqref{eq:Cramer_Rao_Multi_weight} with the weight matrix $\mathcal{W} {=} \text{diag}(1,0)$, we can immediately derive that the variance of our estimation for the parameter of interest, namely  $\theta$, should satisfy
\begin{equation}
    \delta\theta^2\ge \frac{1}{I_{\theta\theta}\left(1-\frac{I_{\theta\theta'}I_{\theta'\theta}}{I_{\theta\theta}I_{\theta'\theta'}} \right)}.
\end{equation}
This clearly shows that the imperfection of control induces a separate error in estimation. The specifics of this correction will depend on the details of the imperfect control, e.g., whether it originates from calibration error or noise. It will be an interesting question to consider on a case-by-case basis whether this noise outweighs any potential advantage of the many-body approach discussed in this review vis-a-vis interferometric approaches. 

\subsection{Machine Learning Based Approaches}

 The ongoing revolution in machine learning has already had a significant impact in the study of many-body quantum physics~\cite{carleo2017solving, gray2018machine, vicentini2021machine}. These range from identification of various exotic phases of matter in hitherto analytically challenging models~\cite{carrasquilla2017machine, greplova2020unsupervised}, to Hamiltonian learning~\cite{wang2017experimental}, and investigation of foundational issues like Bell-nonlocality~\cite{deng2018machine}. Machine-learning algorithms, which learn from previously acquired data, also assist in solving optimization problems that arise in quantum sensing. For a general overview of machine-learning assisted quantum metrology, we refer the interested reader to the recent reviews by Huo \emph{et al}~\cite{huo2022machine} and Huang \emph{et al}~\cite{huang2024quantum}. As we describe below, quantum sensing has already been shown as an especially fruitful area of application for machine learning methods.

\subsubsection{Classical machine learning strategies }

Critical state preparation is of practical relevance for quantum sensing in many-body systems, where the enhanced sensitivity of critical states to external fields is leveraged to perform more precise measurements. \textit{Reinforcement learning} (RL) methods have started showing significant promise in this task, since it does not assume any prior knowledge and only has a cumulative reward function which it seeks to optimize. RL was first employed to optimize over driving protocols to suitably initialize a nonintegrable quantum many-body probe in minimum time and high fidelity by Bukov \emph{et al} in  Ref.~\cite{bukov2018reinforcement}. This was specialised for matrix product states (MPS) by Metz and Bukov in Ref.~\cite{metz2023self}, which shows that RL can adapt to control protocols in real time in the presence of stochastic perturbations as well. Meanwhile, RL has also identified nonlinear measurements as another resource for state preparation in optical cavities~\cite{porotti2022deep}. For the estimation part, RL was used for general time-dependent sensing problems by Xu et al~\cite{xu2019generalizable} where the reward function is the QFI given by the control, and by  Xiao \emph{et al}~\cite{xiao2022parameter}, where they employ a reward function in terms of the ultimate QFI bound as well as design a general reward function. Control pulse engineering via a RL algorithm, viz., the Asynchronous Advantage Actor-Critic (A3C) algorithm, has been demonstrated for atomic spin ensembles~\cite{qiu2022efficient}, resulting in attainment of Heisenberg limit. In yet another approach, Cimini \emph{et al}~\cite{cimini2023deep} employed RL to train on the experimental data to learn Bayesian updates in a multiparameter setting. These general approaches show RL is competitive with GRAPE or other optimal control methods while being more resource-efficient, robust, and not being restricted by veracity of prior assumptions. Thus RL in particular holds much promise for being adapted specifically to many-body sensing probes in future.

Apart from RL, one can exploit machine learning as an estimation algorithm. In Ref.~\cite{nolan2021frequentist}, the authors explore regression for inferring a machine-learning estimation of an unknown parameter. The method is necessarily
frequentist, i.e.~relying on repeated estimates to build up statistics, and eventually converges to the maximum likelihood Bayesian estimation algorithm. In a similar approach, 
parameter estimation can be considered as a sequence of classification tasks, for which one can use artificial neural networks to efficiently perform Bayesian estimation~\cite{nolan2021machine}.

\subsubsection{Quantum machine learning strategies}
The above approaches seek to solve the quantum metrological control problem through classical machine-learning algorithms. Thus, a natural question arises: can we use recent advances in \emph{quantum} machine-learning algorithms for improved sensing? Variational quantum algorithms, seeking to optimize a cost function via parametric quantum circuits, is perhaps the most important of these advances. Variational quantum metrology (VQM) is a direct offshoot of these algorithms. The generic recipe for VQM in many-body systems is to build up the Hamiltonian metrological evolution as a sequence of Trotter steps represented by the quantum circuits and then consider a QFI-like cost function to optimize over. Koczor et al.~\cite{koczor2020variational} used VQM to search for the optimal probe state for an interferometry task. Similar works were reported in Refs.~\cite{ma2021adaptive, maclellan2024end} and carried over to the multiparameter setting~\citep{kaubruegger2023optimal, le2023variational}, with a complete treatment available in Ref.~\cite{meyer2021variational} and experimental photonic implementation in Ref.~\cite{cimini2024variational}. Experimentally, programmable quantum sensors based on VQM has been demonstrated for ion-trap~\cite{marciniak2022optimal}, superconducting qubit~\cite{gong2023quantum}, and Rydberg interaction based neutral atomic array~\cite{kaubruegger2019variational} platforms, with concrete tasks like building atomic clocks~\cite{kaubruegger2021quantumvar} promising to become very important. Specialized computational packages like MetQuan are also becoming available~\cite{sinha2024metquan} recently. 

{\color{black} 
\section{Experimental Realizations} \label{sec:Experimental_Realization}

The main aim of this article is to review the theoretical advancements of quantum many-body probes, as experimental developments have been explored in previous review articles, e.g. see Ref.~\cite{degen2017quantum}. Therefore, we keep this section very brief, mainly hinting the physical platforms in which quantum many-body probes are developed. 

Quantum sensors have been developed across a diverse range of physical platforms, each harnessing unique properties of these systems. This diversity enables versatility in probe size, spanning atomic to microscales, and operational temperatures, ranging from ultra-cold atoms and ions to room temperature. These sensors have been used for detecting electric, magnetic, and gravitational fields, as well as measuring force, acceleration, and mass. Each platform offers distinct advantages, making them suitable for a wide array of sensing applications—from fundamental scientific research to practical technologies in navigation, medical imaging, and environmental monitoring. The key physical platforms for implementing quantum sensors include: ultra-cold atoms~\cite{kasevich1992measurement, peters1999measurement, bongs2019taking, el2020aedge, stray2022quantum, peters2001high, fixler2007atom, tino2021testing, bloom2014optical, hinkley2013atomic, takamoto2005optical}, ion traps~\cite{leibfried2004toward, maiwald2009stylus, biercuk2010ultrasensitive, sawyer2012spectroscopy, brownnutt2015ion, baumgart2016ultrasensitive, baumgart2016ultrasensitive, huntemann2016single}, atomic vapors~\cite{budker2007optical, kominis2003subfemtotesla, dang2010ultrahigh, balabas2010polarized, shah2007subpicotesla, fernholz2008spin, wasilewski2010quantum, xia2006magnetoencephalography}, nuclear magnetic resonance devices~\cite{kitching2011atomic, fang2012advances,Jiang2021search, wang2022limitsonaxions, wu2022enhanced, jiang2024long}, solid state defects such as nitrogen vacancies in diamond~\cite{taylor2008high, clevenson2015broadband, jensen2014cavity, le2012efficient, chernobrod2005spin, balasubramanian2008nanoscale, dolde2011electric, zhao2012sensing, barson2017nanomechanical, wu2021nanodiamond, vetter2022zero, zhou2014quantum, xie2020dissipative, schirhagl2014nitrogen,patel2020subnanotesla}, superconducting circuits~\cite{bal2012ultrasensitive, danilin2018quantum, wang2019heisenberg, yu2025experimental}, photonic systems~\cite{holland1993interferometric, mitchell2004super, pezze2007phase, higgins2007entanglement, nagata2007beating, ono2013entanglement, Xiao2024Non}, and optomechanical devices~\cite{hu2024picotesla, gavartin2012ahybrid, liu2022roomtemperature, westerveld2021sensitive, sansa2020optomechanical,  li2018characterization, fardianmelamed2025infrared,  fogliano2021ultrasensitive, pikovski2012probing, chowdhury2023membrane, gosling2024sensing, forstner2012cavity, mccormick2019quantum, aspelmeyer2014cavity, moser2013ultrasensitive, guzman2014high, krause2012high, chaste2012nanomechanical, forstner2014ultrasensitive, degen2009nanoscale, rugar2004single, li2021cavity, liu2021progress, xia2023entanglement}.
For a comprehensive review on experimental implementation of quantum sensors, one can see Ref.~\cite{degen2017quantum}. 

Thanks to recent advances on quantum technologies implementing quantum many-body probes are now viable. Quantum criticalities have been exploited for achieving quantum-enhanced sensitivity in several experiments. Nuclear Magnetic Resonance (NMR) systems have been exploited to harness quantum criticality with Heisenberg scaling~\cite{liu2021experimental} for sensing an external magnetic field. Quantum criticality in Rydberg atoms have also been exploited for sensing electric fields~\cite{ding2022enhanced}. Furthermore, dipolar-coupled $^{13}$C nuclear spins in diamond have been employed to measure the frequency of time-varying periodic magnetic fields within a discrete time crystal framework~\cite{moon2024discrete}. By leveraging the time crystalline phase in these sensors, the probe's lifetime can be exponentially enhanced by up to three orders of magnitude, enabling highly precise and selective frequency measurements.
In a recent experiment, non-equilibrium quantum probes have also been realized in superconducting quantum processors to mitigate criticality slowing down mechanism and achieve Heisenberg scaling with respect to the encoding time~\cite{yu2025experimental}. All these experiments demonstrate  the potential of quantum many-body systems to serve as quantum probes.

}

\section{Summary and Outlook} \label{sec:Summary}

Exploiting quantum features, such as entanglement, can enhance the precision of a sensor well-beyond the capacity of a classical sensor, the so-called quantum-enhanced sensing. Such enhanced in precision has placed quantum sensing as one of the pillars of emerging quantum technologies. In addition to such enhancement, the size versatility of quantum probes allows for achieving extremely high spatial resolution \cite{ockeloen2013quantum,kose2023superresolution}. The potential applications are immense covering from biological monitoring \cite{taylor2016quantum} to mining \cite{liu2024quantumcivil}, atomic clocks \cite{katori2011optical} , navigation \cite{baili2019quantum} and space exploration \cite{kohlrus2017quantum}.

Interferometric quantum sensing has been the first demonstration of entanglement as a resource for sensing purposes. {\color{black} Some implementation of these types of sensors might be susceptible to decoherence as well as unwanted interaction between particles}. An alternative approach is to exploit quantum many-body systems for sensing. Unlike interferometry-based quantum sensors, in many-body probes interaction plays a crucial role. 
In this review, we have explored different aspects of quantum sensing using many-body systems.

Quantum many-body sensors have been used in two scenarios, namely equilibrium and non-equilibrium. In the former, quantum criticality has been identified as a resource for achieving quantum-enhanced sensitivity. There are several types of quantum criticalities, each with their own characteristics. However, in order to achieve quantum-enhanced sensitivity the criticality has to be accompanied by some sort of gap closing, for instance, Hamiltonian or Liouvillian spectral gap closing. This remarkable feature can be a useful guideline for identifying potential quantum sensors in new materials, which might be even designed by artificial intelligence \cite{krenn2023artificial}. Although criticality-based quantum sensors offer several advantages, they also come with some drawbacks. First, the region over which quantum-enhanced precision is achievable can be very narrow, requiring probe's fine tuning \cite{montenegro2021global}. Second, preparing the probe near its critical point can be very resource-consuming because the preparation process slows down significantly as it approaches criticality \cite{rams2018limits}. In order to avoid these drawbacks, non-equilibrium quantum probes have also been proposed. In such systems quantum superposition and entanglement can be generated during the dynamics and thus complex initialization is not needed. Moreover, in non-equilibrium quantum probes, the evolution time acts as an additional parameter that can be used as a resource to enhance precision. In fact, the precision is often enhanced monotonically as time goes on \cite{mishra2022integrable}. Nonetheless, unlike criticality-based sensors, in non-equilibrium probes the criteria for achieving quantum-enhanced sensitivity with respect to resources, e.g.~system size, is not well-characterized. We summarize the advantages and disadvantages of each approach in Table~\ref{table:final-summary}.

Despite current progress on many-body quantum sensing, there are still several open problems to explore. While criticality has been identified as a resource for sensing in equilibrium scenarios, the criteria which results in quantum-enhanced sensing is not specified in non-equilibrium probes. 
Another problem which requires further investigation is the performance of quantum sensors under imperfect situations, such as the presence of decoherence \cite{matsuzaki2011magnetic}. In general terms, while quantum sensors are naturally expected to be sensitive to the desired parameters, we would like them to be robust against other system's parameters or noisy imperfections. Despite several works on noisy quantum metrology~\cite{escher2011quantum,kolodynski2013efficient,rossi2020noisy,demkowicz2014using,demkowicz2017adaptive,dur2014improved,nichols2016practical,chaves2013noisy,alushi2024optimality,faist2023time,reiter2017dissipative,mihailescu2023thermometry,mihailescu2024multiparameter},  the notion of robustness has not yet been formulated quantitatively for quantum sensors. \textcolor{black}{A related approach is the use of error-correction codes for quantum sensing \cite{dur2014improved,kessler2014quantum,zhou2018achieving,unden2016quantum,zhou2020optimal,layden2019ancilla}. For 1D interacting many-body systems, presence of even minimal disorder may induce Anderson localization \cite{lahini2008anderson}, therefore such approaches may become extremely relevant for many-body sensing.} 

Furthermore, a general issue for quantum sensors arise in the multi-parameter Cram\'{e}r-Rao inequality as the bounds are not tight and thus saturating them may not be achievable  \citep{jiang2021multiparameter,rubio2020bayesian,difresco2022multiparameter,demkowicz2020multiparameter,chen2021hierarchical,di2022multiparameter,liu2020quantum,albarelli2019evaluating, mihailescu2024multiparameter,yang2024quantum,yang2025overcoming}. Developing tighter bounds and strategies towards achieving them in many-body sensors require closer connections between quantum metrology and control theory. One important problem in multi-parameter sensing is the situation where Fisher information matrix is singular and thus sensing becomes impossible \citep{mihailescu2024multiparameter,mihailescu2024uncertain,yang2025overcoming,mihailescu2025metrological,he2025scrambling,frigerio2024overcoming,bressanini2024multi,mukhopadhyay2025stepwise}. Specifying the criteria which results in singularity of the Fisher information matrix in many-body systems and recipes to fix them requires further attention. 

\textcolor{black}{Another direction that holds immense promise is that of using multiple spatially separated sensors in a distributed manner, similar to classical sensor networks~\cite{rubio2020quantum,bringewatt2021protocols,proctor2018multiparameter}. So far, this work has mostly been performed for GHZ type~\cite{zang2024quantum} and interferometric quantum sensors, see Ref.~\citep{zhang2021distributed} and references therein. However, very recently, proposals of distributed quantum sensing with many-body systems have also been put forward~\cite{hassani2025privacy} focusing on privacy, i.e., individual sensor nodes having access to limited information as well as correlated noise estimation~\cite{brady2024correlated}}. 

Finally, experimental realization of quantum many-body sensors is a big objective which ultimately defines the success of these probes. Many physical platforms can implement many-body sensors, including nuclear magnetic resonance systems, Rydberg atoms, optical lattices, superconducting simulators, nitrogen vacancies and levitated objects. 
progress on these setups as well as new physical platforms open further opportunities for developing many-body quantum sensors with wide range of applications.

\begin{acknowledgments}

AB acknowledges support from the National Natural Science Foundation of China (Grants No.~12050410253, No.~92065115, and No.~12274059), and the Ministry of Science and Technology of China (Grant No.~QNJ2021167001L). 
VM acknowledges funding by the National Natural Science Foundation of China (Grants No.~W2432005 and No.~12374482).
RY acknowledges support from the National Natural Science Foundation of China for the International Young Scientists Fund (Grant No.~12250410242).
SS acknowledges support from National Natural Science Foundation of China (Grant No.~W2433012).

\end{acknowledgments}


\bibliography{Review_MB_Probes}

\end{document}